\documentclass[10pt,a4paper]{article}
\usepackage[margin=1in]{geometry}
\pdfoutput=1
\usepackage{amsmath}
\usepackage{amsfonts}
\usepackage{multirow}
\usepackage{graphicx}
\usepackage{dsfont}

\providecommand{\keywords}[1]{\textbf{Keywords:} #1}

\begin{document}

\title{Reducing complexity and unidentifiability when modelling human atrial cells.
\vspace{0.5cm}
}

\author{
    C. Houston$^{1,2}$, 
    B. Marchand$^{2}$, 
    L. Engelbert$^{2}$, 
    C. D. Cantwell$^{1,2}$
}

\date{
\small ${^1}$ElectroCardioMaths Programme, Centre for Cardiac Engineering, Imperial College, London.\\
${^2}$Department of Aeronautics, Imperial College, London.
}

\maketitle

\begin{abstract}

Mathematical models of a cellular action potential in cardiac modelling have become increasingly complex, particularly in gating kinetics which control the opening and closing of individual ion channel currents. 
As cardiac models advance towards use in personalised medicine to inform clinical decision-making, it is critical to understand the uncertainty hidden in parameter estimates from their calibration to experimental data. 
This study applies approximate Bayesian computation to re-calibrate the gating kinetics of four ion channels in two existing human atrial cell models to their original datasets, providing a measure of uncertainty and indication of potential issues with selecting a single unique value given the available experimental data.
Two approaches are investigated to reduce the uncertainty present: re-calibrating the models to a more complete dataset and using a less complex formulation with fewer parameters to constrain. 
The re-calibrated models are inserted back into the full cell model to study the overall effect on the action potential.
The use of more complete datasets does not eliminate uncertainty present in parameter estimates. 
The less complex model, particularly for the fast sodium current, gave a better fit to experimental data alongside lower parameter uncertainty and improved computational speed.
\end{abstract}


\section{Introduction}

A central component in simulations of cardiac electrophysiology is a model of
    an action potential (AP) for a representative cardiomyocyte. These models
    describe how the transmembrane potential, and other physiological properties of a cardiac cell, vary through time due to changing  environmental conditions or applied stimuli. Since the development of the
    initial, relatively simple, model of a
    neuron~\cite{hodgkin_quantitative_1952}, AP models have grown in scope and
complexity as new experimental data have become
available~\cite{niederer_meta-analysis_2009}.

Uncertainty is an unavoidable aspect of scientific experiments, particularly in
the biological sciences. Further understanding in this area has been designated
a priority as cardiac models advance towards safety-critical
applications~\cite{mirams_uncertainty_2016}.  Uncertainty is introduced by
variability inherent in the biological system (e.g. differences in AP waveforms
between different myocytes), the stochastic nature of biological processes
(e.g.  opening and closing of ion channels) and imperfect recording systems
(e.g. noise in voltage patch clamp experiments)~\cite{mirams_uncertainty_2016}.
It is common for AP models to use equations with a large number of parameters
to describe the flow of ions across the cell membrane.
This is a particular challenge for model calibration in which parameters
are tuned to reflect experimental observations, usually by comparing model
output to experimental measurements using a regression method (e.g.
least-squares).  The consequence of the uncertainty and high number of
parameters 
is that multiple different
parameter combinations may produce the same residual to
experimental data; a unique optimum parameter set may not exist.

Traditional fitting techniques, such as simple least squares regression,
implicitly assume that a single point estimate exists for each parameter in a
model~\cite{daly_hodgkin-huxley_2015}. 
In cardiac modelling, this is unlikely to be the case for reasons outlined
above.
Bayesian methods can quantify uncertainties in parameter estimates by
determining a posterior distribution over parameter values given the available
data~\cite{johnstone_uncertainty_2016}. 
These distributions can highlight unidentifiable
parameters, those without a single unique optimum, 
and make it possible to capture
the effect of experimental and biological variability on model parameters.
Unidentifiability may be either structural, caused by an overly complex model
where parameters can be varied simulataneously without a change in model output,
or practical, where insufficient data is available to determine a parameter's
value.
In cardiac cell modelling, it can be difficult to obtain an exact likelihood function
necessary for exact Bayesian inference when summary statistics are used, as is
commonplace in studies of cardiac ion channels.
This is further complicated by the high-dimensional parameter space, nonlinear nature and
indirect observation properties of AP models~\cite{daly_comparing_2017}.
Instead, one can employ approximate Bayesian methods, such as approximate
Bayesian computation (ABC), which provide a reasonable estimate to the
posterior distributions of parameter values~\cite{sunnaker_approximate_2013}.

When developing a model of a new cell type, the
most common approach is to `inherit' gating kinetic formulations from existing
models and tune channel conductances to data from multiple
sources~\cite{fink_cardiac_2011}.
However, equations describing gating kinetics of ion channels can be extremely
complex and often contribute the majority of the parameters in AP models.  In
contrast to channel conductances, which can be adjusted to 
measurements of APs, gating kinetics are typically
calibrated to data from voltage patch clamp experiments. In these experiments,
channels are isolated, by pharmacological means or the use of
specific voltage step protocols, to take measurements of individual current traces. 
Previous studies~\cite{csercsik_identifiability_2012,sher_local_2013,walch_parameter_2016} have been valuable in developing
approaches to investigate parameter
identifiability in both generalised Hodgkin-Huxley models
and more detailed widely-used channel models. We build on this
field of work to include consideration of the available
experimental data across a range of simple and complex
channel models of a human atrial cell.

In this study, we apply an approximate Bayesian method to investigate the
uncertainty and parameter unidentifiability present in channel gating kinetics in a human atrial
cardiomyocyte.
Computational experiments are carried out on two human atrial cell models, the
Nygren~\cite{nygren_mathematical_1998} and
Courtemanche~\cite{courtemanche_ionic_1998} models, henceforth referred to as
the N and C models respectively. These were the first two biophysical models
developed to simulate the AP from a human atrial myocyte and have proved
influential in the development of subsequent models and whole-heart tissue-scale
modelling.
The N and C models are detailed cell models, each including descriptions of
twelve ion currents which contribute to the AP in human atrial myocytes. We
thus only focus our study on the four major ion currents which are %known to be
prominent determinants of their AP morphology~\cite{lombardo_systematic_2017,jaeger_detecting_2019}: %in this cell type: 
the fast sodium channel ($I_\text{Na}$), L-type calcium channel
($I_\text{CaL}$), transient outward potassium channel ($I_\text{to}$)
and ultra-rapid delayed rectifier potassium channel ($I_\text{Kur}$).

We first re-calibrate parameters in each channel model to experimental datasets
used in the original publications to investigate the existing level of
uncertainty and parameter unidentifiability. To explore whether these issues
can be alleviated from inclusion of more data, which
would suggest practical unidentifiability, a `unified' dataset is formed 
and the models re-calibrated to these
data. To investigate whether these models suffer from structural unidentifiability, a model of reduced complexity~\cite{beattie_sinusoidal_2018} is
calibrated to the same unified dataset, and parameter posterior
distributions and the overall goodness-of-fit of the model
compared to the re-calibrated N and C models.
These re-calibrated channel models are then inserted
into the full N and C cell models to study the effect of the
re-calibration on AP morphology.
We conclude by discussing the relative advantages and
drawbacks of these approaches and limitations of the study.

\section{Methods}

\subsection{Action potential models}

The AP models studied in this work follow the commonly-used Hodgkin-Huxley
gating form~\cite{hodgkin_quantitative_1952}. The changing transmembrane
voltage is calculated from the solution of several coupled ordinary
differential equations describing individual ion currents. 
Each current is of the common form

\begin{align}\label{eq:ion-current}
    \begin{split} 
        I_i &= g_i \prod_{j}\gamma_{j}^{k_j} f(V_m),\qquad\frac{d\gamma_j(t)}{dt} =
        \alpha_{\gamma_j}(V_m;\boldsymbol{\lambda})\left[1-\gamma_j\right] -
        \beta_{\gamma_j}(V_m;\boldsymbol{\lambda})\gamma_j,
    \end{split} 
\end{align}

\noindent where $g_i$ is the maximum channel conductance which scales the current amplitude
($S/F$); $\gamma_j$ are gates of the channel determined by voltage-dependent
forward and backward transition rates between open and
closed states, $\alpha$ and $\beta$, characterised by gating parameters $\boldsymbol{\lambda}$; $k_j$ is an exponent term that may be
applied to represent multiple identical gates in parallel; and $f$ is some
voltage-dependent forcing function (most commonly
the difference between $V_m$ and the ion Nernst potential).
The gating equation may
equivalently be transformed into a form explicitly specifying steady-state
values, $\gamma_\infty$ and time constants, $\tau_\gamma$

\begin{align}\label{eq:gates}
    \begin{split} 
        \frac{d\gamma(t)}{dt} =
        \frac{\gamma_\infty(V_m)-\gamma}{\tau_\gamma(V_m)},\qquad
        \tau_\gamma(V_m) &= \left[\alpha_\gamma(V_m;\boldsymbol{\lambda}) +
        \beta_\gamma(V_m;\boldsymbol{\lambda})\right]^{-1},\\
        \gamma_\infty(V_m) &= \alpha_\gamma(V_m;\boldsymbol{\lambda})
        \tau_\gamma\qquad\text{if activating gate},\\
        \gamma_\infty(V_m) &= \beta\gamma(V_m;\boldsymbol{\lambda})
        \tau_\gamma\qquad\text{if inactivating gate},\\
    \end{split} 
\end{align}

\noindent (where we omit the indexing subscript for clarity).
There are no
standard formulations for the voltage-dependent transition rates
$\alpha_\gamma(V_m;\boldsymbol{\lambda})$
and $\beta_\gamma(V_m;\boldsymbol{\lambda})$, and each model implements a
different set of
equations~\cite{nygren_mathematical_1998,courtemanche_ionic_1998}. 
For
$I_\text{Na}$, and for the C model also $I_\text{CaL}$, the structure of these equations was inherited
directly from the parent model (of a rabbit atrial cell~\cite{lindblad_model_1996} and guinea pig
ventricular cell~\cite{luo_dynamic_1994} for the N and C model respectively),
while $I_\text{to}$, $I_\text{Kur}$ and, for the N model, $I_\mathrm{CaL}$ were introduced as new formulations
in each model.
The equations are included in Section S4.
In this work, we are interested in the ability of the gating kinetics to
reflect the experimental data, and the identifiability of parameters
$\boldsymbol{\lambda}$
with respect to this data. 

The standardised formulation, henceforth referred to as the S model, is used as a relatively
simple baseline to compare to the more complex formulations in the N and C
models. In this formulation, the transition rates between open and closed states have a structure
based on free energy
arguments~\cite{tsien_transition_1969,keener_mathematical_2009}, which has been shown as sufficient
to capture the kinetics of a rapid delayed rectifier potassium current~\cite{beattie_sinusoidal_2018}.
The transition rates are given by

\begin{align}\label{eq:standardised}
\begin{split}
    \alpha(V_m) &= \lambda_1\exp(\lambda_2 V_m),\qquad
    \beta(V_m) = \lambda_3\exp(-\lambda_4 V_m),
\end{split}
\end{align}

\noindent where the parameters requiring calibration are ${\lambda_1,...,\lambda_4}$ for
each gate in the channel model. For $I_\text{Na}$ and $I_\text{CaL}$ which have
two components of inactivation, we add another inactivation gate in parallel
for the S model, which is related directly to the existing
inactivation gate by a scale parameter on its magnitude, e.g. $\tau_{\gamma 2} =
a\tau_{\gamma 1}$ where $a$ is the scale parameter. Only the activation gate of
the S model for $I_\text{Na}$ has a power of 3 applied to remain
consistent with both N and C models.

\subsection{Datasets and calibration}

Parameters underlying gating kinetics are calibrated to experimental data from
voltage patch clamp experiments conducted on isolated cardiomyocytes. Though
more complex protocols may be better able to explore the entire range of
kinetics exhibited by different ion channels~\cite{beattie_sinusoidal_2018},
the majority of available data were generated through the use of `traditional'
voltage stepping protocols.  In these experiments, the transmembrane potential
is held fixed and subsequently clamped to a series of voltage steps while the
current across the membrane is recorded. Specific features of the recorded
current
can then be calculated and summarised across different cells or experiments.

Data from voltage patch clamp experiments in human atrial myocytes for
$I_\text{Na}$~\cite{sakakibara_characterization_1992,schneider_characterization_1994},
$I_\text{CaL}$~\cite{mewes_l-type_1994,li_properties_1997,sun_mechanisms_1997},
$I_\text{to}$~\cite{shibata_contributions_1989,wang_sustained_1993,firek_outward_1995}
and $I_\text{Kur}$~\cite{wang_sustained_1993,firek_outward_1995} were digitised
(including any error measurement). A virtual voltage clamp protocol was created
to replicate \textit{in-silico} each of the \textit{in-vitro} experiments.
A full description of data sources and voltage clamp protocols 
are included in Section S1.
%Supplementary Table 1 gives a comprehensive list of data sources for each component of each channel.
The N model did not
include calibration data for any time constants in $I_\text{Na}$, for
activation time constants in $I_\text{to}$ and for deactivation of
$I_\text{Kur}$. The C model did not calibrate to voltage-dependent recovery
data in $I_\text{CaL}$ and $I_\text{Kur}$. Neither model included an activation
time constant measurement available for $I_\text{CaL}$.

We use approximate Bayesian computation (ABC) to calibrate each channel
model to the experimental data. 
ABC replaces an exact likelihood function by sampling parameter values from a
chosen prior distribution and simulating the model under the specific voltage clamp
protocol. These simulated data are processed into summary statistics
which can be compared to experimental data
using a distance function. 
The prior distribution for each parameter is set to a uniform distribution.
For the N and C models, the width of the prior is set based
on the published value of the parameter and its position in
the model structure. The width was increased if it was noted
that during calibration the distribution was being restricted
by the upper or lower prior limit. For the S model, the prior
ranges were set as previously~\cite{beattie_sinusoidal_2018,clerx_four_2019-1}.

The summary statistics are calculated from the output of a function that makes
specific measurements, for example peak current or decay rate from fitting an
exponential equation, of the current trace in response to the voltage clamp
protocol replicated from the experimental publication.
The summary statistic functions are assumed invariant to the magnitude
of current, and thus channel conductance is not included as 
a calibration parameter.
A low distance value generated by the distance
function indicates that a
particular sample from the parameter space is more likely to be from the `true'
distribution. This behaviour is captured algorithmically by using a threshold value
which is used to decide whether to accept or reject a specific sample. 
For a more detailed overview of ABC, see
e.g.~\cite{sunnaker_approximate_2013}.

An advantage of ABC is prior knowledge of
the experimental data can be embedded in the distance function during calibration.
Data from voltage-clamp experiments includes error bars to account for the different
results from experimental repeats due to observational noise and other sources of
experimental uncertainty~\cite{mirams_uncertainty_2016}. We are more certain of the value of data
points with low variance (small error bars) from the experimental data sets. 
To account for this, we use a weighted least squares distance function with weights proportional to
the inverse of the standard deviation at experimental data points. A regularisation
parameter is included to avoid divide-by-zero errors and set to $0.05$. To
avoid bias to individual experiments with more data points, this
weighting is also proportional to
the number of data points in an experiment. Further details are included in Section S2.

\subsection{Implementation}

Voltage clamp experiments were simulated using the myokit Python
library~\cite{clerx_myokit:_2016}. The ABC sampling process uses the pyABC Python
library~\cite{klinger_pyabc:_2018} to implement the Toni ABC sampler based on
sequential Monte Carlo~\cite{toni_approximate_2009}. In this sampler, the ABC
process above is repeated through a number of iterations with reducing
threshold value. Further details are included in Section S2. We
created the \textit{ionchannelABC} Python library for applying ABC in this context
which integrates pyABC and myokit for voltage patch clamp ABC calibration (see
Data Accessibility).

When comparing the relative computational speed to solve different channel models, we
apply a voltage pulse train protocol of 100 pulses (using channel-dependent
voltage steps indicated in the text). We record the time taken for a
simulation from each of the 100 samples from the posterior parameter distributions to
account for variability.

To simulate the effect on the AP of re-calibrated channels,
each new parameterisation was inserted into either the entire N or
C model. Channels were tested one at a time, and
parameters in other channels left at their published values. 100 samples were
taken from the unified posterior distribution and
a pulse train protocol applied to generate AP samples from the full
model. The pulse train consisted of 1ms current stimuli at a rate of
1 Hz and with amplitude 40 pA/pF. The AP elicited during the 100th
pulse was recorded for analysis. 

S channel models were then used in place of the published formulas to
study the effect of a reduction in complexity on the overall AP. In all cases,
the conductance of the channel was set by matching the peak current magnitude from
each sample to
the peak from the published channel model (peak current was assumed
to occur at 60mV for $I_\mathrm{to}$ and $I_\mathrm{Kur}$ models).
This experiment could not be conducted for $I_\text{Na}$
as both the N and C modellers positively shifted the steady-state curves
according to macro measurements of the AP (such as velocity of the
upstroke), and in the previous section this channel was calibrated
to the original experimental data. As in this study voltage clamp
protocols were replicated exactly as described in the experimental
publications, it is not clear how these would be adapted to fit the artificially-shifted
data, or how to manually shift the steady-state curves.

All figures display experimental
measurements as mean $\pm$ standard deviation reported in the experimental
publication. The calibrated model is displayed by taking $100$ samples from the
posterior distribution of parameters and plotting the output from simulations as median $\pm$
89\% high density posterior intervals (HDPI)~\cite{mcelreath_statistical_2018}.

\section{Results}

\subsection{Existing gating parameter uncertainties}

We first sought to study uncertainty and unidentifiability present in gating parameters of the
existing models using the datasets originally cited for
calibration~\cite{nygren_mathematical_1998,courtemanche_ionic_1998}. Only
$I_\text{Na}$ and $I_\text{to}$ channels of the C model are calibrated to the
full range of data available (Table S1). It would be expected that a higher level of
uncertainty is present in kinetics of the channel within voltage ranges that
have not been explicitly tested. A high level of uncertainty
in the parameter value is indicative of potential structural
or practical unidentifiability.

\begin{figure}[!h]
\centering\includegraphics[width=\textwidth]{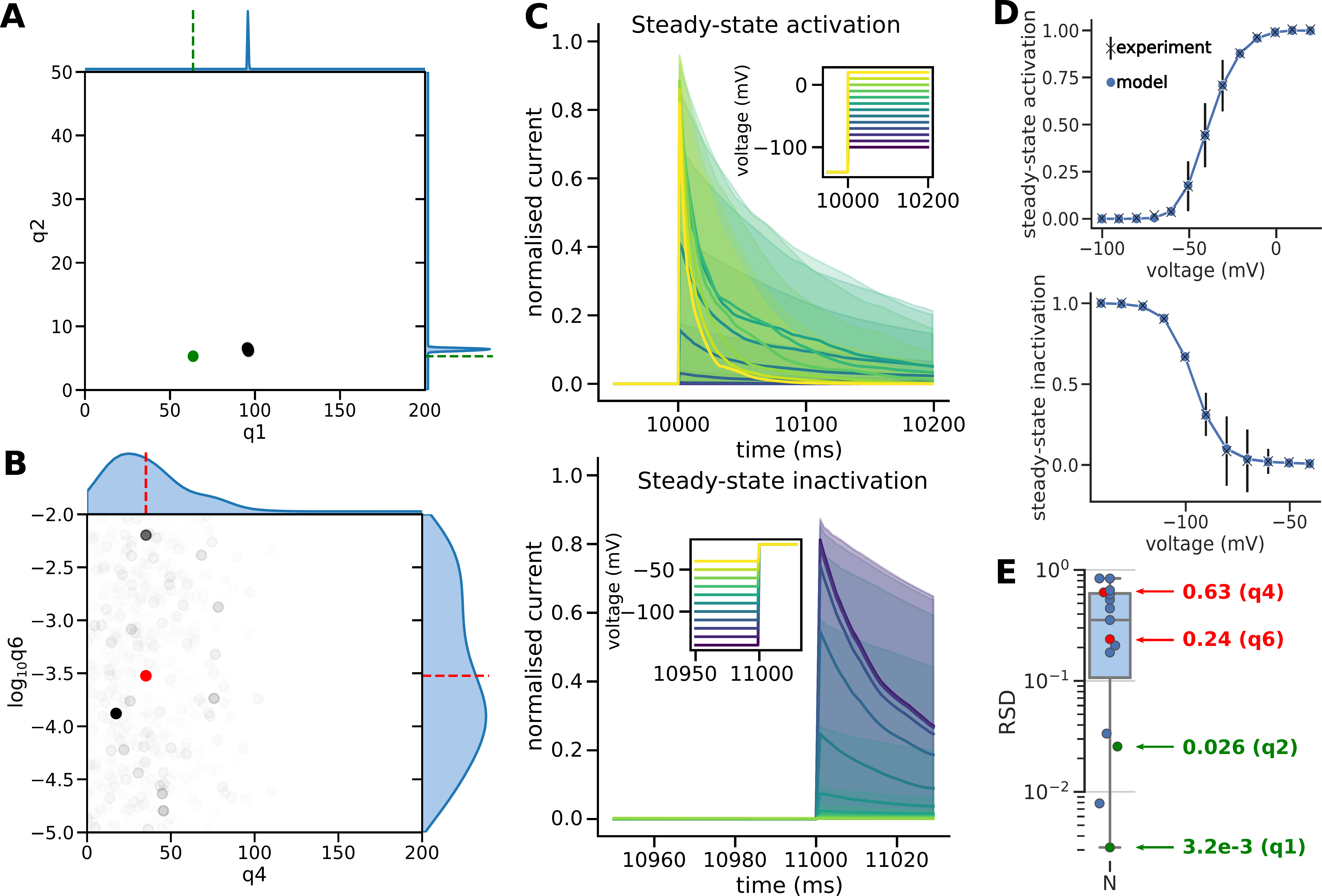}
\caption{
\textbf{A,B} Examples of parameter posterior distributions from the
    N model of $I_\text{Na}$ after calibration to the
    original dataset which only included steady-state data. Scatter plots show
    the population of posterior ABC particulars with weights indicated by
    opacity. 1D KDEs of each parameter posterior
    distribution are displayed along the corresponding axes. 
    The original parameter values from the model are
    indicated as dots and dashed lines. \textbf{C} Normalised current traces of
    posterior model response to inset voltage protocol displayed as median line with
    shading indicating 89\% HDPI from 100 samples
    from parameter posterior distributions.
    The high variability in decay rate is the result
    of no time constant data included in the dataset used
    for calibration. \textbf{D} Summary statistics generated from
    results in C. \textbf{E} RSD values of all parameter posteriors with
    representative parameters from A and B highlighted.} 
\label{fig:nyg_reported}
\end{figure}

For example, the $I_\text{Na}$ channel in the N model was only directly
compared to steady-state experimental data~\cite{sakakibara_characterization_1992}.
Figures~\ref{fig:nyg_reported}A and~\ref{fig:nyg_reported}B show representative posterior distributions and
kernel density estimates (KDEs) of those distributions following ABC calibration
for this channel. Parameters underlying steady-state components of the channel
(Figure~\ref{fig:nyg_reported}A) exhibited narrow posterior distributions
indicating they were well-constrained by the data. In contrast, parameters
underlying time constants had relatively wide posterior distributions implying
they are poorly constrained (Figure~\ref{fig:nyg_reported}B) and suggesting
there is a higher level of uncertainty surrounding their value in the model
(see Section S4.1 for equations). The effect
of the poorly constrained time constant parameters can be seen in
Figure~\ref{fig:nyg_reported}C which shows the response of the calibrated N
model of $I_\text{Na}$ to the voltage clamp protocols. The decay
rate of the current is highly variable, consistent with the observation of poorly constrained time constant parameters. Figure~\ref{fig:nyg_reported}D shows how this
uncertainty is `hidden' by the steady-state summary statistics function used to
process the traces in experiments.
Note that the non-physiological error bars of the
experimental data are a result of plotting as mean $\pm$
standard deviation.

Relative standard deviation (RSD; defined as $\sigma / |\mu|$) is a
scale invariant measure of the width of the parameter posterior distributions
and used to provide a comparison of the parameter uncertainty
between models. Higher RSD values can indicate that
a particular parameter is unidentifiable with 
respect to the model structure or available data.
Figure~\ref{fig:nyg_reported}E shows the RSD values for all
parameter posterior distributions in the calibrated N model, and highlights the
values for distributions shown in Figures~\ref{fig:nyg_reported}A
and~\ref{fig:nyg_reported}B.
The parameters with narrower posterior distributions have RSD values orders of
magnitude lower than those with wide posteriors (note the log scale on the
y-axis). In this case, the four
parameters with an RSD less than $10^{-1}$ can be
interpreted as governing the shift and steepness of the
steady-state activation and inactivation curves in
Figure~\ref{fig:nyg_reported}D, and it is thus not surprising
that these were more identifiable than parameters involved in
rise and decay rates of the current.

\subsection{Re-calibrating to a unified dataset}

Having observed a range of poorly and well-constrained parameters when
calibrating to the original datasets, we next sought to investigate the effect
of re-calibrating each model to a different `unified' dataset. This dataset is
assembled from a union of the original experimental data sources.

\begin{figure}[!h]
\centering\includegraphics[width=\textwidth]{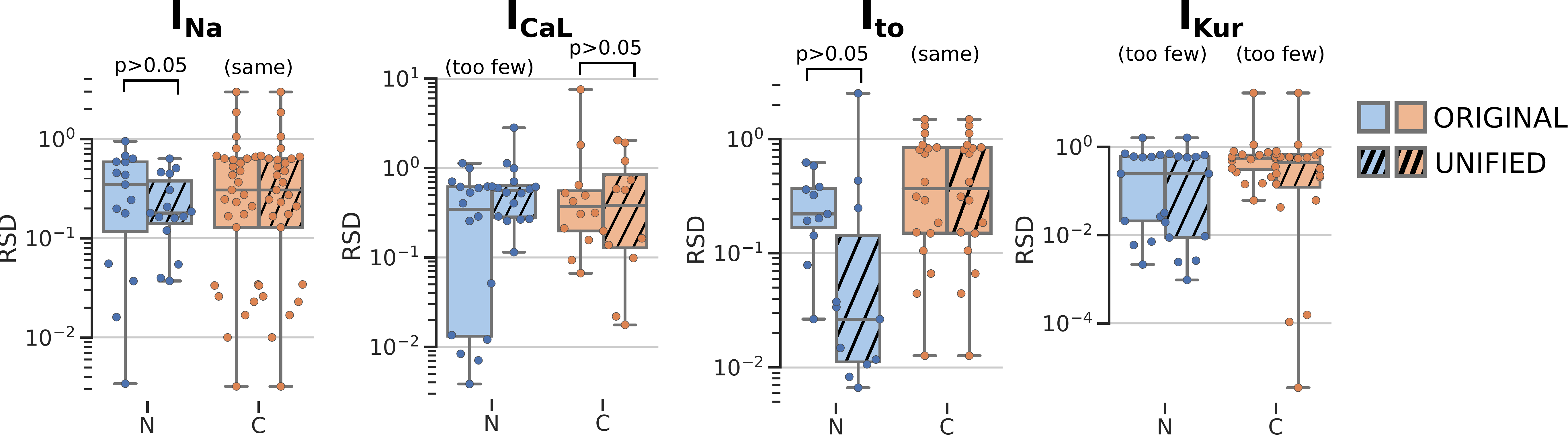}
    \caption{
    RSDs of each parameter
    in each channel model for original and unified experimental datasets in both N and
    C models. Significance tested with Wilcoxon signed rank test.
    Text `too few' above a pair of boxplots indicates that the
    sample size was too small to conduct the statistical test
    after eliminating differences of zero.}
\label{fig:summary_boxplots}
\end{figure}

Figure~\ref{fig:summary_boxplots}A shows the RSD of parameter posteriors for all
channels studied in the N and C models when calibrated to the original and unified
datasets. In all models, a wide range of RSD values is observed for the original datasets
which confirms each model has a combination of parameters which are
well-defined and parameters which are potentially unidentifiable with respect
to the given data. 
In the N model, no significant differences in the
parameter posterior RSDs were observed between original and unified dataset calibrations
for $I_\text{Na}$ and $I_\text{to}$ using a Wilcoxon signed-rank test. 
In both $I_\text{Na}$ and $I_\text{CaL}$ the minimum RSD
increased after calibrating to the unified dataset. In $I_\text{CaL}$ and
$I_\text{to}$, we also noted an increase in the maximum RSD. No significant
differences were observed for the C model. Note that a Wilcoxon signed-rank
test was not carried out for $I_\text{CaL}$ (for the N model) or $I_\text{Kur}$
as this statistical test requires discarding
differences between pairs of zero and the resulting sample size was too small
for a normal approximation. This is a result of only re-calibrating parameters
of one gate to unified data in these models as the unified data for the other
gate was the same as the original dataset (thus the RSD values of 
parameters in the other gate remains constant between the original and unified dataset).

Regions of high uncertainty in gating functions generally
corresponded to behaviour or voltage ranges not tested by
the experimental voltage-clamp protocols.
Adding the additional data in the unified dataset generally reduced the variability in these
regions, though often at the expense of other aspects of the model.
For example, the N model of $I_\mathrm{Na}$, showed high
uncertainty in the posterior estimates for time constant
functions with the original dataset, which was reduced on
calibration to the unified dataset. However, this reduction
came at the expense of greated variability in the steady-state
behaviour of the channel. A similar effect was noticed in the 
inactivation processes of the C model of $I_\mathrm{CaL}$.
For the N model of $I_\mathrm{to}$ and the inactivation of
the C model of $I_\mathrm{Kur}$, calibrating to the unified
dataset resulted in noticeable changes to the shift and steepness
of the steady-state functions. Full graphs of the posterior
gating functions are included in Section S3.1.

\subsection{Comparing to a standardised model}

No statistically significant reduction in RSD values for parameter posterior distributions
was observed when re-calibrating the N and C models to more complete datasets. 
This implies that the additional data covering a wide
range of kinetics for each channel was not sufficient to reduce unidentifiability observed in
parameters. We hypothesised that this could be due to
problems with structural unidentifiability caused by the complex
form of the equations in either model.
To test this, we next studied whether a simpler model structure of the S model could be used to reflect
the same experimental data with reduced parameter uncertainty for each channel. 

\begin{figure}[!h]
\centering\includegraphics[width=\textwidth]{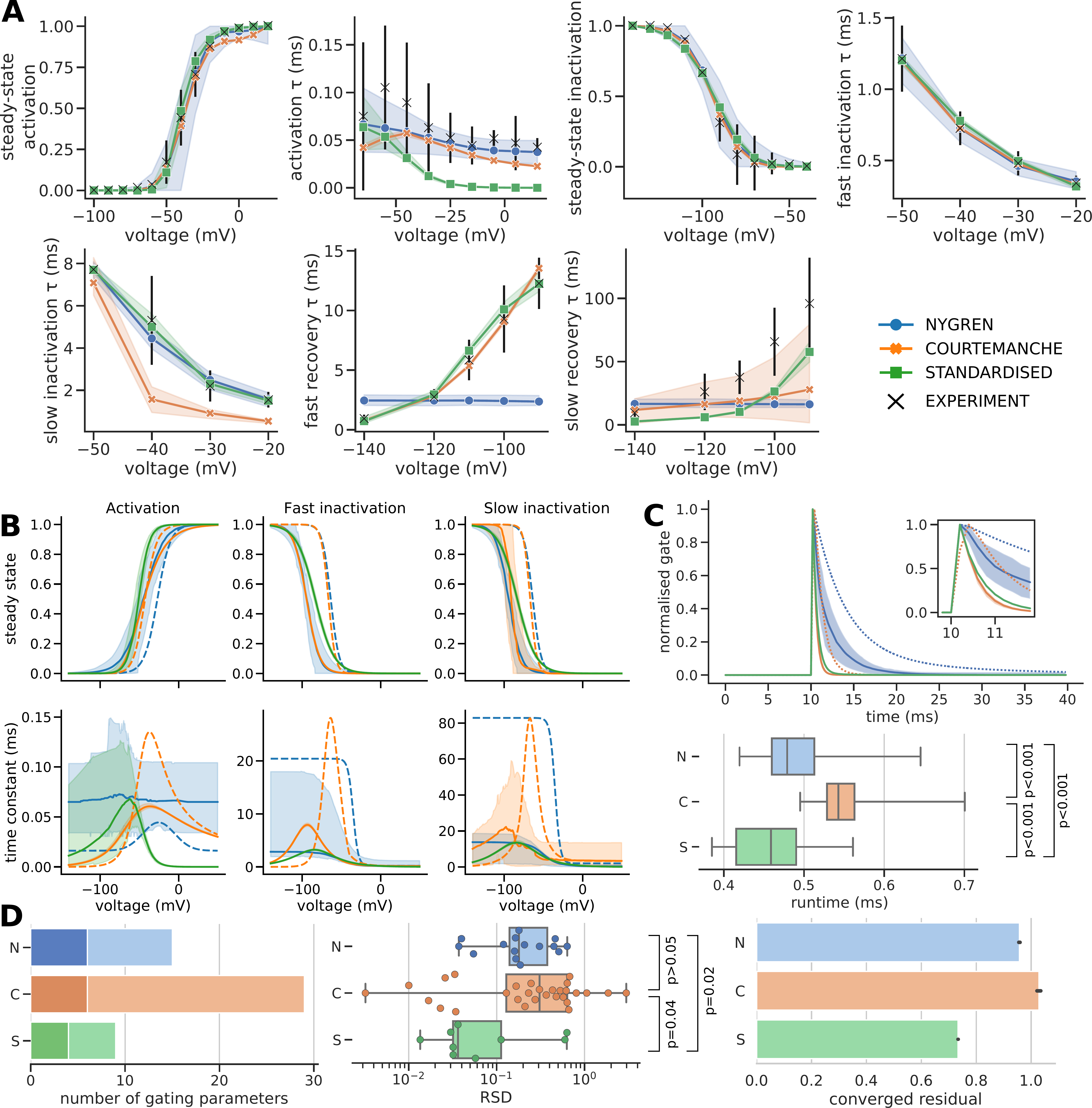}
    \caption{
    \textbf{A} Results of calibrating each $I_\text{Na}$ model to the
    unified dataset. Model output is plotted as median line and 89\%
    HDPI from 100 parameter posterior samples. Experimental data is plotted as black crosses and bars representing mean and SD.
    \textbf{B} Steady-state and time constant functions for each gate from 
    samples in A. Dashed lines indicate the published N and C
    models. \textbf{C} Example traces from each model
    generated from the last step of a pulse train of 100 steps from -140 mV to -30 mV
    for 100 ms at a rate of 1 Hz using samples from A. Dashed lines indicate 
    the published N and C models. Boxplot compares time taken to
    run the protocol for each model.
    \textbf{D} Number of gating parameters in each model
    (left). Dark and light shading correspond to activation/inactivation gating parameters respectively. 
    RSD of parameter posteriors in each model (centre). 
    Goodness of
    fit assessed by the converged residual of ABC (right). Significance 
    tested using Mann-Whitney U-test.}
\label{fig:ina_all}
\end{figure}

\textbf{Fast sodium channel.}
Figure~\ref{fig:ina_all}A compares $I_\text{Na}$ models calibrated to the unified
experimental dataset. 
The goodness-of-fit of each model can be assessed by
comparing the final converged residual obtained
during ABC, shown in the rightmost graph of Figure~\ref{fig:ina_all}D (a per-experiment version of this measure for
all channels is given in Section S3.5).
In most experiments, the S model qualitatively reflects the
experimental data to a comparable degree as the N and C models, improving
notably over the N model in recovery experiments. 
The overall converged residual of the S model is $23.5\%$ lower
than the N model and $30.0\%$ lower than the C model. It notably
deviates from experimental data in the upper voltage range of the activation time constant
where it falls too quickly towards zero.
We observe that although the S model has in total 9 parameters compared to 15
in the N model and 29 in the C model (Figure~\ref{fig:ina_all}D, left), the S model has more tightly constrained
parameter posterior distributions exhibiting lower RSD
(Figure~\ref{fig:ina_all}D, centre). 
This reduction of RSD values was statistically significant when
tested using a Mann-Whitney U-test against the N model ($p=0.02$)
and C model ($p=0.04$). 

Figure~\ref{fig:ina_all}B compares the underlying gate
functions of each $I_\text{Na}$
model.  The S model exhibits generally well-constrained behaviour other than in
the region of a gap in experimental data (deactivation data in lower voltages
of the activation gate time constant). Only the C model exhibits low
uncertainty in this region. 
In this case, the uncertainty may not be beneficial
as it could imply undue confidence in the C model's behaviour
in model space without experimental data to compare.
There is a distinct difference in the
form of the N model inactivation gate time constants which have a sigmoid shape
rather than the peaked curve exhibited by both C and S models. Comparing the
current trace of each model at the end of a pulsetrain
(Figure~\ref{fig:ina_all}C), there is little difference between the fully
calibrated traces of the C and S models, and the S model has a significantly
reduced runtime for this protocol.

\textbf{L-type calcium channel.}
$I_\text{CaL}$ has calcium- as well as voltage-dependent components of inactivation.
The N and C models differ in how they
formulate this channel; the former including two voltage-dependent inactivation
gates and the latter including a single voltage-dependent inactivation gate. Both include a single calcium-dependent
inactivation gate that was held at a constant value to isolate
the voltage-dependent features of the channel. Given the data typically
shows a fast and slow component of inactivation~\cite{li_properties_1997}, we include two voltage-dependent
inactivation gates in the S model of $I_\text{CaL}$. Thus comparisons are
more meaningful between the N and S models in this case, as the structure of
the C model differs substantially and relies more directly on intracellular
calcium concentration to modulate the rate of current decay.

\begin{figure}[!h]
\centering\includegraphics[width=\textwidth]{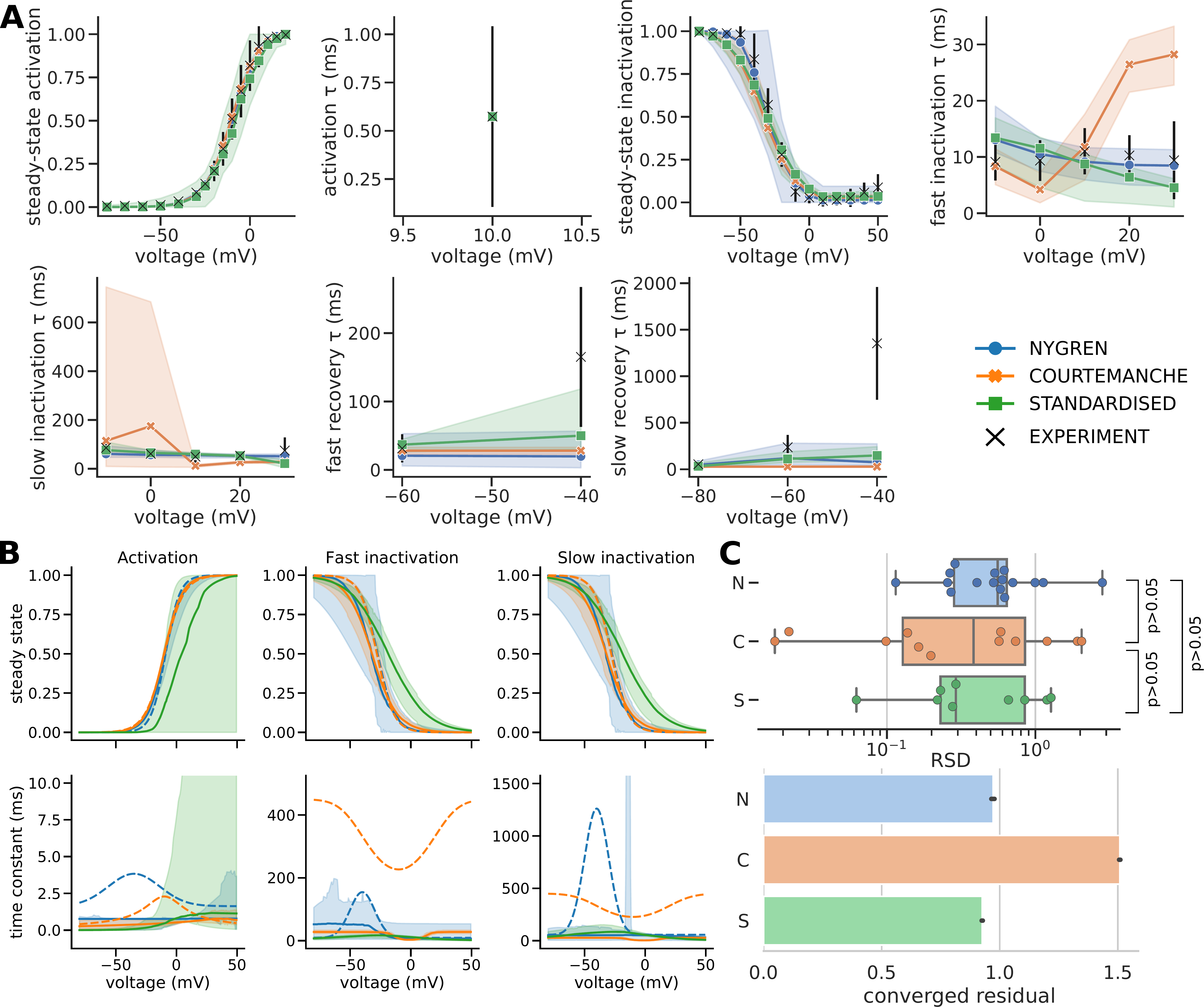}
    \caption{
    \textbf{A} Results of calibrating $I_\text{CaL}$ models to the
    unified dataset. Plotted as described in Figure~\ref{fig:ina_all}. \textbf{B}
    Steady-state and time constant functions for each gate from the samples 
    in A. Dashed lines indicate published N and C models.
    \textbf{C} 
    RSD of parameter posteriors (upper).
    Goodness of fit assessed by converged residuals from
    ABC (lower).}
\label{fig:ical_all}
\end{figure}

Figure~\ref{fig:ical_all} summarises the results for $I_\text{CaL}$. None of the models appear able to calibrate to the slowest
components of recovery from inactivation (Figure~\ref{fig:ical_all}A). The
gating functions in Figure~\ref{fig:ical_all}B show high uncertainty in the
steady-state function of the S model and differs from the N and C models at
higher voltages of the inactivation steady-state curve. Each model has reduced
uncertainty around the voltages of time constants which are explicitly tested
(inactivation and recovery $\tau$ measurements) with, particularly in the N
model, wide uncertainty outside of these ranges.
Figure~\ref{fig:ical_all}C shows there were no significant differences between RSD values
between the models, highlighting that each model suffers from
issues with parameter identifiability in some parts.
The N and S models provided better fits to the data when
assessed by the final converged residual though, as noted
above, the C model relies more heavily on the calcium-dependent
inactivation processes held constant in this experiment.

\begin{figure}[!h]
\centering\includegraphics[width=\textwidth]{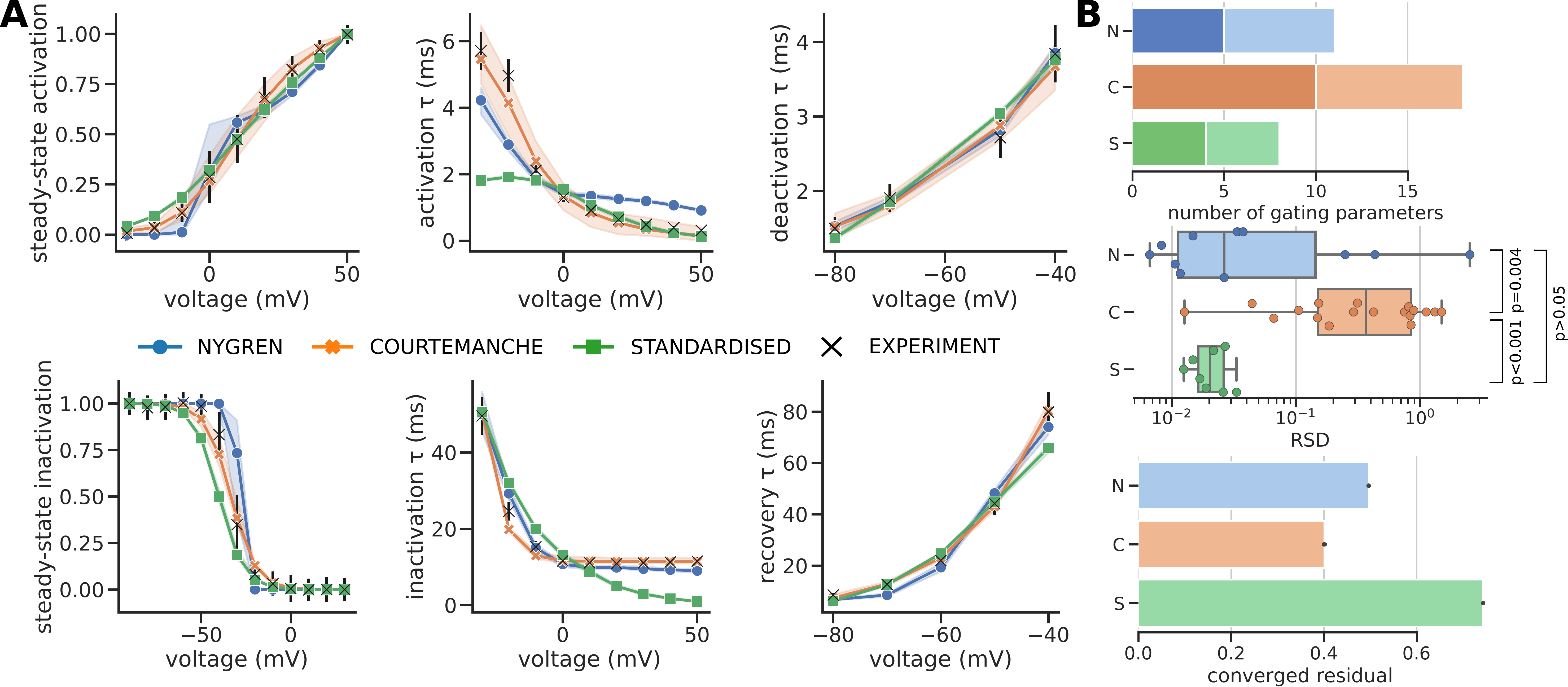}
    \caption{\textbf{A} Results of calibrating each $I_\text{to}$ model to the
    unified dataset. Plotted as described in Figure~\ref{fig:ina_all}.
    \textbf{B} Number of gating parameters (top). 
    RSD of parameter posteriors (centre). Goodness of fit
    assessed by converged residuals from ABC (bottom).}
\label{fig:ito_all}
\end{figure}

\textbf{Potassium channels.}
Figure~\ref{fig:ito_all} summarises the results from the
calibration for $I_\mathrm{to}$. Although the N and S models
both show parameter posteriors with significantly lower RSD
values than the C model, this is balanced by the lower goodness-of-fit to the
experimental data (Figure~\ref{fig:ito_all}B, lower). In particular, the
S model was unable to capture kinetics such as the plateau region of the upper voltage range for the inactivation time
constant. In contrast, the C model has the most parameters and
appears to suffer from unidentifiability in a subset of these
parameters (suggested by their high RSD values), but it also produces
the best fit to the experimental dataset.

The underlying gating
functions also reveal clear differences between
models in the midpoint and slope of steady-state activation, and peak
activation time constant (Figure S9). The peak time constants of each model are
approximately at the midpoint voltage of activation and thus also differ from
each other.

$I_\text{Kur}$ exhibits a very slow voltage-dependent component of inactivation
and only partially inactivates in the available voltage clamp experimental
data~\cite{firek_outward_1995}. As a result of these factors, each channel
model showed distinct differences when calibrated to experimental data using
the experimental voltage protocols rather than comparing gating functions to
experimental data directly. Each channel deviates from experiment data points
at lower voltage ranges of steady-state activation, and the S model converged
to a substantially different model output for the steady-state inactivation gate
(Figure S10).

\subsection{Effect on action potential}

\begin{figure}[!h]
\centering\includegraphics[width=\textwidth]{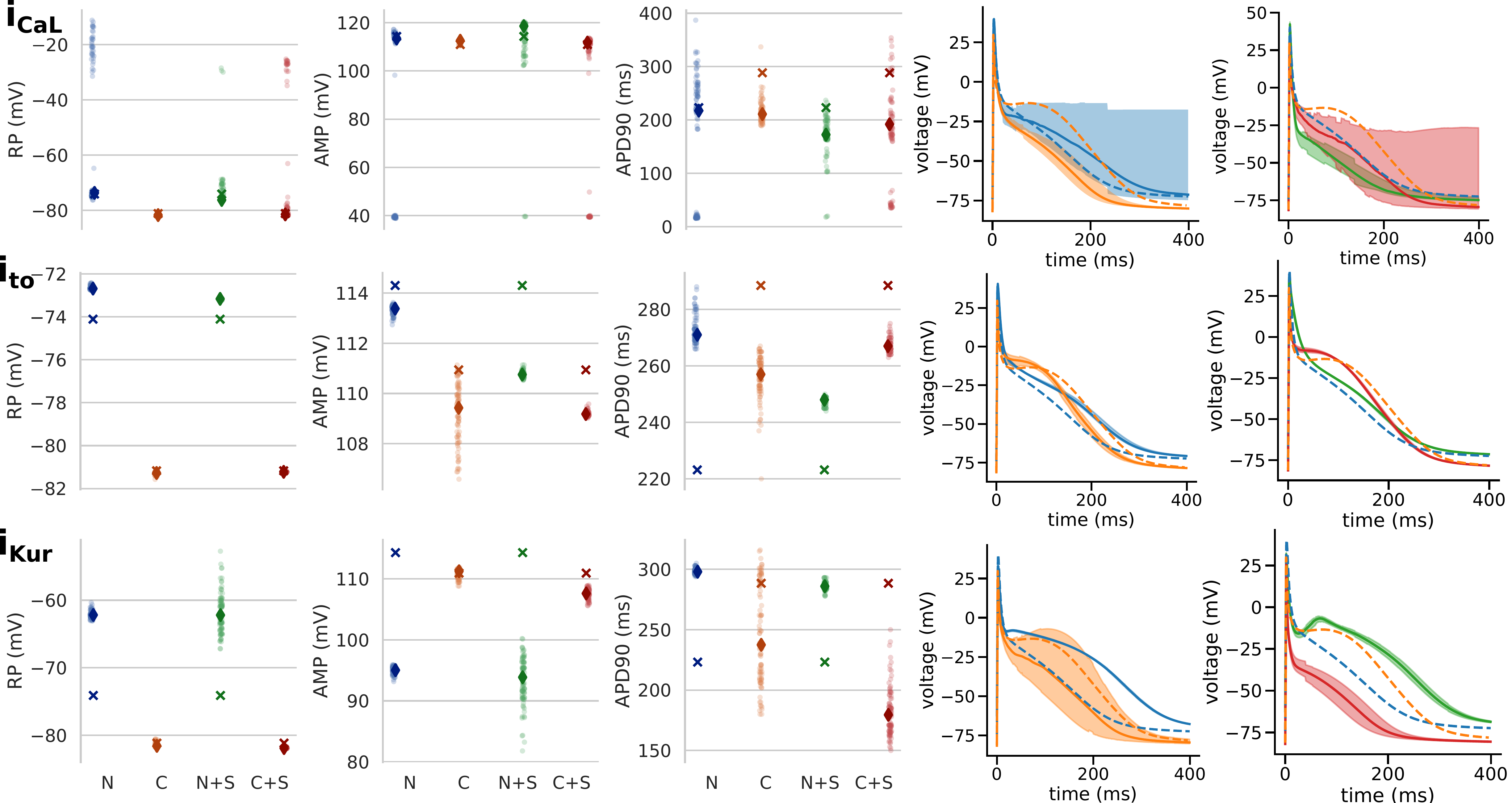}
    \caption{
    Strip plots showing measurements of resting potential (RP), action potential
    amplitude (AMP) and AP duration to 90\% 
    repolarisation from APs elicited from the
    full N and C models. Crosses indicate values from the
    models run at published settings. Each point is a measurement
    from an AP generated from a sample from the posterior
    parameter distribution of the channel indicated in bold. The
    diamond is the median of the samples. N: using N
    unified posterior distribution in the full N model, C: using C unified
    posterior distribution in the full C model, +S: indicates the
    standardised model replaced the corresponding channel in the
    full model and used the S posterior distribution.
    Line plots in the two rightmost columns display a summary
    of the AP traces in each case. Traces are displayed as
    median lines with shading representing 89\% HDPI from
    100 posterior samples. The published N and C models
    are plotted as dashed lines in all plots for comparison.}
\label{fig:ap_measures}
\end{figure}

We next studied how the inclusion of uncertainty
in the gating of these channels would impact the full
AP of the cell models.
Figure~\ref{fig:ap_measures} shows how the AP changed for the
channel models for the re-calibrated N channels, re-calibrated
C channels, and inserting the re-calibrated S channel into
either model (referred to as N+S and C+S). As noted in the Methods, this experiment could not
be completed for $I_\mathrm{Na}$ models. More
detailed results are in Section S3.6.

For $I_\mathrm{CaL}$ a portion of posterior samples in the N
model resulted in an elevated resting potential, though the median was still close to the published value. This causes the
wide 89\% HDPI observed in the corresponding trace, while the median
line is much closer to the dashed line of the original model. This
also occurred to a lesser degree when the S model inserted into
the C cell model. The re-calibrated C model was more stable with
the action potential duration (APD) being mainly reduced from the
published value as a result of quicker and more complete inactivation
of the channel during the AP (Figure 13B).

Traces generated by using the posterior parameterisations for
$I_\mathrm{to}$ generally showed small differences from the published
values for resting potential and AP amplitude, though
larger changes in the APD. For the full N model (both with the original channel form and S form), an increased APD is observed while the opposite was observed for the C model. This makes the two AP models, which
have quite different published AP morphology, more similar to one
another. A small degree of variability is apparent from the posterior
intervals around the traces and, in contrast to $I_\mathrm{CaL}$ all
samples resulted in `normal' APs.

Using the re-calibrated form of $I_\mathrm{Kur}$ resulted in more
noticeable changes to the AP than $I_\mathrm{to}$. For both the N and N+S experiments, this increases the resting potential and
AP amplitude accompanied by an increase in APD to a similar level
as the published C model. In both cases, the triangular morphology of the
published N model was altered to have a more prolonged plateau phase. For the C model, the resting potential
and amplitude of the AP from close to the published values,
while a large degree of variability is observed in the APD.
When using the C form of the channel, this variability encompassed
the published value, which is apparent from the shading of
the AP trace including the dashed trace of the published C model. When using the S form of the channel, the
cell repolarises more quickly as a result of an
increased $I_\mathrm{Kur}$ current density.

\section{Discussion}

Understanding the uncertainty and unidentifiability of parameters in AP models
is critical to the development of trustworthy cardiac models for the era of
personalised medicine~\cite{johnstone_uncertainty_2016,mirams_uncertainty_2016}. In this
study, we have applied an ABC method based on sequential Monte
Carlo to characterise the existing uncertainty in gating kinetics of four
major ion channel types in human atrial cell models. The wide posterior
distributions for a subset of parameters in all models was indicative of
potential unidentifiability, which may be structural or practical. We then sought to assess
whether the poorly constrained parameters could be more clearly identified by re-calibrating to complete
datasets or through using a standardised gating formulation
with less complex structure and fewer parameters to constrain. We finally looked at the changes
introduced by using the re-calibrated $I_\mathrm{CaL}$, $I_\mathrm{to}$ and $I_\mathrm{Kur}$ channels in
the full cell models.

Figure~\ref{fig:nyg_reported} shows uncertainty in parameter estimates
for the $I_\text{Na}$ channel current in the N model manifest in a range of
possible outputs of the current trace in response to voltage steps. The medians of
the posterior parameter distributions related to time constant behaviour were
close to the original reported values but showed a high degree of uncertainty.
This was the expected result as only steady-state data were used to calibrate
$I_\text{Na}$ of the N model. 
As a consequence, the time constant parameters in this  example likely
exhibit practical unidentifiability, as they cannot be constrained
given the provided experimental data.
It should be noted that the well-constrained
parameter $q_1$ in Figure~\ref{fig:nyg_reported}A is offset from the published
value due to a constant offset applied to the steady-state curve
in the N model justified by time- and/or temperature-dependent drifts in
steady-state characteristics of the current~\cite{nygren_mathematical_1998}.

In comparison,
the C model 
used a more complete dataset to calibrate
kinetics for this channel. Nevertheless, this model also exhibits
parameters with high RSD values suggesting unidentifiability 
(Figure~\ref{fig:summary_boxplots}). 
In contrast to the practical unidentifiability observed
in the case of the N $I_\mathrm{Na}$ model, this is likely due
to structural identifiability issues. The high number of
parameters in the C $I_\text{Na}$ model (Figure~\ref{fig:ina_all}D)
facilitates over-fitting to experimental data and leads to redundancy of some
parameters.  
In~\cite{daly_hodgkin-huxley_2015}, Daly et al
investigated the uncertainty in parameter estimates of the Hodgkin-Huxley AP
model. They reported wide posteriors around the estimates of certain parameters
in the potassium and sodium channels of that far simpler AP model. It therefore
is unsurprising to observe these results in more complex models of a sodium
channel where there is more opportunity for parameters to covary within the structure of the model. Additionally,
the use of conventional voltage-clamp protocols 
further aggravates this issue as it only provides
an indirect measure of the underlying gating kinetics~\cite{carro_limitations_2017}.

We next re-calibrated the N and C models to the unified datasets. 
In Figure~\ref{fig:summary_boxplots} it was apparent that the addition of data
sources, most often relating to the time constant of gates, would reduce the
uncertainty around the gating functions though often at the expense of
increased uncertainty in other regions of the model behaviour.
In the case of the N model of $I_\text{Na}$, the addition of
time constant data led to a reduction in the RSD value of
time constant parameters at the expense of increased RSD values
of the steady-state parameters. This may be a result of the
structure of the N model $I_\mathrm{Na}$, with sigmoid
functions for inactivation time constants, not providing
a good fit to the experimental data. As a result, the ABC
calibration reaches the sampling rate stopping criteria earlier
because the steady-state experimental data now represents only
a small portion of the overall calibration dataset.

The C model of $I_\mathrm{Na}$ has 29 parameters
to constrain across three gates, the most of any of the 
channels investigated. Although using a complete dataset for
calibration, the high RSD values of parameters show that
this model still suffers from parameter unidentifiability.
The complex form of the equations makes over-fitting possible when
using conventional calibration processes such as simple least-squares,
which would not highlight the consequences of doing so as with
a Bayesian method such as ABC.
Often these complex forms of
equations are initially based on direct comparisons between the gating
functions and the experimental data and may be tailored to 
the specific data sources selected for calibration, rather than through forward
evaluations of the model.  
In~\cite{carro_limitations_2017}, the authors
highlight the importance of replicating the experimental conditions and voltage
protocols as closely as possible when calibrating AP models, despite the
inherent difficulties of doing so. The complex form of model gating functions,
which are often combined in parallel, means that the behaviour of the gating
function itself may not be representative of the behaviour of the full channel
model when tested with experimental protocols.

Reducing the structural complexity of the model equations
by adopting a standardised gating formulation showed channel-dependent success. For $I_\text{Na}$, the
S formulation resulted in a significant reduction in RSD values
for parameter posterior distributions (p=0.02 compared to N model, p=0.04 compared to C model) which suggests it is
partly alleviating unidentifiability concerns. The $I_\mathrm{Na}$ S model
output also gave a goodness-of-fit measure of 
converged residual lower than either other model (Figure~\ref{fig:ina_all}D). Despite this, for some experiments the goodness-of-fit was clearly worse
than either N or C model, such as for activation and slow recovery time constants. In addition, there was 
variability around the lower voltage range of the activation gate
time constant for which there is an absence of experimental data, highlighting
this model is not immune to practical identifiability issues.
For $I_\text{Na}$, there appears to be an
advantage to the less complex formulation which provides more confidence in the
identifiability of its parameters without sacrificing representation of the
experimental data. Fewer parameters also allow more direct reasoning about the
effects of altering gating parameters on the overall channel behaviour. As with any model, considerations should
to be taken of the specific goals of the modelling study, though the
S $I_\mathrm{Na}$ presents a less complex foundation from which to build on
with additional, context-specific data.

For $I_\text{CaL}$, the S model produced approximately the same
goodness-of-fit as the more complex N model. 
However, in this case, the similar RSD values across
each model imply that the use of a simpler model structure
did not alleviate issues relating to parameter unidentifiability.
This is particularly noticeable in the steady-state
activation gate of the channel, where there is a greater variation in the posterior behaviour compared to the other
two models (Figure~\ref{fig:ical_all}B). This is potentially
a result of the way steady-state summary statistics are calculated through
normalising to a reference value in the output. As a consequence, the gate 
is less constrained to be fully open at the maximum activation as the normalisation hides
this behaviour. The unidentifiability of parameters across
all models for this channel is likely a consequence of the 
relative paucity of experimental data relating to the voltage-dependent
time constant behaviour of $I_\mathrm{CaL}$, which has fewer voltages
tested in the unified dataset compared to the other channels studied.

For both potassium channel models ($I_\mathrm{to}$ and
$I_\mathrm{Kur}$), simplifying the model structure is a balance
between reduced unidentifiability of parameters and reduced
goodness-of-fit to experimental data.
The inactivation $\tau$ plot of
Figure~\ref{fig:ito_all}A shows an example where both the N and C
models are able to capture the plateau region at upper voltages ranges while the S model tends to zero. Similarly, the form of the S model
requires the peak of the time constant curve to occur at the mid-point of
activation or inactivation which appears inappropriate for the
activation gate of this channel. 
This is reflected in the higher converged residual of the
S model compared to the N and C models (Figure~\ref{fig:ito_all}B lower).
The parameters
of the less complex model have lower RSD values than the C model ($p<0.001$) and a smaller range than the N model. This suggests
that the closest fit of this model to the experimental data is
identifiable, despite also being a poorer fit
than either other model.
This highlights the fact that
low uncertainty in posterior parameters does not necessary imply a model fits
the data well (and vice-versa), and a trade-off exists
which may depend on the goals and particular use-case of a
modelling study.

Beattie et al~\cite{beattie_sinusoidal_2018} used the same standardised gating
formulation to model the behaviour of a rapid delayed rectifier potassium
current using an information-rich voltage protocol. In their case, the studied
current appears to satisfy the requirement of peak time constant of the gate at
mid-point of the steady-state curve (see for example Figure 5
in~\cite{beattie_sinusoidal_2018}). Based on these observations and our
results, this standardised approach may therefore be appropriate in cases when
experimental data suggests particular requirements, such as this one, are met.
Using a standard gating formulation alleviates problems associated with the
high number of parameters in very detailed models, without sacrificing the
biophysical basis for the model. In contrast to
purely phenomenological models, the form of the S model is based on
Eyring-derived transition rates giving its parameters a physical
interpretation~\cite{tsien_transition_1969,beattie_sinusoidal_2018}.

The S model leads to a reduction
in simulation times, which is an important consideration as patient-specific
modelling is further explored in whole-heart tissue simulations. However, in
contrast to simpler models
which have been shown to reproduce patient-specific AP
morphologies~\cite{lombardo_comparison_2016}, the
standardised formulation retains information about specific ion channel
currents which provides a stronger body of evidence in terms of model
validation.
It would be encouraging if future ion channel modelling 
promoted the use of common
forms of equations in cell models rather than the current heritage of complex
equations. Another promising proposition is a model reductionist
approach to reduce uncertainty in parameter estimates by eliminating parameters
which have little effect on the model output, e.g.  manifold boundary
approximation~\cite{lombardo_systematic_2017}. 

When testing the new parameterisations in the full AP models, it
was apparent that a proportion of samples led to non-physiological
behaviour. This was also the case when combining samples from all new
parameterisations, which generally caused simulations to fail.
This highlights the importance of a feedback process in the
development of AP models, where the form of the full AP also
informs the design of the underlying currents. A future step for
this work could follow a similar approach to Kernik et al.~\cite{kernik_computational_2019} with multiple stages of
calibration. For example, the full AP samples in these results
could be used to further constrain the posterior distribution of the
channel model parameters by eliminating non-physiological cases.
These results also demonstrate an important consideration
in the development of full AP models that tends to be omitted
from modelling papers: the sheer challenge and achievement of combining a
variety of nonlinear models of individual channels usually
developed in isolation into a single model of a cardiac cell and tuning
to produce a physiological action potential.

\textbf{Limitations.} RSD was used as a measure of the width of parameter posterior
distributions and as an indicator of parameter unidentifiability
in this study. This measure tends to become inflated when the
mean value in the denominator is close to zero.
This approach does not allow us to separate structural and
practical unidentifiability and the likely form was inferred
from the availability of experimental data and structural
form of the model in experiments.

Our ABC stopping criterion for all experiments was set to halt execution once an
iteration had dropped below a 1\% particle acceptance rate. This was based on
preliminary experiments on the C model where it was assumed the algorithm is
close to the optimum solution once sampling became too difficult. However, in
some cases this criterion may be excessive or insufficient, as was observed for
example for the original N and S model of the $I_\text{CaL}$ channel. Investigations into more
appropriate stopping criterion were outside the scope of this work.
Parameters involved in the calcium-dependent gate in
$I_\text{CaL}$ of all models were omitted from calibration
and the gate set to a constant value
due to the lack of specific data and difficulty in isolating calcium
handling components of the cell model. Particularly for the C model, which
relies on the calcium transient to modulate the inactivation rates, it is
perhaps inappropriate to attempt calibration using this approach.

The calibration process relied on summary statistics of the model responses to
the virtual voltage clamp protocols. Reducing the data in this fashion was
necessary to obtain the same form as in the experimental dataset. However, as
highlighted in~\cite{sunnaker_approximate_2013}, it is generally not possible
to obtain a finite dimensional set of summary statistics that are sufficient to
fully capture all relevant information obtained from a voltage clamp protocol.

Channel models with greater than 14 parameters (N: $I_\text{CaL}$, $I_\text{Kur}$; C:
$I_\text{Na}$, $I_\text{to}$, $I_\text{Kur}$) could not be calibrated as a complete
model due to the large number of particles required to sample the high-dimensional parameter hyperspace.
In these cases, we calibrated the behaviour of parameter subsets for each gate
separately to the relevant experimental data while leaving the remainder at
their published values. It is possible that the original gates could affect the
calibration of the chosen gate, for example in exponential fitting to decay
traces for channels with fast and slow inactivation. Despite this, it should be
noted that conventional calibration techniques (e.g. least squares regression)
do not restrict the modeller from applying the method in the case of this kind
of sparse sampling
space and will not explicitly convey the implications of doing so.

\textbf{Conclusions.} In this work, we have applied ABC to re-calibrate the gating kinetics in
detailed ion channel models of human atrial myocytes. We calibrated these
models to the experimental datasets used in the published calibration and showed a portion of parameters
exhibited wide posterior distributions indicative of
unidentifiability. 
Calibration to more complete experimental datasets did not
reduce the unidentifiability present, which suggested that
it may be both structural and practical.
Reducing the structural complexity of the model through a
common gating form was successful in reducing unidentifiability
in $I_\mathrm{Na}$ without sacrificing goodness-of-fit. 
Experiments with other channels suggested that a trade-off
exists between tailoring a model to provide a good fit to
experimental data, and identifiability of parameters as models
become more complex.
The technique employed in this work is
general and could be applied to any model of an action potential.\vskip6pt

\enlargethispage{20pt}

\subsection*{Data access}
Supplementary Material is available with the article. All code is
available in the project repository at
\url{https://github.com/charleshouston/ion-channel-ABC} under the human-atrial project.
Results database files: Dryad doi:10.5061/dryad.p2ngf1vmc~\cite{houston_data_2019}.

\subsection*{Author contributions}
CH and CDC conceived of and designed the study and critically revised the
manuscript. CH carried out the experiments and performed the data
analysis. BM and LE carried out preliminary experiments and data analysis. CH
drafted the manuscript. All authors gave final approval for publication and agree
to be held accountable for the work performed therein.

\subsection*{Competing interests}

The authors declare that they have no competing interests.

\subsection*{Funding}
This work was supported by the British Heart Foundation [grant numbers PG/15/59/31621, RE/13/4/30184].

\subsection*{Acknowledgements}
The authors would like to thank members of the ElectroCardioMaths
programme of the Imperial Centre for Cardiac Engineering for discussions
pertaining to this research.

\bibliographystyle{ieeetr}
{\small
\bibliography{references}}

\end{document}

\maketitle

\tableofcontents
% \maketitle
%\thispagestyle{firstpage}
%\newpage

%\renewcommand{\thefigure}{S\arabic{figure}}
\renewcommand{\figurename}{Supplementary Figure}
\renewcommand{\tablename}{Supplementary Table}
\renewcommand{\thesection}{S\arabic{section}}

\clearpage
\newpage
\section{Datasets and simulations}
\label{sec:datasets}

\subsection{Data sources}

Table~\ref{tab:data_sources} contains a complete list
of all experimental data sources for both original and unified
datasets.

\begin{table}[!h]
\centering
    \begin{tabular}{@{}ccccccc@{}}
\toprule
    Channel & Gate & & Data &  N~\cite{nygren_mathematical_1998} &
    C~\cite{courtemanche_ionic_1998} & Unified\\
\midrule
    $I_\text{Na}$ & activation, $m$ & $m_\infty$ &Fig. 2~\cite{sakakibara_characterization_1992}&\cmark&\cmark&\cmark\\
    & & $\tau_m$ &Fig. 3C~\cite{schneider_characterization_1994}&\xmark&\cmark&\cmark\\
    & inactivation$^\ast$, $\{h,j\}$ & $\{h,j\}_\infty$ &Fig. 7~\cite{sakakibara_characterization_1992}&\cmark&\cmark&\cmark\\
    & & $\tau_{\{h,j\}}$ (inact.) &Fig. 5B~\cite{sakakibara_characterization_1992}&\xmark&\cmark&\cmark\\
    & & $\tau_{\{h,j\}}$ (recov.) &Fig. 9~\cite{sakakibara_characterization_1992}&\xmark&\cmark&\cmark\\
\midrule
    \multirow{2}{*}{$I_\text{CaL}$} & \multirow{2}{*}{activation, $d$} &
    \multirow{2}{*}{$d_\infty$} & Fig. 5C~\cite{mewes_l-type_1994}$^a$ & \multirow{2}{*}{\cmark $^a$} & \multirow{2}{*}{\cmark $^b$} & \multirow{2}{*}{\cmark $^b$}\\
    & & & Fig. 2B~\cite{li_properties_1997}$^b$ & & &\\
    & & $\tau_{d}$ & Pg. H233~\cite{li_properties_1997} & \xmark & \xmark & \cmark \\
    & inactivation$^\dagger$, $f$ & $f_\infty$ & Fig. 2B~\cite{li_properties_1997} & \cmark & \cmark & \cmark\\
    & & \multirow{2}{*}{$\tau_f$ (inact.)} & Fig.
    3B~\cite{li_properties_1997}$^a$
    &\multirow{2}{*}{\cmark $^a$} & \multirow{2}{*}{\cmark $^b$} &
    \multirow{2}{*}{\cmark$^a$}\\
    & & & Fig. 4B~\cite{sun_mechanisms_1997}$^b$ & & &\\
    & & $\tau_f$ (recov.) & Pg. H230~\cite{li_properties_1997} & \cmark &
    \xmark & \cmark\\
\midrule
    \multirow{2}{*}{$I_\text{to}$} & \multirow{2}{*}{activation, $r$} &
    \multirow{2}{*}{$r_\infty$} & Fig. 3A~\cite{shibata_contributions_1989}$^a$
    & \multirow{2}{*}{\cmark$^a$} & \multirow{2}{*}{\cmark$^b$} & \multirow{2}{*}{\cmark$^b$} \\
    & & & Fig. 2A~\cite{wang_sustained_1993}$^b$ & & & \\
    & & $\tau_r$ (act.) & Fig.
    5D~\cite{courtemanche_ionic_1998}$^\ddagger$ & \xmark & \cmark & \cmark \\
    & & $\tau_r$ (deact.) & Fig.
    5D~\cite{courtemanche_ionic_1998}$^\ddagger$ & \xmark &
    \cmark & \cmark \\
    & \multirow{2}{*}{inactivation, $s$} & \multirow{2}{*}{$s_\infty$} &
    Fig. 3C~\cite{firek_outward_1995}$^a$ & \multirow{2}{*}{\cmark$^a$} & \multirow{2}{*}{\cmark$^b$} & \multirow{2}{*}{\cmark$^b$} \\
    & & & Fig. 2C~\cite{wang_sustained_1993}$^b$
    & & &\\
    & & \multirow{2}{*}{$\tau_s$ (inact.)} & Fig.
    4C~\cite{nygren_mathematical_1998}$^{a,\ddagger}$ & \multirow{2}{*}{\cmark$^a$} &\multirow{2}{*}{\cmark$^b$} &\multirow{2}{*}{\cmark$^b$}\\
    & & & Fig.
    5D~\cite{courtemanche_ionic_1998}$^{b,\ddagger}$ & & &\\
    & & \multirow{2}{*}{$\tau_s$ (recov.)} & Fig.
    4C~\cite{nygren_mathematical_1998}$^{a,\ddagger}$& \multirow{2}{*}{\cmark$^a$} &\multirow{2}{*}{\cmark$^b$} &\multirow{2}{*}{\cmark$^b$}\\
    & & & Fig.
    5D~\cite{courtemanche_ionic_1998}$^{b,\ddagger}$ & & &\\
\midrule
    $I_\text{Kur}$$^\ast$ & activation, $a$ & $a_\infty$ & Fig.
    8E~\cite{wang_sustained_1993} & \cmark & \cmark & \cmark \\
    & & $\tau_a$ (act.) & Fig. 8F~\cite{wang_sustained_1993} & \cmark & \cmark &
    \cmark \\
    & & $\tau_a$ (deact.) & Fig. 5B~\cite{courtemanche_ionic_1998}$^\ddagger$ & \xmark & \cmark & \cmark \\
    & \multirow{2}{*}{inactivation, $i$} & \multirow{2}{*}{$i_\infty$} &
    Fig. 3C~\cite{firek_outward_1995}$^{a,\S}$ & \multirow{2}{*}{\cmark$^a$} &\multirow{2}{*}{\cmark$^b$} &\multirow{2}{*}{\cmark$^a$}\\
    & & & Fig. 7A~\cite{wang_sustained_1993}$^{b,\S}$ & & &\\
    & & \multirow{2}{*}{$\tau_i$ (inact.)} & Fig.
    4D~\cite{nygren_mathematical_1998}$^{a,\ddagger}$ & \multirow{2}{*}{\cmark$^a$} & \multirow{2}{*}{\cmark$^b$} & \multirow{2}{*}{\cmark$^a$} \\
    & & & Fig. 5B~\cite{courtemanche_ionic_1998}$^{b,\ddagger}$ & & &\\
    & & $\tau_i$ (recov.) & Fig. 4D~\cite{nygren_mathematical_1998}$^\ddagger$ &
    \cmark & \xmark & \cmark\\
\bottomrule
\end{tabular}
\caption{Summary of patch clamp experimental datasets used in modelling papers
in human atrial myocytes for each channel studied. Ticks and crosses are used
to indicate which datasets are included in the original model calibration and
which compose the unified dataset.
$^\ast$There are some differences in the terminology used in each model. The N model refers to the inactivation gates of $I_\text{Na}$ as $h_1$
and $h_2$, and the $I_\text{Kur}$ channel as $I_\text{sus}$.
$^\dagger$The L-type calcium current has a calcium-dependent inactivation process
which is not calibrated in this experiment (discussed further in results).
$^\ddagger$In some cases it was not clear from the modelling paper where the
comparison data plotted were obtained from. In these cases, the data points from the modelling paper
itself are used and a protocol assumed based on the experimental paper cited.
$^\S$In some cases it was not explicit which figure among a number of
possibilities within a cited data source was used; the choice was inferred from the modelling paper.}
\label{tab:data_sources}
\end{table}

\clearpage\newpage
\subsection{Temperature adjustment}
\label{ssec:temp-adjust}

\begin{figure}[!h]
\centering\includegraphics[width=0.8\textwidth]{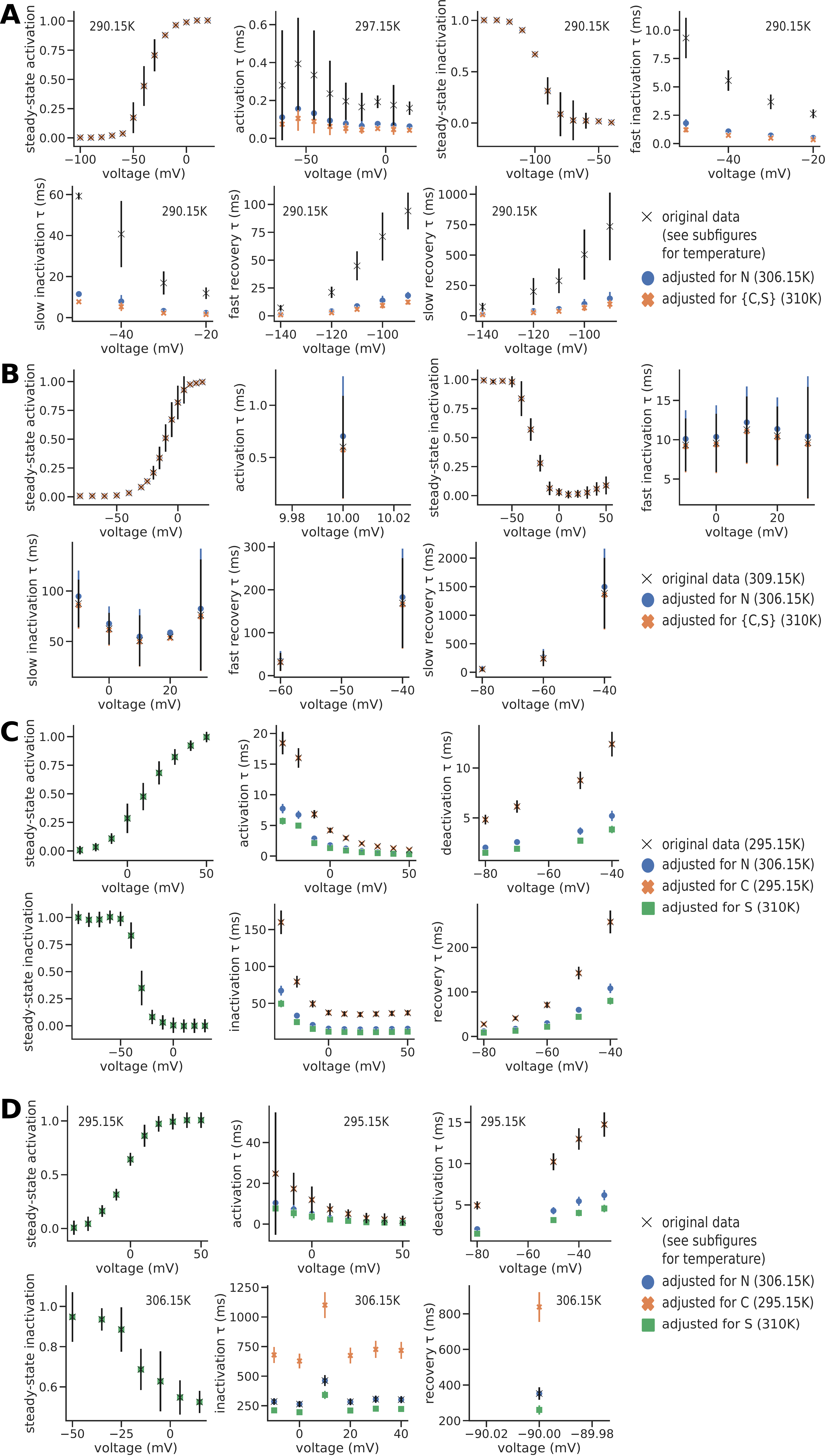}
\caption{Unified datasets for each channel model showing adjustments to experimental calibrating data at each model
    temperature. The adjustment is made using a Q10
    factor as indicated in the text. \textbf{A} $I_\text{Na}$. \textbf{B}
    $I_\text{CaL}$. \textbf{C} $I_\text{to}$. \textbf{D} $I_\text{Kur}$.}
\label{fig:temp-adjust}
\end{figure}

The N model was created to simulate a human atrial action potential at
306.15K (33$^\circ C$), whereas the C model was created to simulate
at 310K ($\sim$37$^\circ C$). Time constant measurements of rise and decay rates are temperature-dependent, and thus it was important
to account for this during the calibration.
During ABC, time constant measurements from the experimental
sources were adjusted to the temperature of the model being
calibrated using a Q10 factor from an experimental source.
The Q10 factors used are: $I_\text{Na}$: 2.79~\cite{ten_tusscher_model_2004}; 
$I_\text{CaL}$: 1.7 (activation), 1.3 (inactivation), both calculated from
values in~\cite{li_properties_1997}; $I_{\{\text{to},\text{Kur}\}}$:
2.2~\cite{wang_sustained_1993}. In the original publication for
the C model, $I_\text{to}$ and
$I_\text{Kur}$ were fitted at room temperature (295.15K) and then
adjusted by the authors to the model temperature (310K) by
dividing time constants by a factor of 3~\cite{courtemanche_ionic_1998}.
To maintain consistency, we also calibrate the C model at
room temperature rather than make any adjustment to the experimental
data. The time constants for this channel are adjusted before
being used to simulate a full action potential.
The S model was always calibrated by adjusting experimental data
to 310K.
Figure~\ref{fig:temp-adjust} shows the temperature-adjusted
datasets for all experiments across channel models.

When comparing channel models on the same figure throughout this work,
their time constants are adjusted for the same temperature of
310K. For full action potential simulations, the N and C model time
constants are kept at their model temperature, and only the S
model time constant adjusted from 310K to 306.15K when it is added to the N model.

\clearpage
\newpage
\subsection{Voltage-clamp protocols and summary statistic functions}
\label{ssec:protocols}

Throughout this section, the lettering in the headers refers to the figure
describing voltage protocols for the channel.
All curve fitting for finding time constants was carried out using the
\textit{scipy.optimize} Python library.

\subsubsection{$I_\mathrm{Na}$}

All experiments by Sakakibara et al.~\cite{sakakibara_characterization_1992}
used equal extracellular and pipette solution sodium concentration of
$5\mathrm{mM}$ at a temperature of $290\mathrm{K}$ ($17^\circ C$). The time
constant of activation experiment from Schneider et
al.~\cite{schneider_characterization_1994} used sodium
concentration of $120\mathrm{mM}$ in the extracellular solution and
$70\mathrm{mM}$ in the pipette solution, and the experiment was conducted at
$297\mathrm{K}$ ($24^\circ C$).

\textbf{A: Steady-state activation}
(p.538~\cite{sakakibara_characterization_1992}).  The standard protocol to
measure steady-state activation of the channel holds the membrane at a
sub-threshold potential of $-140\mathrm{mV}$ and then steps the channel to a
series of voltage steps between $-100\mathrm{mV}$ and $20\mathrm{mV}$, with
intervals of $10\mathrm{mV}$. The steps last for $1\mathrm{s}$ and there are
$10\mathrm{s}$ between each step.

The degree of steady-state activation is measured by recording the peak current
during the voltage step (normalised to cell capacitance). The conductance is
calculated by dividing the peak current by the forcing term, usually assumed to
be the potential difference to the Nernst potential of the primary ion carrier,
$g = \frac{I}{V-E_\mathrm{X}}$. In the computational protocols, we directly
measure the conductance of the channel to bypass this calculation. To plot
the activation curve, the conductance is plotted against the voltage step normalised
to its maximum value in any voltage step.

\textbf{B: Time constant of activation}
(p.85~\cite{schneider_characterization_1994}). The time constant constant of
inactivation is measured by fitting an equation to the current trace from a
standard steady-state activation protocol as described in the previous protocol. In this
case, the holding potential is $-135\mathrm{mV}$ and the steps are from
$-65\mathrm{mV}$ and $15\mathrm{mV}$ in steps of $10\mathrm{mV}$. The time at
holding potential between each step was not given and so assumed to be
$10\mathrm{s}$ (more than enough for the $I_\mathrm{Na}$ channel to return to
steady-state) and each test pulse lasted for $12\mathrm{ms}$.

The activation time constant was measured by fitting the entire current trace
at a pulse to the equation $I_\mathrm{Na} =
I_\mathrm{Na,max}\left[1-e^{-t/\tau_\mathrm{m}}\right]^3
e^{-t/\tau_\mathrm{h}}+\text{constant}$
(p.87~\cite{schneider_characterization_1994}). In this equation, $t$ is the time in $\mathrm{ms}$,
$I_\mathrm{Na,max}$ is the peak of the current trace, $\tau_\mathrm{m}$ is the
activation time constant and $\tau_\mathrm{h}$ is the inactivation time
constant.

\textbf{C: Steady-state inactivation}
(p.541~\cite{sakakibara_characterization_1992}). The protocol used to measure
steady-state inactivation is often also referred to as an availability
protocol. The membrane is held at a holding potential of $-140\mathrm{mV}$ for
$10\mathrm{s}$, then stepped to a conditioning potential for $1\mathrm{s}$ to
activate and inactivate the channel. The membrane potential is returned to the
holding potential for $2\mathrm{ms}$ before stepping to a test
pulse at $-20\mathrm{mV}$ for $30\mathrm{ms}$. A series of different
conditioning pulses between $-140\mathrm{mV}$ and $-40\mathrm{mV}$ in steps of
$10\mathrm{mV}$ is used to test the amount of inactivation of the channel at different
voltages.

The steady-state inactivation is measured by recording the peak current during
the test pulse, normalised to the current in a test pulse when no conditioning step
is applied (usually the maximum current amplitude). In the virtual voltage
clamp experiment, we measure conductance directly in this step (which is
equivalent as the forcing is the same during each test pulse and eliminated
during the normalisation).

%In the summary table below, the steady-state activation protocol will be referred to
%as a function:
%\begin{verbatim}
%ss_inact(vhold, vsteps, vtest, tpre, tstep, twait, ttest)
%\end{verbatim}
%where \texttt{vhold} is the holding potential, \texttt{vsteps}
%is the level of conditioning steps, \texttt{tpre} is the time at
%holding potential before each conditioning step, \texttt{tstep} is the time
%at each step, \texttt{twait} is the time at holding potential between the
%conditioning step and test step and \texttt{ttest} is the time at the test step.

\textbf{D: Fast/slow inactivation time constant}
(p.539~\cite{sakakibara_characterization_1992}). The protocol to determine
inactivation time constants is a simple steptrain of test pulses from a holding
potential of $-140\mathrm{mV}$ for $10\mathrm{s}$ to a series of
$100\mathrm{ms}$ test pulses to voltages from $-50\mathrm{mV}$ to
$-20\mathrm{mV}$ in steps of $10\mathrm{mV}$.

The fast and slow inactivation constants are determined by fitting the
\textit{decay} part of the current trace (after the peak current) to the
equation
$I_\mathrm{Na}=A_1e^{-t/\tau_\mathrm{f}}+A_2e^{-t/\tau_\mathrm{s}}+A_0$
(p.538~\cite{sakakibara_characterization_1992}) where
$A$ are amplitude variables, $t$ is the time and $\tau_\mathrm{f}$ and
$\tau_\mathrm{s}$ are the fast and slow time constants of inactivation
respectively.

\textbf{E: Fast/slow recovery time constant}
(p.538~\cite{sakakibara_characterization_1992}). The time constants of recovery
from inactivation were determined using a double pulse protocol. The first
pulse is a conditioning pulse for $1000\mathrm{ms}$ to $-20\mathrm{mV}$
followed by a recovery period to a holding potential between $-140\mathrm{mV}$
and $-90\mathrm{mV}$ in steps of $10\mathrm{mV}$, of varying length between $2-1000\mathrm{ms}$
(this was not specified and we assumed a series of $t_\mathrm{r} = 2^i$
where $i=1,2...10$). The recovery period is followed by a test pulse identical
to the conditioning pulse. We assumed $10\mathrm{s}$ at holding potential between each pair
of pulses.

The recovery time constant is measured through a series of processing steps.
Firstly, the peak current in the test pulse is normalised to the peak current
in the preceding conditioning pulse to give a measurement of proportion of the
channel recovery for each recovery time period. This recovery measure is plotted against the recovery time
period and the resulting curve is fit to a double exponential equation $r =
A_0-A_1e^{-t_\mathrm{r}/\tau_\mathrm{r(f)}}-A_2e^{-t_\mathrm{r}/\tau_\mathrm{r(s)}}$ where $r$ is
the proportion of recovery, $A$ are amplitude parameters, $t$ is the recovery
time period and $\tau_\mathrm{r(f)}$ and $\tau_\mathrm{r(s)}$ are the fast and
slow recovery time constants respectively. These values are calculated for each
holding potential.

\begin{figure}[!h]
\centering\includegraphics[width=\textwidth]{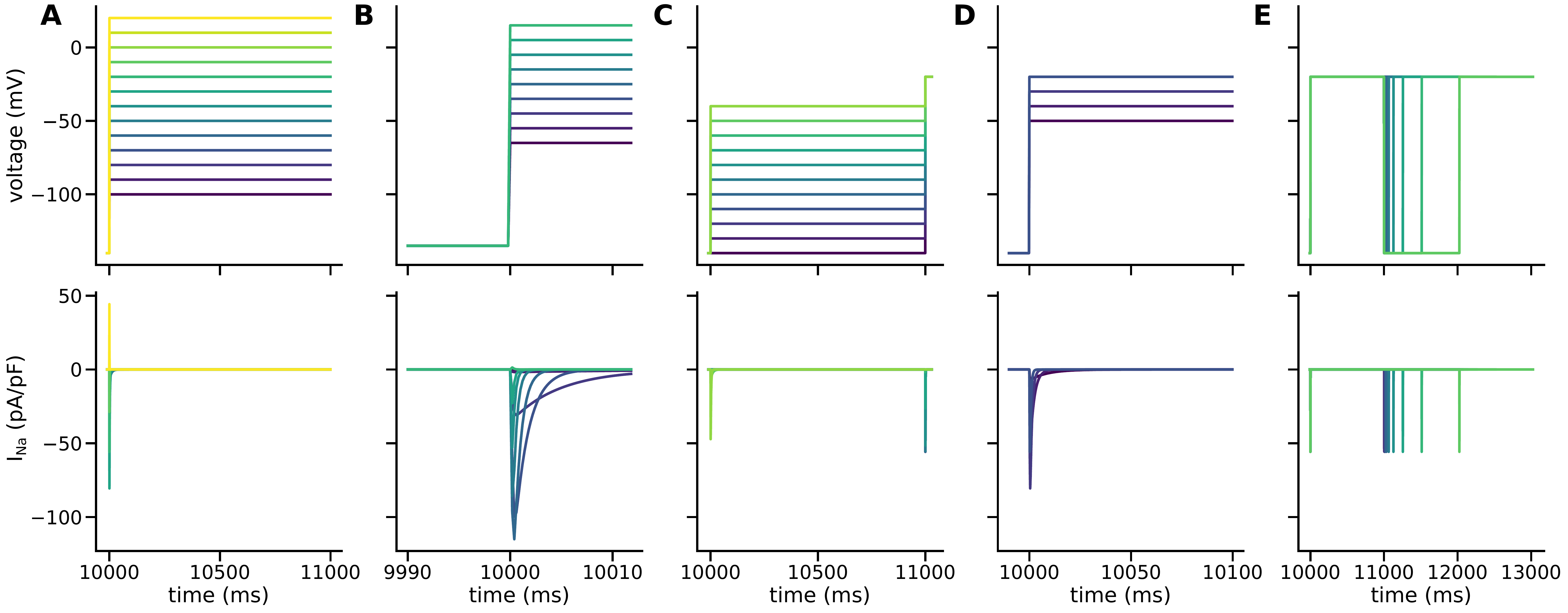}
    \caption{Voltage steps and current response (from C model) of
    $I_\mathrm{Na}$ protocols. A: steady-state activation,
    B: time constant of activation, C: steady-state inactivation, D: fast and slow time
    constants of inactivation, E: fast and slow time constants of recovery from
    inactivation. The recovery protocol is repeated at multiple holding
    potentials (only one shown). See text for details of how the current traces are processed
    into summary statistics.} 
    \label{fig:ina-protocols}
\end{figure}

\subsubsection{$I_\mathrm{CaL}$}

Experiments by Mewes and Ravens~\cite{mewes_l-type_1994} were carried out at
room temperature (assumed by authors to be $295\mathrm{K}$) and external solution with calcium
concentration of $1.8\mathrm{mM}$. Those by Li and Nattel~\cite{li_properties_1997}
were conducted at $309\mathrm{K}$ with $2.0\mathrm{mM}$ external calcium
concentration.  Experiments by Sun et al.~\cite{sun_mechanisms_1997} were
completed at room temperature (assumed by authors to be $296\mathrm{K}$) and
external calcium concentration of $1\mathrm{mM}$. It is difficult to estimate
the level of intracellular calcium concentration during these experiments due
to the mechanisms of the intracellular calcium stores and buffering. In our
virtual voltage clamp experiments, the level of the intracellular calcium was
kept constant at the resting value from the published N and C models 
(72.5nM and 101.3nM respectively) and, for the S model, set to the same
value as the C model (101.3nM).

\textbf{A, B: Steady-state activation} (p.1309~\cite{mewes_l-type_1994},
p.H228~\cite{li_properties_1997}). In Mewes and Ravens~\cite{mewes_l-type_1994},
steady-state activation was assessed using a conventional steptrain protocol
from a holding potential of $-40\mathrm{mV}$ to $450\mathrm{ms}$ steps between
$-35\mathrm{mV}$ to $15\mathrm{mV}$ in intervals of $5\mathrm{mV}$, with
$10\mathrm{s}$ between each pulse.  In Li and
Nattel~\cite{li_properties_1997}, the activation curve was generated from the
IV curve dataset which used a similar step train protocol. This time the
holding potential was $-80\mathrm{mV}$ and the $300\mathrm{ms}$ test pulses
ranged from $-80\mathrm{mV}$ to $20\mathrm{mV}$ in steps of $10\mathrm{mV}$.
Both activation curves were generated as
described in $I_\mathrm{Na}$ steady-state activation.

\textbf{B: Activation time constant} (p.H233~\cite{li_properties_1997}). The
single activation time constant value was determined from the current trace
evoked during the $10\mathrm{mV}$ test pulse in the activation protocol
from~\cite{li_properties_1997}. The activation time constant was estimated by
fitting the upstroke of the normalised current trace to
$I_\mathrm{CaL}=1-Ae^{-t/\tau_\mathrm{a}}$ where $A$ is an amplitude parameter,
$t$ is the time and $\tau_\mathrm{a}$ is the activation time constant.

\textbf{C: Steady-state voltage-dependent inactivation}
(p.H229~\cite{li_properties_1997}). The voltage-dependent steady-state
inactivation was assessed using a standard availability protocol.
Conditioning pulses to a series of voltages between $-80\mathrm{mV}$ and
$50\mathrm{mV}$ in steps of $10\mathrm{mV}$ were followed immediately by a test
pulse to $10\mathrm{mV}$ for $300\mathrm{ms}$. 
In~\cite{li_properties_1997}, there are three datasets using different length
of conditioning pulses of either $150\mathrm{ms}$, $300\mathrm{ms}$ or
$1000\mathrm{ms}$. We use the $1000\mathrm{ms}$ prepulse dataset as this was
used for calibration in both modelling papers. The inactivation curve was
calculated as in $I_\mathrm{Na}$ steady-state inactivation.

\textbf{D, E: Fast/slow voltage-dependent inactivation time constant}
(p.H229-H230~\cite{li_properties_1997}, p.H1628~\cite{sun_mechanisms_1997}).
Fast and slow inactivation time constants were calculated by fitting a
biexponential equation to the decay portion of the current trace during test
pulses from a holding potential.  In~\cite{li_properties_1997}, three different
holding potentials were used and we use the same as the modelling papers:
$-80\mathrm{mV}$. Test pulses lasted $300\mathrm{ms}$ to steps between
$-10\mathrm{mV}$ and $30\mathrm{mV}$ in intervals of $10\mathrm{mV}$ (we assume
$10\mathrm{s}$ between pulses).  In~\cite{sun_mechanisms_1997}, the holding
potential was $-80\mathrm{mV}$, test pulses had duration $1000\mathrm{ms}$ with
the same levels as above, and were preceded by a $500\mathrm{ms}$ pulse to
$-40\mathrm{mV}$. In both cases, time constants of inactivation were calculated
by fitting the decay portion of the current trace during the test pulses to
$I_\mathrm{CaL}=A_0+A_fe^{-t/\tau_\mathrm{f}}+A_se^{-t/\tau_\mathrm{s}}$ were
$A$ are amplitude parameters, $t$ is the time and $\tau_\mathrm{f}$ and
$\tau_\mathrm{s}$ are the fast and slow time constants respectively.

\textbf{F: Fast/slow voltage-dependent recovery time constant}
(p.H229-H230~\cite{li_properties_1997}). Recovery time constants were assessed
with a two-pulse protocol as in $I_\mathrm{Na}$ recovery experiments.  In this
case, the holding potentials were $-80\mathrm{mV}$, $-60\mathrm{mV}$ and
$-40\mathrm{mV}$ with conditioning and test pulses both to $10\mathrm{mV}$ for
$300\mathrm{ms}$. Recovery periods were generated from $t_\mathrm{r}=2^i$
where $i=1,2,...,11$ based on the range of data points in Fig
4B~\cite{li_properties_1997}. The data was processed as in $I_\text{Na}$
recovery experiments with the exception that a single exponential function was
used to fit the $-80\mathrm{mV}$ recovery curve to give a single (slow)
recovery time constant.

\begin{figure}[!h]
\centering\includegraphics[width=0.7\textwidth]{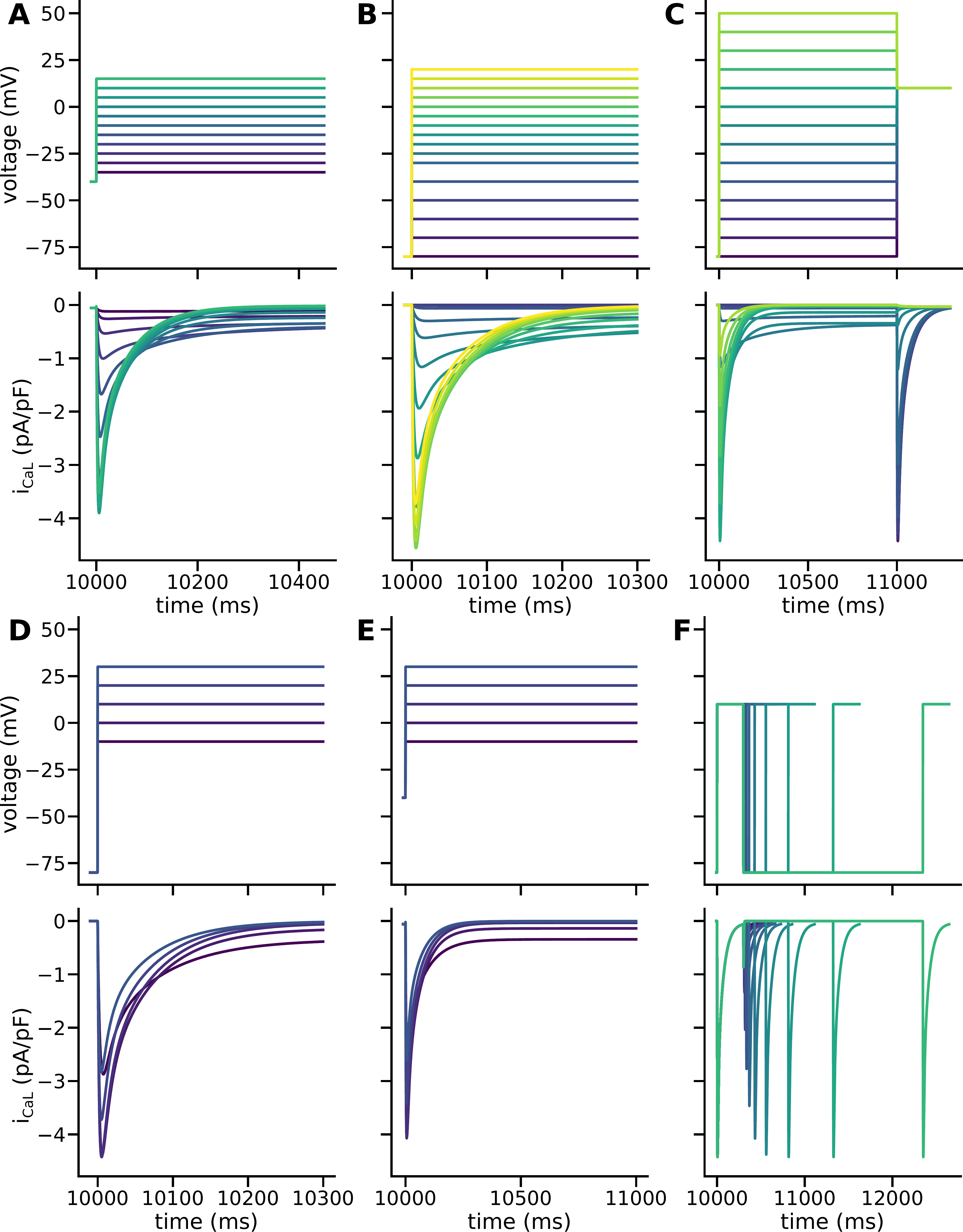}
    \caption{Voltage steps and current response (from N model) of
    $I_\mathrm{CaL}$ protocols. A: steady-state
    activation~\cite{mewes_l-type_1994}, B: steady-state and time constant
    of activation~\cite{li_properties_1997}, 
     C:steady-state inactivation~\cite{li_properties_1997}, D: fast and slow time
    constants of inactivation~\cite{li_properties_1997}, E: fast and slow time
    constants of inactivation~\cite{sun_mechanisms_1997}, F: fast and slow time constants of recovery from
    inactivation~\cite{li_properties_1997}. The recovery protocol is repeated at multiple holding
    potentials (only one shown). See text for details of how the current traces are processed
    into summary statistics.} 
    \label{fig:ical-protocols}
\end{figure}

\subsubsection{$I_\mathrm{to}$}

Experiments by Shibata et al.~\cite{shibata_contributions_1989} and Wang et
al.~\cite{wang_sustained_1993} were conducted at room temperature (assumed to
be $295\mathrm{K}$). Firek and Giles~\cite{firek_outward_1995} used temperature of
$306\mathrm{K}$ ($33^\circ C$). Shibata et al. used an extracellular and
pipette potassium concentration of $4.5\mathrm{mM}$ and $150\mathrm{mM}$
respectively, Wang et al. used $5.4\mathrm{mM}$ and $130\mathrm{mM}$ and Firek
et al. used $5.4\mathrm{mM}$ and $140\mathrm{mM}$. 

Some data for $I_\mathrm{to}$ was extracted directly from the N and C modelling
papers~\cite{nygren_mathematical_1998,courtemanche_ionic_1998} as the source of
the comparison experimental data plotted was not clear from the text. In these
cases, conditions were assumed to be the same as the lab which produced the
model.  Thus, conditions for the assumed experimental data from Nygren et
al.~\cite{nygren_mathematical_1998} was set to the same conditions as Firek and
Giles~\cite{firek_outward_1995} above, and data from Courtemanche et
al.~\cite{courtemanche_ionic_1998} was assumed to have been
collected at the conditions in Wang et
al.~\cite{wang_sustained_1993}.

\textbf{A, B: Steady-state activation} (p.
H1776~\cite{shibata_contributions_1989}, p.1065~\cite{wang_sustained_1993}). In
Shibata et al.~\cite{shibata_contributions_1989}, steady-state activation was
determined by holding at a potential of $-60\mathrm{mV}$ for $20\mathrm{s}$,
then depolarising for $15\mathrm{ms}$ to a step between $-30\mathrm{mV}$ to
$80\mathrm{mV}$, and finally stepping to $-40\mathrm{mV}$ for $100\mathrm{ms}$.
The activation curve was generated by recording the peak current amplitude in
the final step (the `tail' current). Wang et al.~\cite{wang_sustained_1993} used a standard
steady-state activation protocol with a holding potential of $-80\mathrm{mV}$
to a series of test potentials between $-40\mathrm{mV}$ and $50\mathrm{mV}$ and
measured the peak current during the $1000\mathrm{ms}$ depolarising step
(before processing as above for $I_\mathrm{Na}$).

\textbf{C: Activation time constant} (p.H305~\cite{courtemanche_ionic_1998}).
Activation time constant data corresponding to that plotted in the modelling
paper could not be found in the cited experimental
source~\cite{wang_sustained_1993}. A simple steptrain protocol as described
in~\cite{wang_sustained_1993} was assumed with a holding potential of
$-50\mathrm{mV}$ and $100\mathrm{ms}$ test pulses. The activaton time course of
$I_\mathrm{to}$ was fitted to a single exponential equation:
$I_\mathrm{to}=A_0-Ae^{-t/\tau_\mathrm{a}}$ where $A$ parameters are
amplitudes, $t$ is the time course and $\tau_\mathrm{a}$ is the activation time
constant.

\textbf{D: Deactivation time constant} (p.H305~\cite{courtemanche_ionic_1998}).
The deactivation time constant data source is also uncertain and it is assumed
to use a similar protocol as in Fig. 9A in Wang et
al.~\cite{wang_sustained_1993}. This protocol holds the membrane potential at
$-50\mathrm{mV}$ for $20\mathrm{s}$ before applying a $10\mathrm{ms}$
conditioning pulse to $50\mathrm{mV}$. This is followed by a test pulse to
elicit a tail current, which is fit to a single exponential (same as the
activation time constant above) to determine the deactivation time constant.

\textbf{E, F: Steady-state inactivation} (p.34~\cite{firek_outward_1995},
p.1065~\cite{wang_sustained_1993}). In Firek and Giles, a standard steady-state
availability protocol was applied as described in more detail in
$I_\mathrm{Na}$ steady-state inactivation. The holding potential is
$-80\mathrm{mV}$, followed by a $400\mathrm{ms}$ conditioning pulse to levels
between $-80\mathrm{mV}$ and $16\mathrm{mV}$, and finally a $400\mathrm{ms}$
test pulse to $0\mathrm{mV}$ to activate the outward current. Wang et
al.~\cite{wang_sustained_1993} similarly uses a standard availability protocol
with the same holding potential followed by $1000\mathrm{ms}$ conditioning
pulse to a range of voltages between $-90\mathrm{mV}$ and $30\mathrm{mV}$
followed by a $1000\mathrm{ms}$ test pulse to $60\mathrm{mV}$. Output was
processed into summary statistics as described in $I_\mathrm{Na}$ steady-state
inactivation.

\textbf{G, H: Inactivation time constant} (p.66~\cite{nygren_mathematical_1998},
p.H305~\cite{courtemanche_ionic_1998}). It was not clear in either N or C model
where experimental comparison data of time constants of inactivation for $I_\mathrm{to}$ were obtained
from. Consequently, simple protocols were assumed based on single-pulse
protocols in experimental papers originating from the same labs.
For~\cite{nygren_mathematical_1998}, a single pulse protocol from a holding
potential of $-80\mathrm{mV}$ to $400\mathrm{ms}$ test pulses between $0\mathrm{mV}$ and
$40\mathrm{mV}$ was applied and the decay phase of the current trace fit to a
single exponential equation as above for activation time constants.
For~\cite{courtemanche_ionic_1998}, a similar protocol was assumed based
on~\cite{wang_sustained_1993} with a holding potential of $-50\mathrm{mV}$ and
$100\mathrm{ms}$ test pulses to between $-40\mathrm{mV}$ and $50\mathrm{mV}$.

\textbf{I, J: Recovery time constant} (p.66~\cite{nygren_mathematical_1998},
p.H305~\cite{courtemanche_ionic_1998}). Similarly to the inactivation time
constant data, it was unclear where the recovery data was obtained from for
both N and C models. For the N model, we assume the recovery protocol
in~\cite{shibata_contributions_1989} was used. This is a standard two-pulse
recovery protocol with holding potentials of $-100\mathrm{mV}$,
$-80\mathrm{mV}$ and $-60\mathrm{mv}$ and $100\mathrm{ms}$ test pulses to
$20\mathrm{mV}$. For the C model, we assume the protocol
in~\cite{wang_sustained_1993} was used with holding potential between
$-60\mathrm{mV}$ and $40\mathrm{mV}$ and $200\mathrm{ms}$ test pulses to
$50\mathrm{mV}$.

\begin{figure}[!h]
\centering\includegraphics[width=\textwidth]{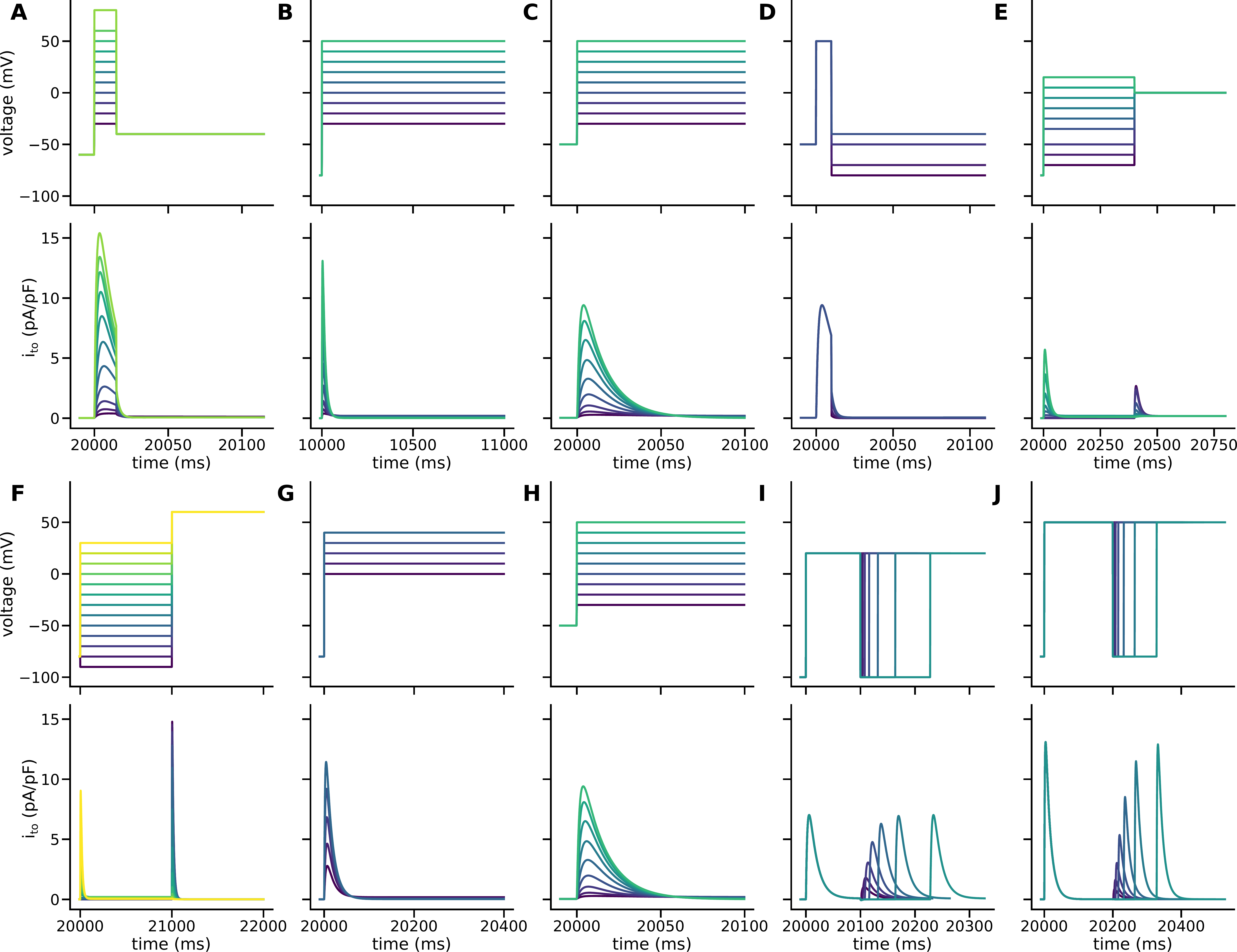}
    \caption{Voltage steps and current response (from N model) of
    all $I_\mathrm{to}$ protocols. From left to right: A: steady-state
    activation~\cite{shibata_contributions_1989}, B: steady-state
    activation~\cite{wang_sustained_1993}, C: activation time
    constants~\cite{courtemanche_ionic_1998}, D: deactivation time
    constants~\cite{courtemanche_ionic_1998},  
    E: steady-state inactivation~\cite{firek_outward_1995}, F: steady-state
    inactivation~\cite{wang_sustained_1993}, G: time
    constant of inactivation~\cite{nygren_mathematical_1998}, H: time constant of
    inactivation~\cite{courtemanche_ionic_1998}, I: time constant of recovery from
    inactivation~\cite{nygren_mathematical_1998}, J: time constant of recovery
    from inactivation~\cite{courtemanche_ionic_1998}. The recovery protocols
    are repeated at multiple holding
    potentials (only one shown for each). See text for details of how the current traces are processed
    into summary statistics.} 
    \label{fig:ito-protocols}
\end{figure}

\clearpage
\newpage
\subsubsection{$I_\mathrm{Kur}$}

Experiments by Wang et al.~\cite{wang_sustained_1993} and Firek and
Giles~\cite{firek_outward_1995} use the same conditions
stated above for $I_\mathrm{to}$. As before, in some cases it was unclear where
the comparison data in the modelling papers was obtained from. Details are
given in the following sections.

\textbf{A: Steady-state activation, activation time constant} (p.1069~\cite{wang_sustained_1993}).
$I_\mathrm{Kur}$ was measured from a holding potential of $-50\mathrm{mV}$ followed by a
$1\mathrm{s}$ prepulse to $50\mathrm{mV}$ (experimentally used to inactivate
the $I_\mathrm{to}$ current which would otherwise interfere with isolating
$I_\mathrm{Kur}$). The potential is returned to $-50\mathrm{mV}$ for
$20\mathrm{ms}$ before being stepped to a range of $100\mathrm{ms}$ test pulses between
$-40\mathrm{mV}$ to $50\mathrm{mV}$ each followed by a repolarising pulse to $-10\mathrm{mV}$.
$I_\mathrm{Kur}$ was measured as the peak current in the final repolarising
pulse and the activation curve determined as previously described for $I_\mathrm{Na}$
steady-state activation. The activation time constants were determined by fitting the time course of
$I_\mathrm{Kur}$ trace during the test pulses to a single exponential function
as described above.

\textbf{B: Deactivation time constant} (p.305~\cite{courtemanche_ionic_1998}).
As it was not clear where the data points were obtained from, the protocol from
Fig. 9 in~\cite{wang_sustained_1993} was assumed to have been used. This is the
same protocol as described in deactivation time constant for $I_\mathrm{to}$.

\textbf{C, D: Steady-state inactivation} (p.34~\cite{firek_outward_1995},
p.1068~\cite{wang_sustained_1993}). For Firek and
Giles~\cite{firek_outward_1995}, the protocol was the same as the steady-state
inactivation protocol described for steady-state inactivation of
$I_\mathrm{to}$ with an increase of the length of the conditioning pulse from
$400\mathrm{ms}$ to $2500\mathrm{ms}$. $I_\mathrm{Kur}$ was measured as the
steady-state current at the end of the test pulse. In Wang et
al.~\cite{wang_sustained_1993}, steady-state inactivation is measured from the
steady-state current at the end of a $2000\mathrm{ms}$ test pulse to
$40\mathrm{mV}$ after a $1000\mathrm{ms}$ conditioning pulse from a range of
voltages. The holding potential was $-60\mathrm{mV}$.

\textbf{E, F: Inactivation time constant} (p.66~\cite{nygren_mathematical_1998},
p.H305~\cite{courtemanche_ionic_1998}). It was unclear where the experimental
data points from the N model paper came from. We assume they are generated from
the protocol used in~\cite{firek_outward_1995} from the same lab. This protocol
is a simple $400\mathrm{ms}$ step to a range of potentials from a holding potential of
$-80\mathrm{mV}$. In~\cite{firek_outward_1995}, the current decay is fit to a
double exponential equation in order to separate $I_\mathrm{to}$ and
$I_\mathrm{Kur}$ decay rates. As in the virtual voltage clamp there is only
$I_\mathrm{Kur}$, we fit the current decay to a single exponential. 

Similary for the C model, we assume a similar protocol as
in~\cite{wang_sustained_1993} was used. This is a simple $2000\mathrm{ms}$ step
to a range of test potentials from a holding potential of $-50\mathrm{mV}$. The
decay portion of the current trace is fit to a single exponential function as
above.

\textbf{G: Recovery time constant} (p.66~\cite{nygren_mathematical_1998}). As
previously for $I_\mathrm{to}$, we assume that the recovery protocol
from~\cite{shibata_contributions_1989} was used to determine a recovery time
constant for $I_\mathrm{Kur}$. The protocol is as described for $I_\mathrm{to}$
recovery time constant.

\begin{figure}[!h]
\centering\includegraphics[width=0.9\textwidth]{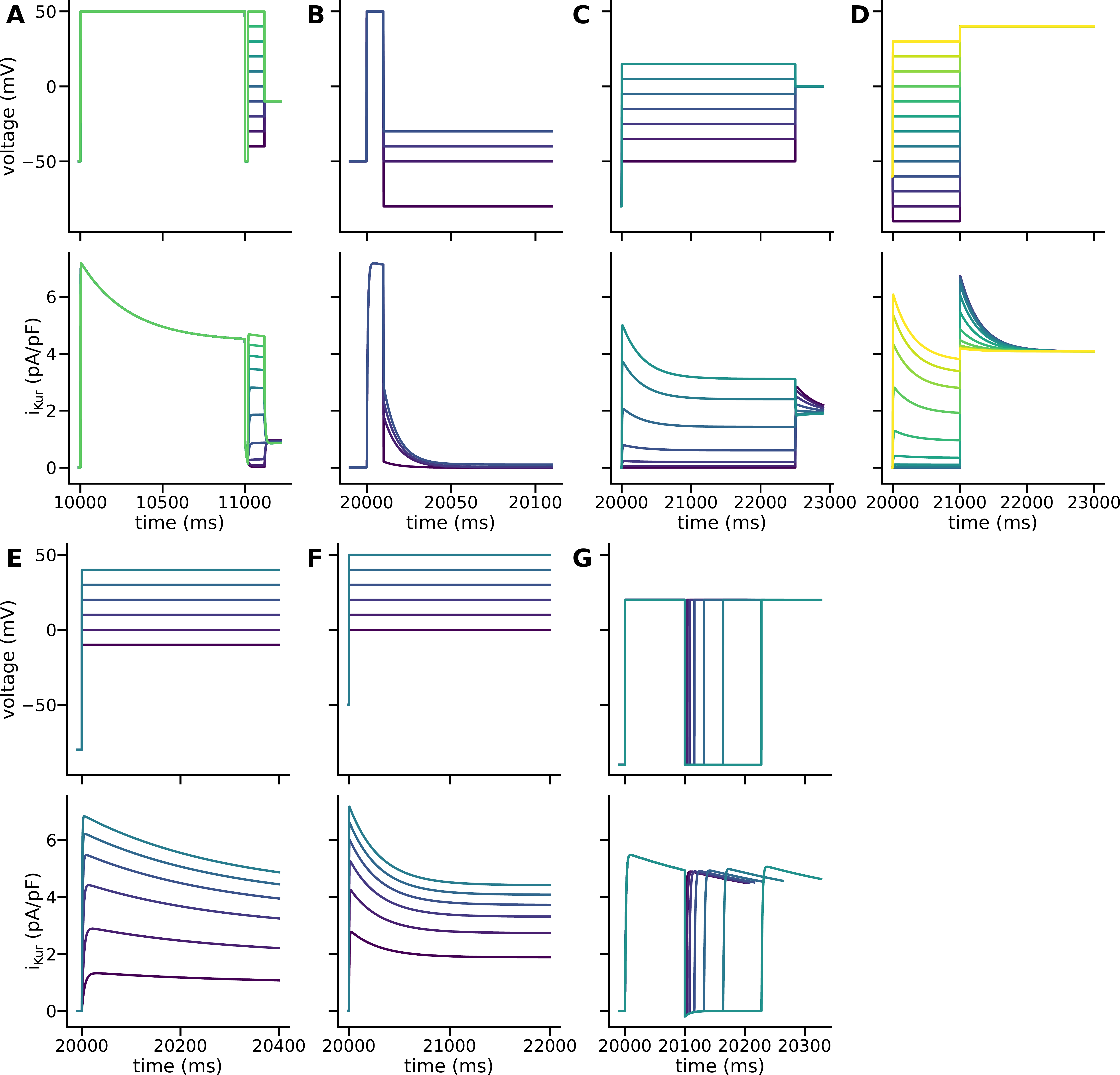}
    \caption{Voltage steps and current response (from N model) of
    all $I_\mathrm{Kur}$ protocols. A: steady-state
    activation and activation time constant~\cite{wang_sustained_1993},
    B: deactivation time constants~\cite{courtemanche_ionic_1998},  
    C: steady-state inactivation~\cite{firek_outward_1995}, D: steady-state
    inactivation~\cite{wang_sustained_1993}, E: time
    constant of inactivation~\cite{nygren_mathematical_1998}, F: time constant of
    inactivation~\cite{courtemanche_ionic_1998}, G: time constant of recovery from
    inactivation~\cite{nygren_mathematical_1998}. See text for details of
    how the current traces are processed into summary statistics.} 
    \label{fig:isus-protocols}
\end{figure}

\clearpage
\newpage
\section{Approximate Bayesian computation}
\label{sec:abc}

Formally, ABC approximates the true parameter posterior distribution
$P(\boldsymbol{\lambda}|\boldsymbol{D})$ by
$P(\boldsymbol{\lambda}|\rho(\boldsymbol{\hat{D}},\boldsymbol{D})\le\epsilon)$
where $\boldsymbol{D}$ are the experimental data, $\rho$ is the chosen distance
function, $\boldsymbol{\hat{D}}$ is the model summary statistics and $\epsilon$
is the threshold value.  We use the Toni ABC sampler based on sequential Monte Carlo to infer our
parameter posterior distributions~\cite{toni_approximate_2009}. In this
sampler, the ABC process above is repeated through a number of iterations with
reducing $\epsilon$. A population of parameter samples, referred to as
`particles', are propagated through each iteration and represent a discrete
surrogate to the continuous posterior distribution. 

At each iteration of the
algorithm, the previous population of particles are perturbed slightly, by a
multivariate Gaussian kernel, and used as the prior distribution for the
current iteration. $\epsilon$ is reduced over iterations and chosen as the
median distance of samples from the previous
iteration. The particle population number is
set according to the number of parameters being constrained in the experiment
by considering the size of the parameter sampling hyperspace and assuming at
least two particles in each dimension. A limit of 10000 particles is enforced
due to computational demands. The algorithm terminates when less than 1\% of
parameter samples are accepted in a given iteration indicating the algorithm is
struggling to improve on the current optimum. This criterion is chosen over
termination at an absolute value of the distance metric termination used in
other studies due to differences in number of model parameters and availability
of data between experiments.

To compare our model summary statistics to
the experimental data (which include error measurements at each point),
we use a weighted Euclidean distance function
\begin{align}
\begin{split}
    \rho(\boldsymbol{\hat{D}}, \boldsymbol{D})
    &=\left[\sum_{i=0}^M \left( \frac{\hat{D}_i - D_i}{w_i} \right)^2
    \right]^{1/2},\\ 
    w_i &= \text{max}\left(\sigma_i,\delta\right) n_{\text{exp}|i},
\end{split}
\end{align}
where $w_i$ are the weights for each data point, $\sigma_i$ is the standard
deviation associated with the error at each data point, $\delta$ is a
regularisation factor applied when weights are close to zero, and $n_{\text{exp}|i}$
is the number of data points within the experiment. In this
work $\delta$ was set to 0.05. Once generated, the weights
are also mean-normalised to improve convergence of the ABC
algorithm. The choice of distance
function reduces the weighting of data points based on the magnitude of
experimental uncertainty. This allows us to use information of which
experimental data points we are relatively more confident about during the
model calibration, propagating some of the experimental uncertainties
through to this stage in the model development~\cite{mirams_uncertainty_2016}.
In some cases for $I_\text{to}$ and $I_\text{Kur}$, 
it was not clear in the modelling paper where the data were obtained. In these
cases, the points displayed in the figure from the modelling paper were
digitised and 10\% standard deviation error assumed (based on the error of
similar measurements given in~\cite{wang_sustained_1993}).

Additionally, we account for the fact that we are simultaneously calibrating
to multiple datasets by weighting according to the number of data points in a
specific experiment. This provides balance between the different types of
channel behaviour, rather than favouring an experiment with a greater number of data
points. Each dataset is normalised to the maximum value in that experiment to avoid
preference towards datasets with measurements at a larger scale in the ABC loss
function.

A uniform prior distribution is used for each model parameter in the first
iteration. The width of this prior is set to be sufficiently wide to cover the
range of physiological possibilities. After convergence, the parameter
posterior distributions are inspected and, if observed to be restricted by the
lower or upper limit of the prior, the calibration is re-started with a wider
prior. For the S model, prior ranges are set as in~\cite{beattie_sinusoidal_2018}.

\clearpage
\newpage
\section{Additional results}
\label{sec:additional-results}

\subsection{Gating functions for calibrations to original and unified datasets}

\begin{figure}[!h]
\centering\includegraphics[width=\textwidth]{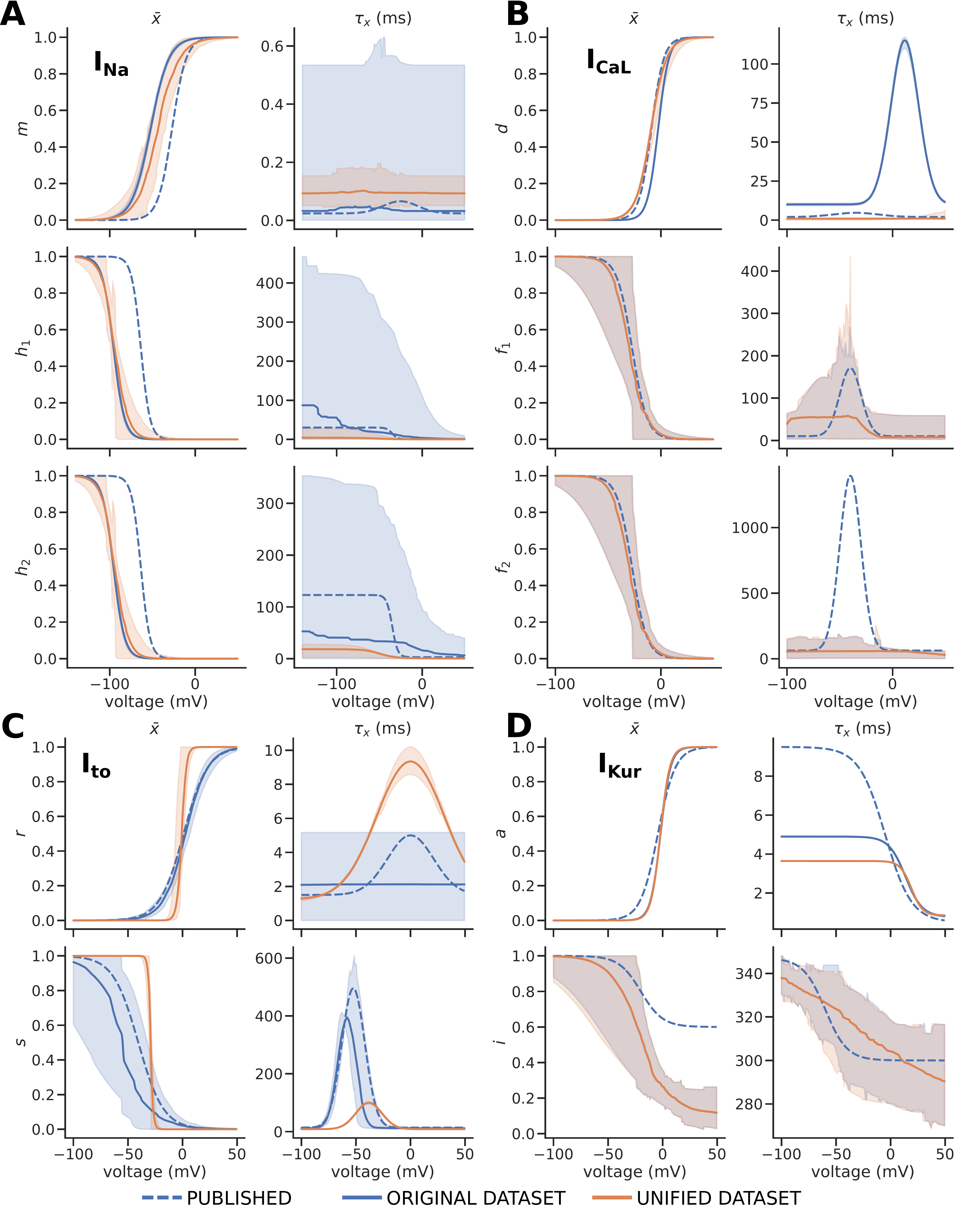}
\caption{
%\DIFF{\bf Figure updated for median +- HDPI.}
\textbf{A-D} Steady-state and time constant functions for each channel of the
    N model using original calibration dataset and unified dataset. Blue
    refers to original dataset and orange to unified dataset.
    Data displayed as median line with shading representing
    89\% HDPI of 100 samples from the parameter posterior distribution.}
\label{fig:nyg_adding_data}
\end{figure}

\begin{figure}[!h]
\centering\includegraphics[width=\textwidth]{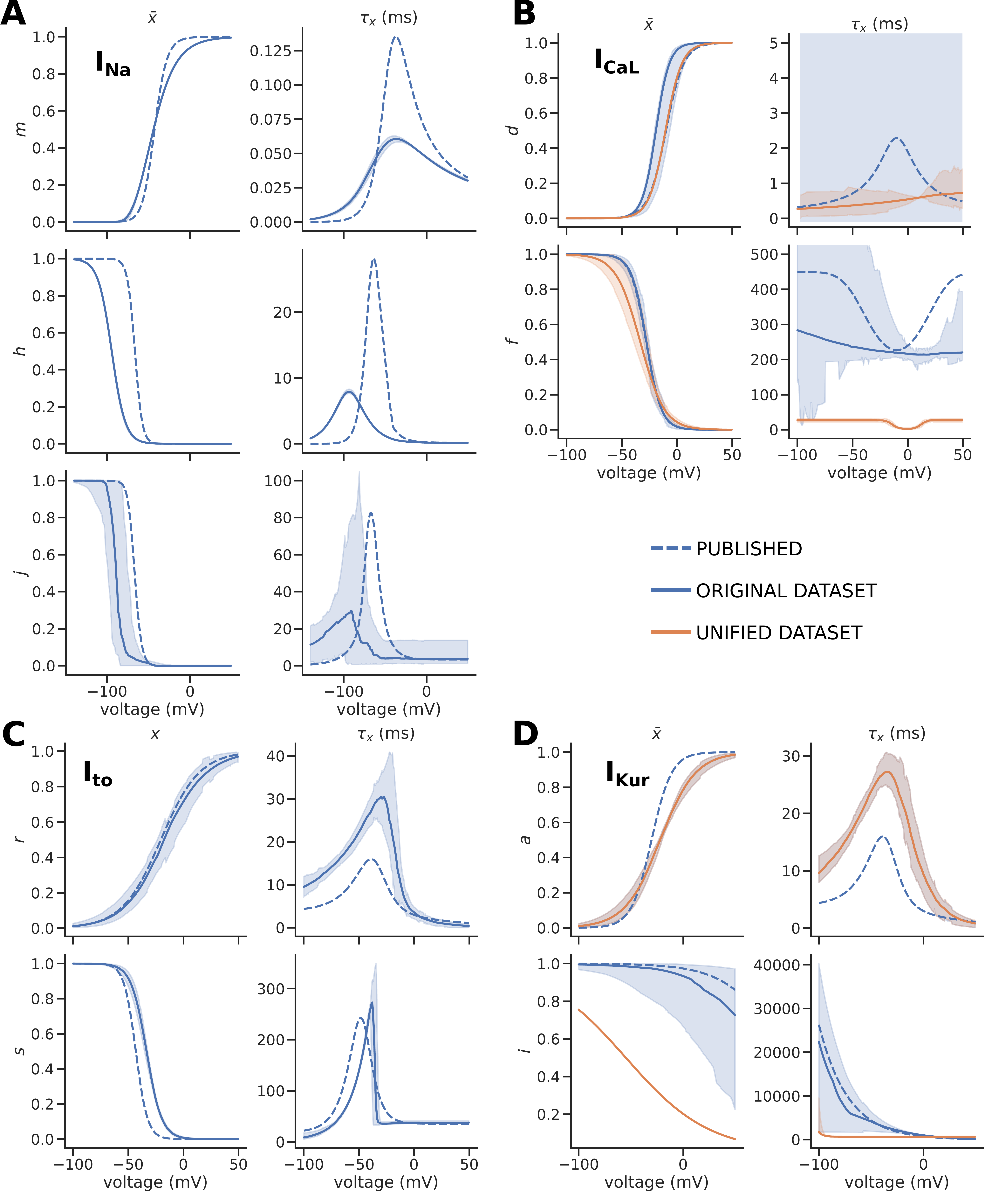}
\caption{
%\DIFF{\bf Figure updated for median +- HDPI.}
\textbf{A-D} Steady-state and time constant functions for each channel of the
    C model using original calibration dataset and unified dataset. Blue
    refers to original dataset and orange to unified dataset. 
    Data displayed as median line with shading representing
    89\% HDPI of 100 samples from the parameter posterior distribution.
    Note for $I_\text{Na}$ and
    $I_\text{to}$ the original and unified dataset are the same.}
\label{fig:cou_adding_data}
\end{figure}

\clearpage
\newpage
\subsection{$I_\mathrm{CaL}$}

\begin{figure}[!h]
\centering\includegraphics[width=\textwidth]{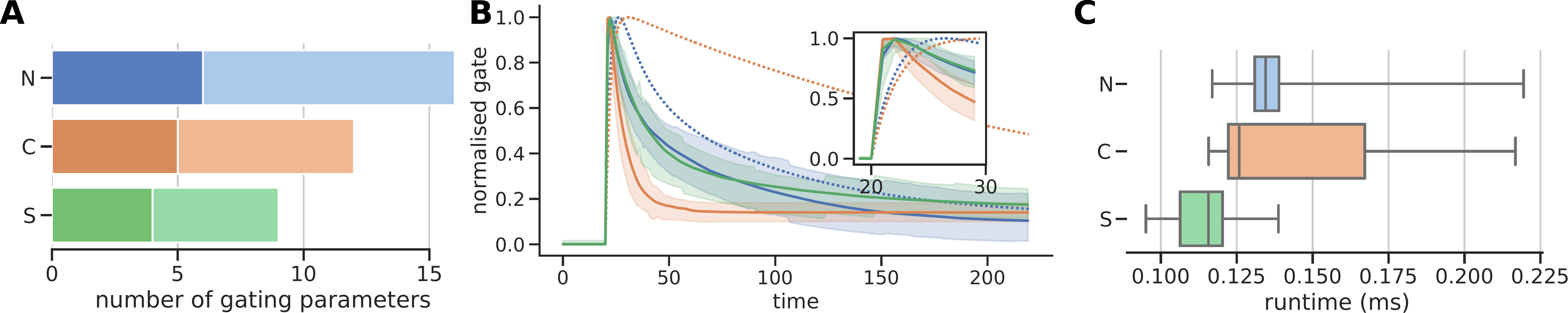}
    \caption{
    %\DIFF{\bf Figure updated for traces to median +- HDPI}
    \textbf{A} Number of gating parameters in equations for
    each $I_\text{CaL}$ model, separated into activation (dark)
    and inactivation (light).
    \textbf{B} Example traces from each model generated from the last
    step of a train of 100 steps from -80 mV to -10 mV for 200 ms at a rate of
    1 Hz using 100 samples from the parameter posterior distributions. Higher detail of the activation portion
    of the trace is shown in
    the inset plot.
    Output is summarised as median line with shading
    representing 89\% HDPI. 
    Dashed lines indicate 
    the response of the published N and C models. 
    \textbf{C} Boxplot comparing runtime of the simulation to
    generate each trace in B for each model. 
}
\label{fig:ical-extra}
\end{figure}

\subsection{$I_\mathrm{to}$}

\begin{figure}[!h]
\centering\includegraphics[width=\textwidth]{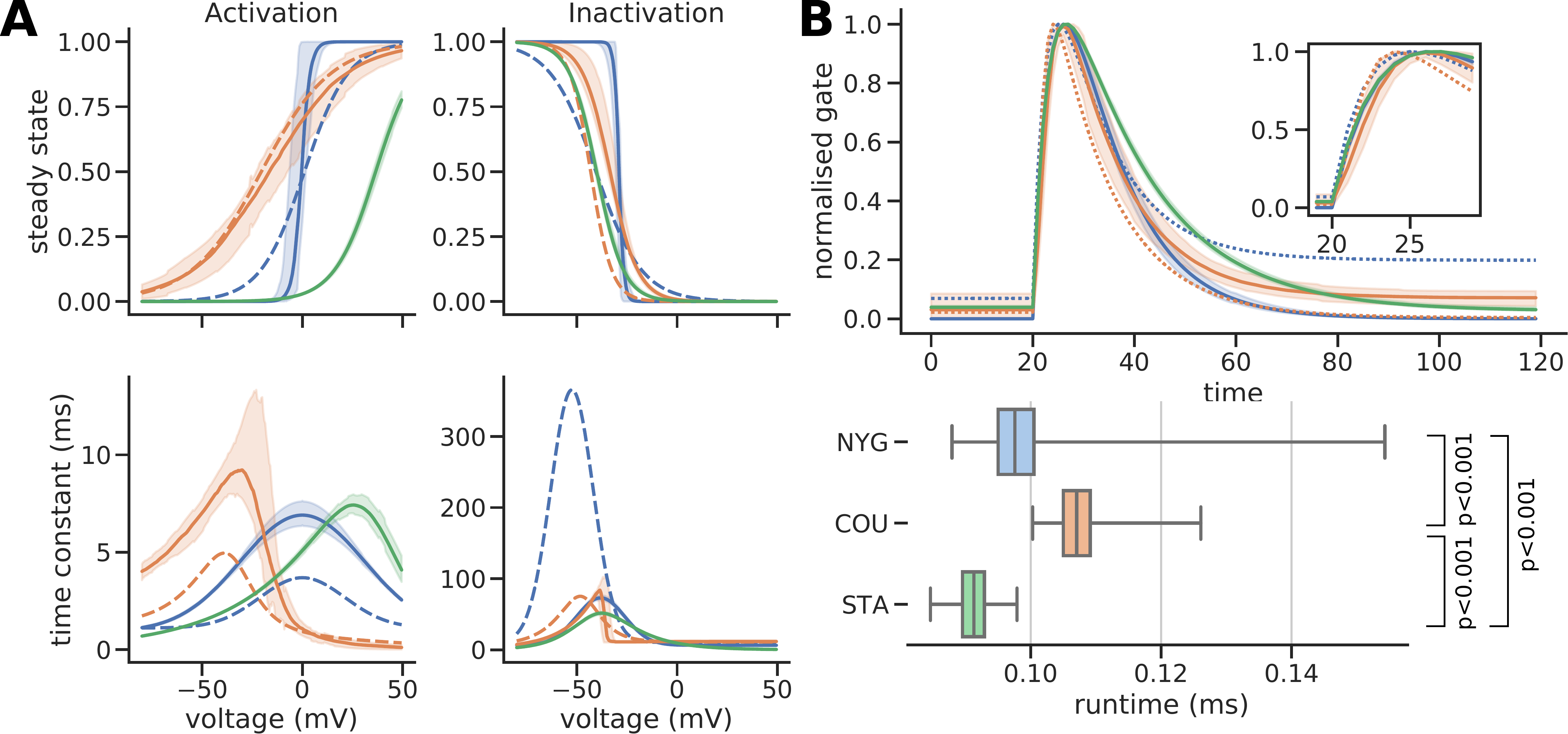}
    \caption{
    %\DIFF{\bf Figure updated for median +- HDPI}
    \textbf{A} Steady-state and time constant functions for
    $I_\text{to}$ model. Plotted as median line
    with shading indicated 89\% HDPI from 100 samples
    from the parameter posterior distribution. Dashed lines indicate the published values.
    \textbf{B} Example traces from each model
    generated from the last step of a pulse train of 100 steps from -50 mV to -10 mV
    for 100 ms at a rate of 1 Hz using samples in A. 
    Higher detail of activation portion of trace is shown in inset.
    Dashed lines indicate 
    the published N and C models. The boxplot below compares runtime of the simulations to generate the above traces for each model.
}
\label{fig:ito-extra}
\end{figure}

\clearpage
\newpage
\subsection{$I_\mathrm{Kur}$}

\begin{figure}[!h]
\centering\includegraphics[width=\textwidth]{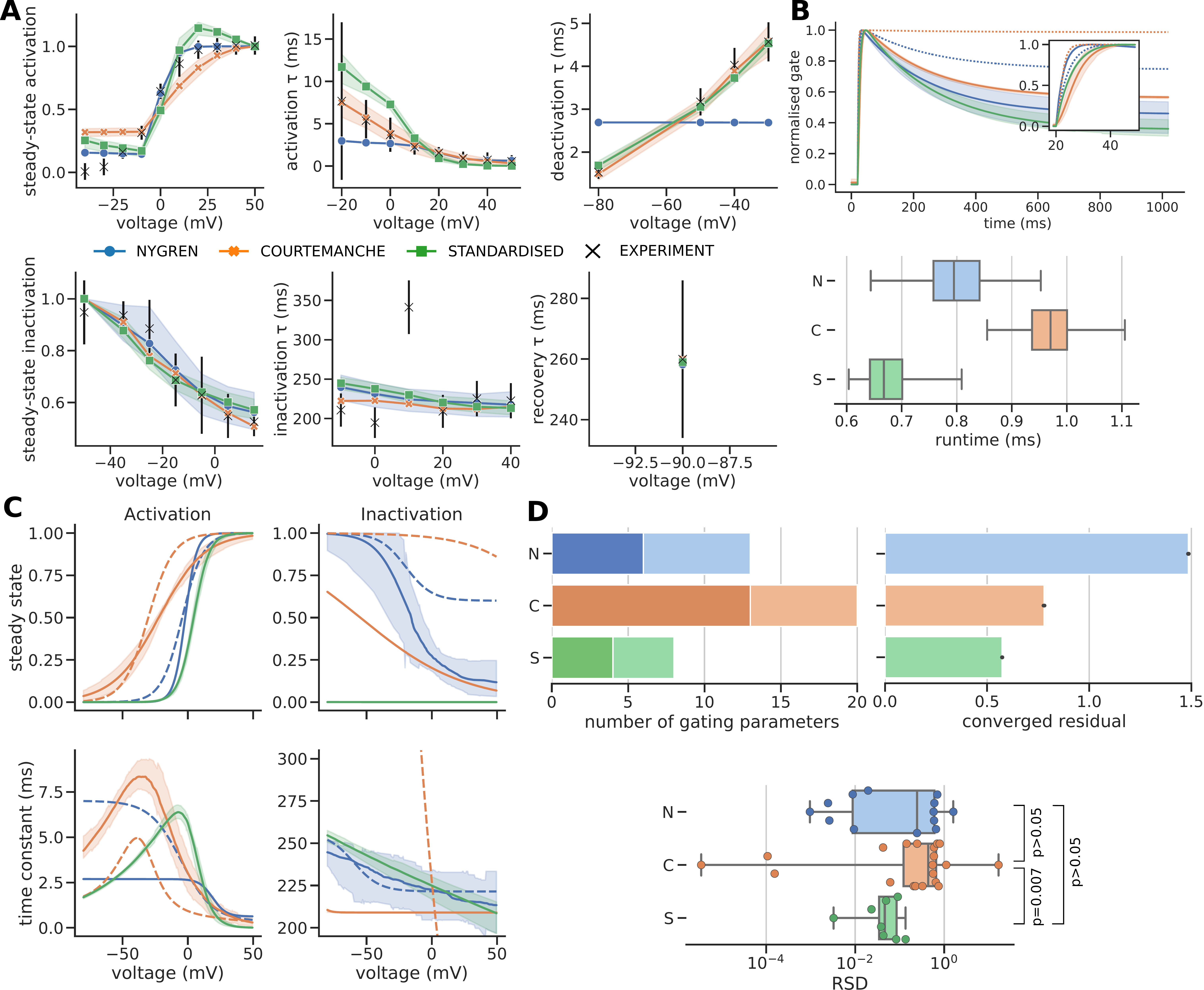}
    \caption{
    % \DIFF{\bf Figure has been updated to show median
    % line with HDPI and rearranged to remove per-experiment
    % residual plot, which is included in a following figure.} 
    \textbf{A} Results of calibrating each $I_\text{Kur}$ model to the
    unified dataset. 
    Model output is plotted as median with shading
    representing 89\% HDPI generated from 100 samples from
    parameter posterior distributions. 
    \textbf{B} Example traces from each model
    generated from the last step of a pulse train of 100 steps from -50 mV to -10 mV
    for 1000 ms at a rate of 0.1 Hz using samples from A. 
    Dashed lines indicate 
    the original N and C models. 
    Higher detail of activation portion of trace is shown in inset plot.
    Boxplot compares runtime of simulations to generate
    the traces above for each model.
    \textbf{C} Steady-state and time
    constant functions for each gate generated from samples in A displayed as median lines with shading showing 89\% HDPI.
    Dashed lines indicate original N and C models.
    \textbf{D} Number of gating parameters (top-left). 
    Dark and light shading correspond to activation gate
    and inactivation gate parameters respectively.
    Goodness of fit assessed by converged residuals
    from ABC (top-right). RSD of parameter posteriors (bottom).
    Mann-Whitney U-test used to test significance.}
\label{fig:isus_all}
\end{figure}

\clearpage
\newpage
\subsection{Goodness-of-fit residuals}
\label{ssec:goodness-of-fit}

\begin{figure}[!h]
\centering\includegraphics[width=\textwidth]{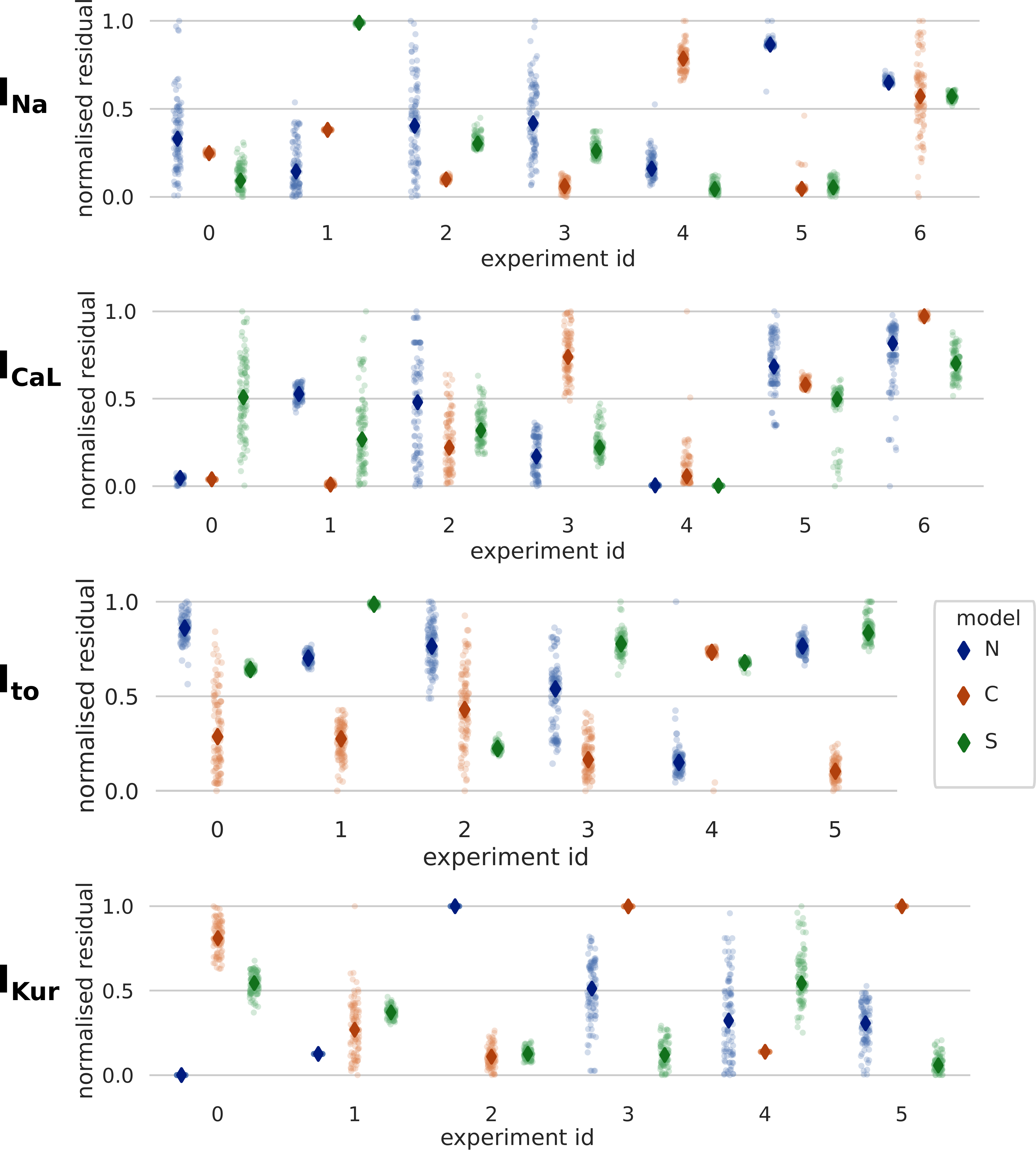}
    \caption{
    % \DIFF{\bf Edited figure to update for new simulations (INa
    % and ICaL).}
    Min-max normalised residuals by experiment and model. Experiment ID follows order of appearance in main
    figure for that channel.}
\label{fig:relative-residuals}
\end{figure}

\newpage
\clearpage
\subsection{Action potential response}
\label{ssec:ap-response}

\begin{figure}[!h]
\centering\includegraphics[width=\textwidth]{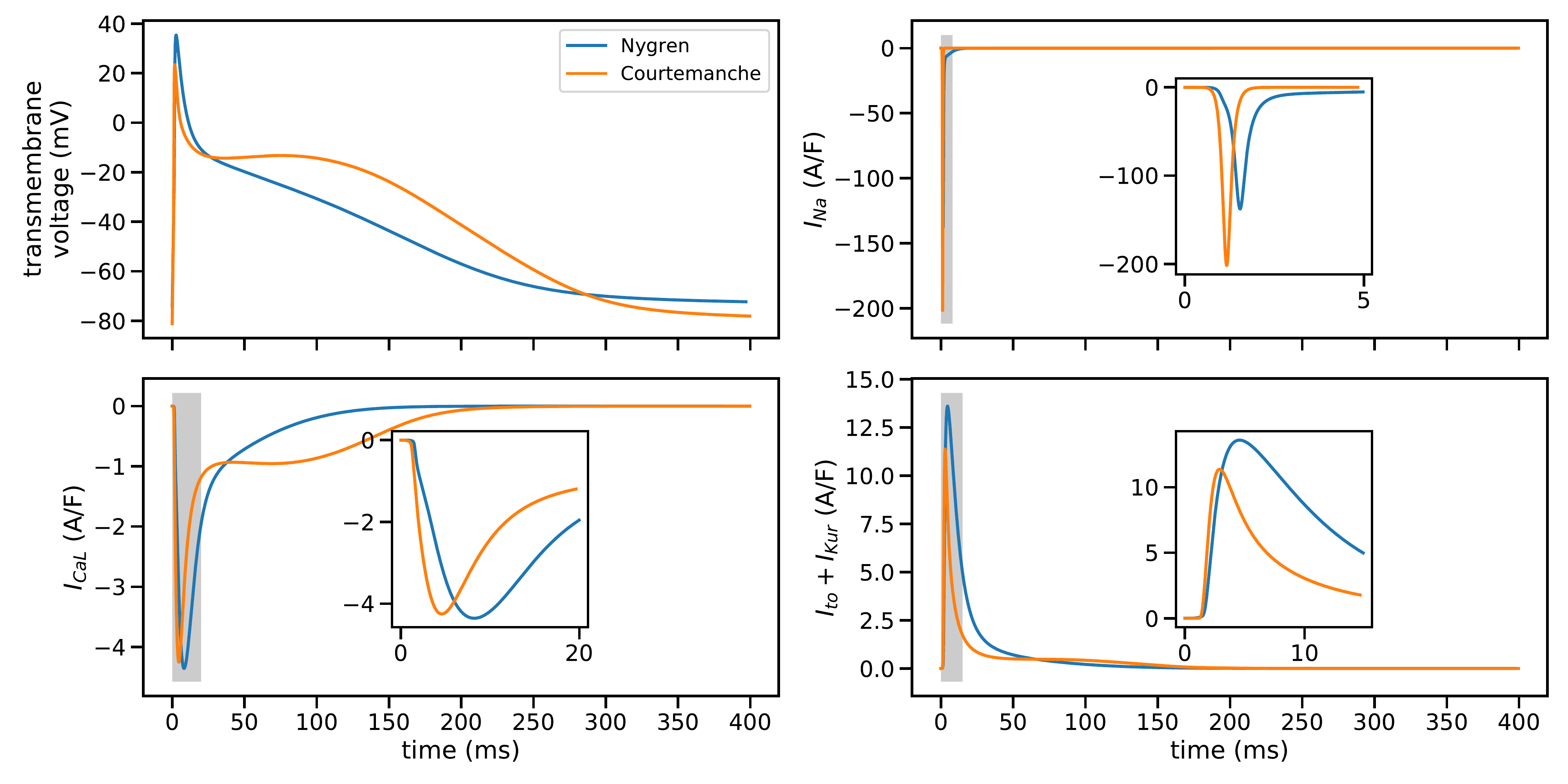}
\caption{Comparison of action potential morphology and major ion currents
    underlying the action potential in published N and C models of the human atrial
    cardiomyocyte. Inset graphs show more detail of the shaded time portion 
    in the main graphs. Action potentials were stimulated by 100s of pacing at a basic cycle length of 1 s using a stimulus current of 40
    pA/pF for 1 ms. The plot shows the final pulse from the
    pulse train protocol.} 
\label{fig:ap-published}
\end{figure}

\begin{figure}[!h]
\centering\includegraphics[width=0.9\textwidth]{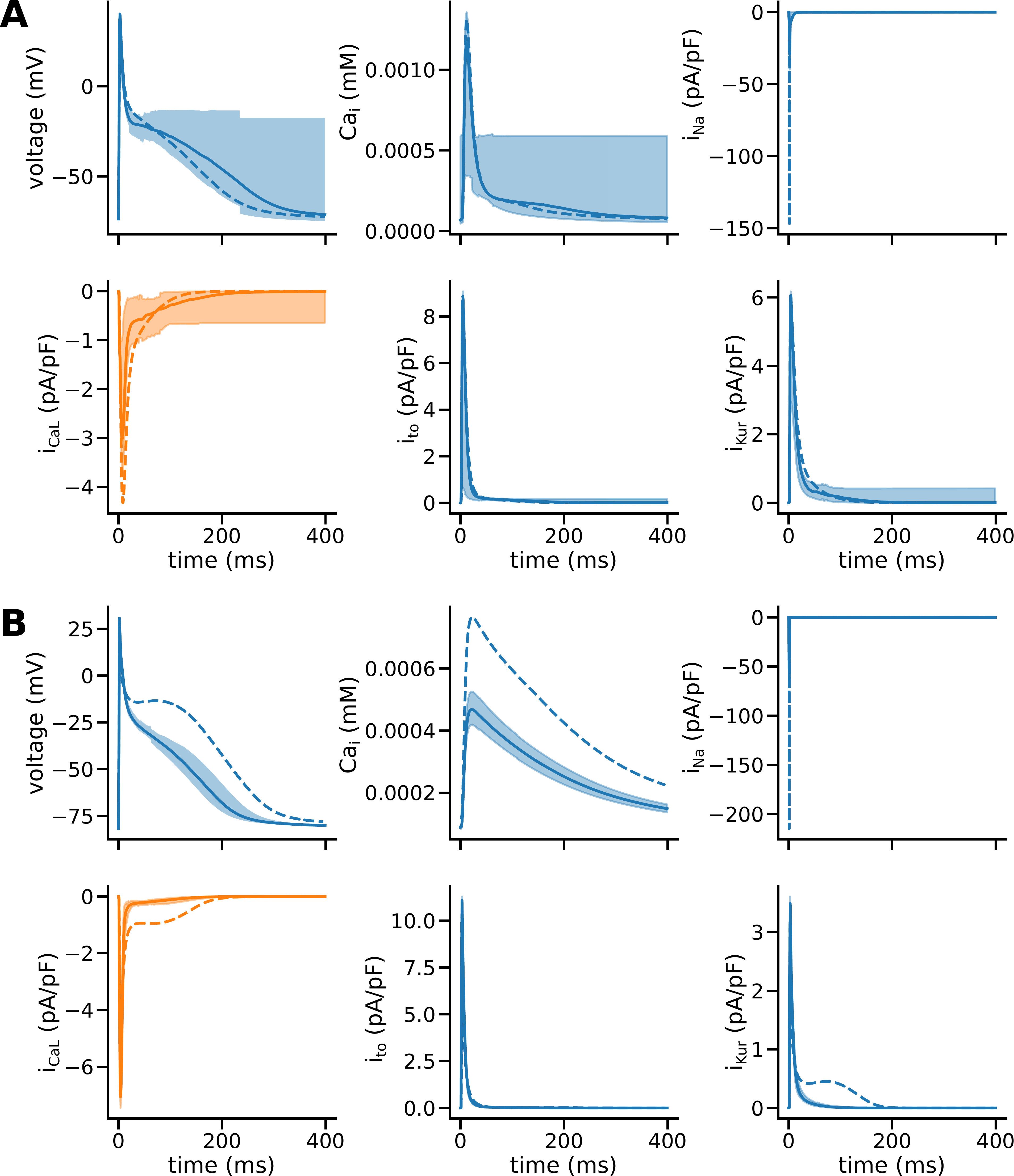}
    \caption{Action potential, major currents and intracellular
    calcium concentrations for A: N model and B: C model. Traces
    using published parameters are represented with dashed lines.
    Samples from the full cell model with $I_\mathrm{CaL}$ 
    replaced by the unified recalibrated parameterisation are displayed as a median line with 89\% high density
    posterior intervals. The orange plot highlights the channel
    that is recalibrated using the parameter posterior distribution
    from ABC. All other model parameters are unchanged from
    published values.} 
    \label{fig:ap-ical}
\end{figure}

\begin{figure}[!h]
\centering\includegraphics[width=0.9\textwidth]{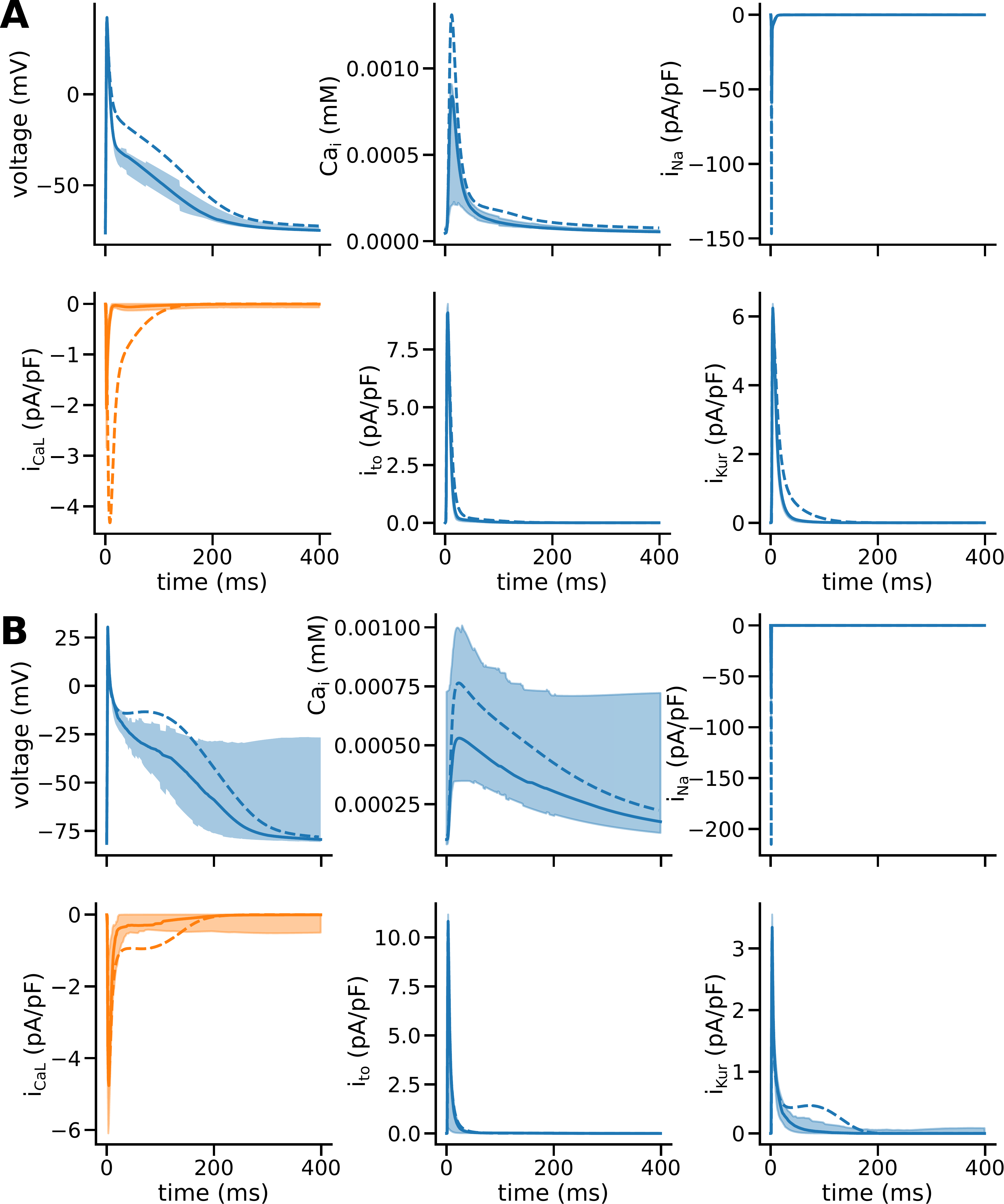}
    \caption{Action potential, major currents and intracellular
    calcium concentrations for A: N model and B: C model, with
    the S form of $I_\mathrm{CaL}$ replacing the original form. Traces
    using published parameters are represented with dashed lines.
    Samples from the full cell model with $I_\mathrm{CaL}$ 
    replaced by the unified recalibrated parameterisation are displayed as a median line with 89\% high density
    posterior intervals. The orange plot highlights the channel
    that is replaced with the standardised model and using the parameter posterior distribution
    from ABC. All other model parameters are unchanged from
    published values.} 
    \label{fig:ap-ical-s}
\end{figure}

\begin{figure}[!h]
\centering\includegraphics[width=0.9\textwidth]{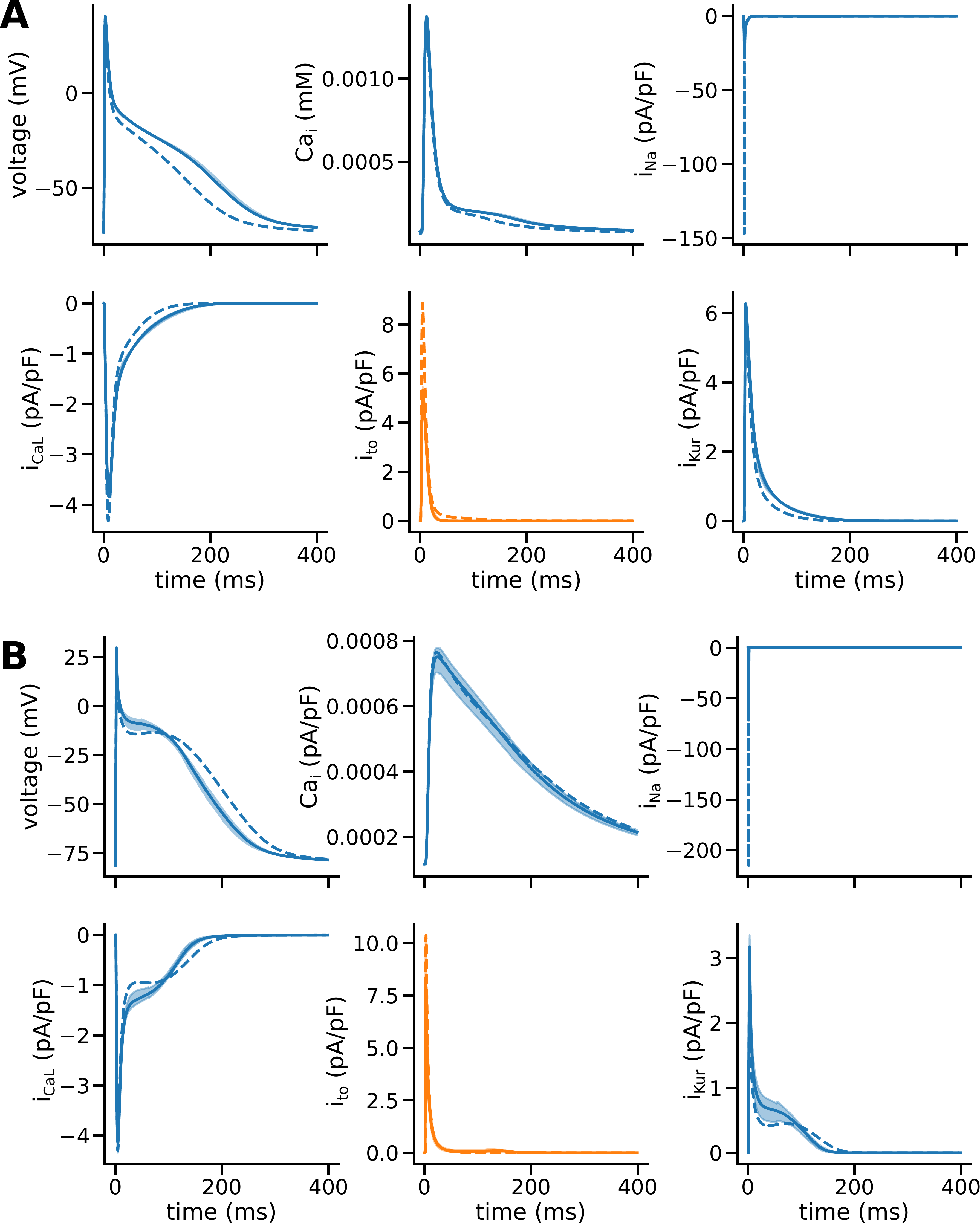}
    \caption{Action potential, major currents and intracellular
    calcium concentrations for A: N model and B: C model. Traces
    using published parameters are represented with dashed lines.
    Samples from the full cell model with $I_\mathrm{to}$ 
    replaced by the unified recalibrated parameterisation are displayed as a median line with 89\% high density
    posterior intervals.. The orange plot highlights the channel
    that is recalibrated using the parameter posterior distribution
    from ABC. All other model parameters are unchanged from
    published values.} 
    \label{fig:ap-ito}
\end{figure}

\begin{figure}[!h]
\centering\includegraphics[width=0.9\textwidth]{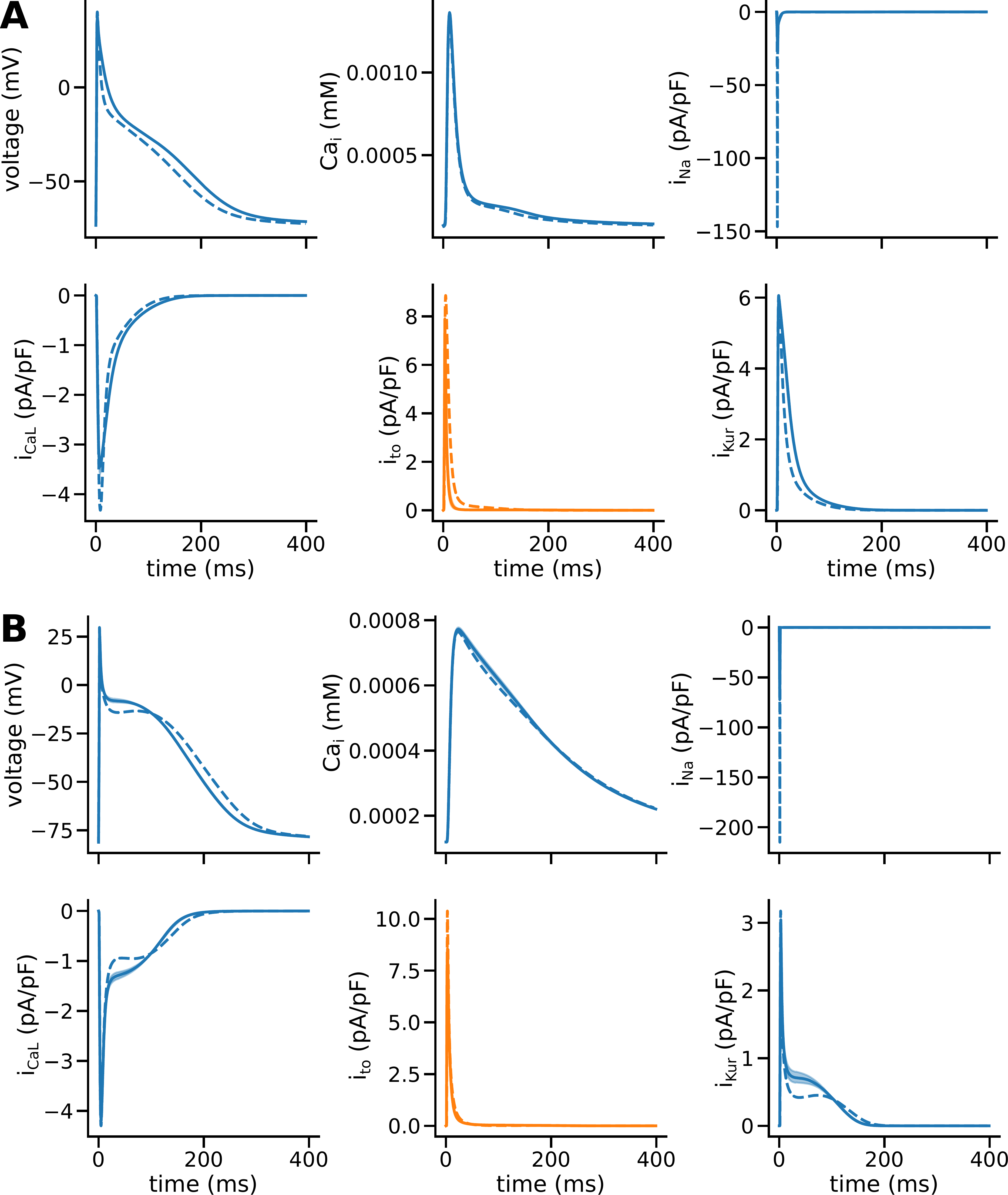}
    \caption{Action potential, major currents and intracellular
    calcium concentrations for A: N model and B: C model, , with
    the S form of $I_\mathrm{to}$ replacing the original form. Traces
    using published parameters are represented with dashed lines.
    Samples from the full cell model with $I_\mathrm{to}$ 
    replaced by the unified recalibrated parameterisation are displayed as a median line with 89\% high density
    posterior intervals. The orange plot highlights the channel
    that is replaced with the standardised model and using the parameter posterior distribution
    from ABC. All other model parameters are unchanged from
    published values.} 
    \label{fig:ap-ito-s}
\end{figure}

\begin{figure}[!h]
\centering\includegraphics[width=0.9\textwidth]{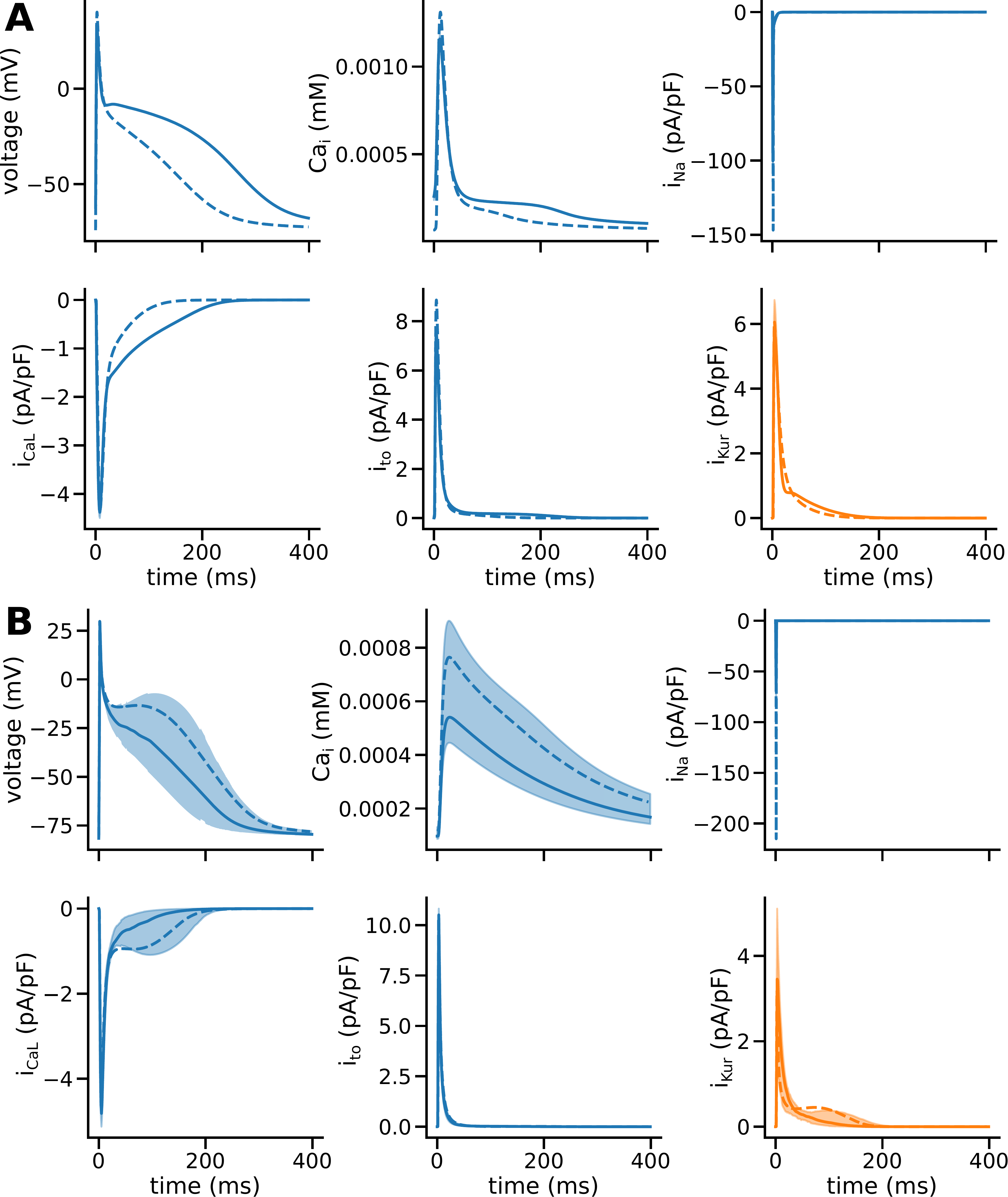}
    \caption{Action potential, major currents and intracellular
    calcium concentrations for A: N model and B: C model. Traces
    using published parameters are represented with dashed lines.
    Samples from the full cell model with $I_\mathrm{Kur}$ 
    replaced by the unified recalibrated parameterisation are displayed as a median line with 89\% high density
    posterior intervals. The orange plot highlights the channel
    that is recalibrated using the parameter posterior distribution
    from ABC. All other model parameters are unchanged from
    published values.} 
    \label{fig:ap-isus}
\end{figure}

\begin{figure}[!h]
\centering\includegraphics[width=0.9\textwidth]{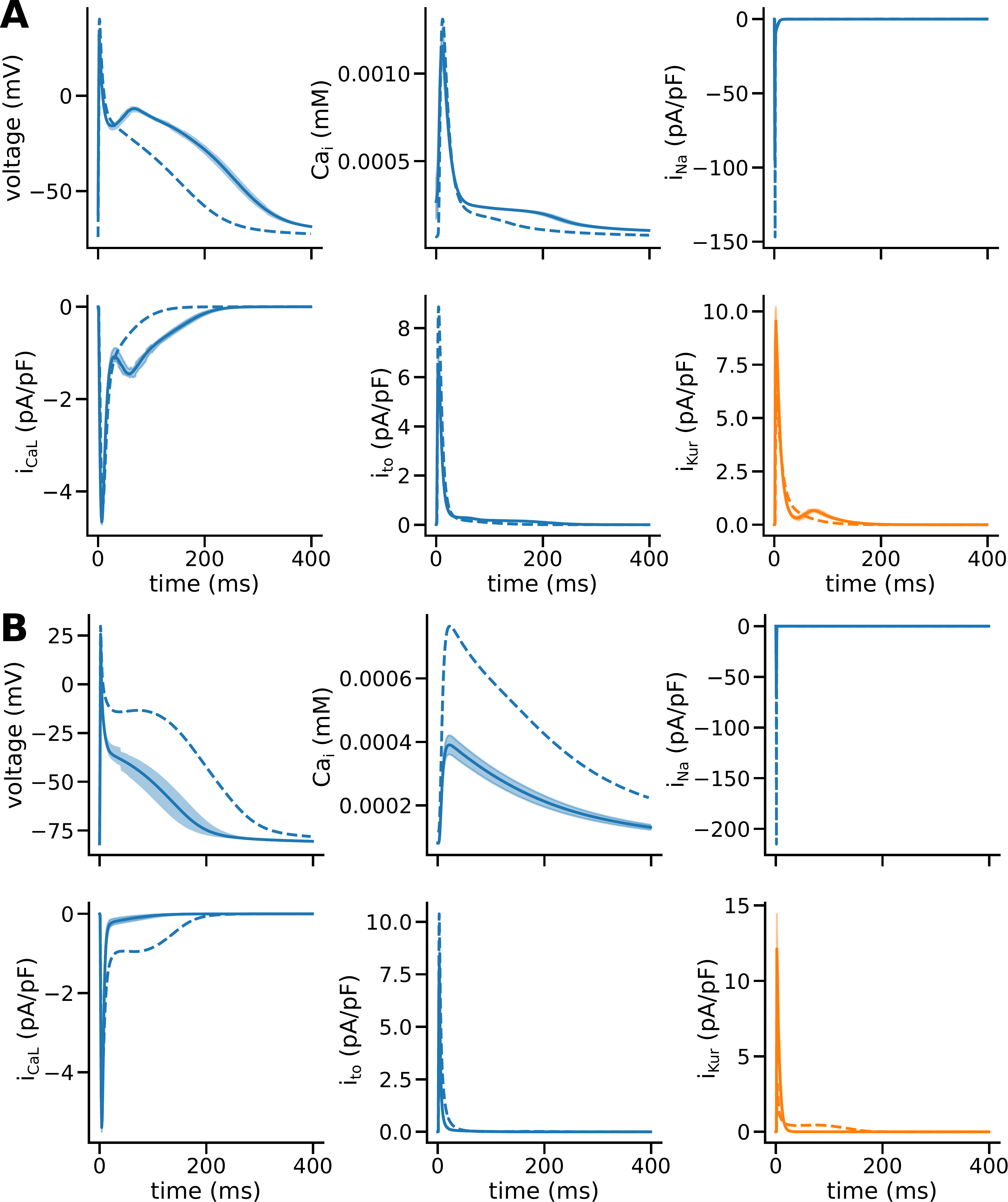}
    \caption{Action potential, major currents and intracellular
    calcium concentrations for A: N model and B: C model, , with
    the S form of $I_\mathrm{Kur}$ replacing the original form. Traces
    using published parameters are represented with dashed lines.
    Samples from the full cell model with $I_\mathrm{Kur}$ 
    replaced by the unified recalibrated parameterisation are displayed as a median line with 89\% high density
    posterior intervals. The orange plot highlights the channel
    that is replaced with the standardised model and using the parameter posterior distribution
    from ABC. All other model parameters are unchanged from
    published values.} 
    \label{fig:ap-isus-s}
\end{figure}

\begin{table}[!h]
\begin{tabular}{@{}llllll@{}}
\toprule
Model & Measure (units) & Published & $I_\mathrm{CaL}$ (89\% HDPI) & $I_\mathrm{to}$ (89\% HDPI) & $I_\mathrm{Kur}$ (89\% HDPI) \\ \midrule
N & RP (mV)         & -74.1    & -73.7 (-76.3, -17.5)        & -72.7 (-72.7, -72.5)       & -62.2 (-63.0, -61.4)        \\
& AMP (mV)        & 114.3     & 113.4 (39.0, 115.8)          & 113.4 (113.1, 113.6)       & 95.0 (94.2, 95.8)           \\
& APD90 (ms)      & 223.2     & 217.5 (16, 295)               & 271 (266, 280)             & 298 (295, 302)              \\ \midrule
C & RP (mV)         & -81.2     & -81.7 (-81.8, -81.6)        & -81.3 (-81.4, -81.2)       & -81.6 (-81.8, -80.9)        \\
& AMP (mV)        & 110.9     & 112.5 (112.3, 112.9)          & 109.4 (107.5, 110.9)       & 111.2 (109.8, 111.8)           \\
& APD90 (ms)      & 288.5     & 211 (189, 246)               & 257 (247, 266)             & 237.5 (190, 304)              \\ \midrule
N+S & RP (mV)         & -74.1    & -76.3 (-76.7, -70.3)        & -73.2 (-73.2, -73.1)       & -62.2 (-66.1, -58.0)        \\
& AMP (mV)        & 114.3     & 118.6 (106.1, 120.3)          & 110.8 (110.6, 111.0)       & 93.9 (88.2, 98.8)           \\
& APD90 (ms)      & 223.2     & 172.5 (131, 233)               & 248 (245, 249)             & 286 (283, 293)              \\ \midrule
C+S & RP (mV)         & -81.2     & -81.5 (-82.0, -27.1)        & -81.2 (-81.2, -81.2)       & -81.9 (-82.0, -81.8)        \\
& AMP (mV)        & 110.9     & 111.7 (39.4, 112.9)          & 109.2 (109.1, 109.3)       & 107.6 (106.4, 108.9)           \\
& APD90 (ms)      & 288.5     & 192 (35, $\infty$)               & 267 (264, 270)             & 179.5 (150, 206)              \\ \bottomrule
\end{tabular}
    \caption{Effect on full AP of using parameter posterior distributions
    to calibrate channel models. Measurements taken are
    resting potential (RP), action potential amplitude (AMP) and
    action potential duration to 90\% repolarisation (APD90).
    N: Nygren model, C: Courtemanche model, +S: represents the
    indicated channel was replaced with the standardised form
    in the full AP model.}
\end{table}

\newpage
\clearpage
\section{Model equations and numerical results}
\label{sec:numerical_results}

\subsection{$I_\mathrm{Na}$}

\begin{table}[!h]
    \[
        \begin{aligned}
            I_{\mathrm{Na}}&=P_{\mathrm{Na}} m^{3}\left(\mathbf{s_1} h_{1}+(1-\mathbf{s_1}) h_{2}\right)\left[\mathrm{Na}^{+}\right]_{\mathrm{c}} \mathrm{V} \frac{\mathrm{F}^{2}}{\mathrm{RT}} \frac{e^{\left(\mathrm{V}-\mathrm{E}_\mathrm{Na}\right) \mathrm{F} / \mathrm{RT}}-1.0}{e^{\mathrm{VF} / \mathrm{RT}}-1.0}\\
            \frac{\mathrm{d}m}{\mathrm{dt}}&=\frac{\widebar{m}-m}{\tau_{m}},\qquad
            \frac{\mathrm{d}h_1}{\mathrm{dt}}=\frac{\widebar{h}-h_1}{\tau_{h_1}},\qquad
            \frac{\mathrm{d}h_2}{\mathrm{dt}}=\frac{\widebar{h}-h_2}{\tau_{h_2}}\\
            \widebar{m} &= \frac{1.0}{1.0+e^{-(\mathrm{V}+\mathbf{r_1})/\mathbf{r_2}}},\qquad
            \tau_m = 1000(\mathbf{r_3} e^{-((\mathrm{V}+\mathbf{r_4})/\mathbf{r_5})^2} + \mathbf{r_6})\\
            %r_1&=27.12, r_2=8.21, r_3=4.2\mathrm{e}{-5}, r_4=25.57, r_5=28.8,
            %r_6=2.4\mathrm{e}{-5}\\
            \widebar{h} &= \frac{1.0}{1.0+e^{(\mathrm{V}+\mathbf{q_1})/\mathbf{q_2}}}\\
            \tau_{h_1} &= 1000\left(\frac{\mathbf{q_3}}{1.0+e^{(\mathrm{V}+\mathbf{q_4})/\mathbf{q_5}}}+\mathbf{q_6}\right),\qquad
            \tau_{h_2} = 1000\left(\frac{\mathbf{q_7}}{1.0+e^{(\mathrm{V}+\mathbf{q_4})/\mathbf{q_5}}}+\mathbf{q_8}\right)\\
            %q_1&=63.6, q_2=5.3, q_3=0.03, q_4=35.1, q_5=3.2,
            %q_6=3\mathrm{e}{-4}, q_7=0.12, q_8=3\mathrm{e}{-3}\\
        \end{aligned}
    \]
    \caption{Gating kinetics in Nygren model of $I_\text{Na}$ channel
    current(see Table 6 in~\cite{nygren_mathematical_1998}). Time constants are multiplied by 1000 to convert from $s$ to
    $ms$.} 
\end{table}

\begin{landscape}
    \begin{table}[!h]
    \centering
\begin{tabular}{@{}lrlrcErcE}
\toprule
     &           & \multicolumn{3}{c}{Original}                    & \multicolumn{3}{c}{Unified}                        \\ \midrule
    \multicolumn{1}{c}{Name} &
    \multicolumn{1}{c}{Published} &
    \multicolumn{1}{c}{Prior} &
    \multicolumn{1}{c}{Median} &
    \multicolumn{1}{c}{89\% HDPI} &
    \multicolumn{1}{c}{$\mathrm{log}_{10}$RSD} &
    \multicolumn{1}{c}{Median} &
    \multicolumn{1}{c}{89\% HDPI} &
    \multicolumn{1}{c}{$\mathrm{log}_{10}$RSD} \\ \midrule
    r1   & 27.12     &  $\mathcal{U}$(0, 100)& 52.25     & (50.93, 53.52)         & -1.795        & 42.83     & (33.91, 55.13)         & -0.7966       \\
    r2   & 8.21      &  $\mathcal{U}$(0, 20) & 10.69     & (9.769, 11.64)         & -1.256        & 13.19     & (7.761, 19.94)         & -0.5127       \\
    r3$^\ast$   & 4.2e-05   &  $\mathcal{U}$(-6, -3) & 2.097e-05 & (1.009e-06, 0.000325)  & -0.749        & 1.371e-05 & (1.002e-06, 0.0002247) & -0.7856       \\
    r4   & 25.57     &  $\mathcal{U}$(0, 100)& 57.99     & (22.82, 99.97)         & -0.3404       & 56.79     & (16.97, 99.79)         & -0.293        \\
    r5   & 28.8      &  $\mathcal{U}$(0, 20) & 9.979     & (0.2856, 17.75)        & -0.2282       & 8.932     & (0.03437, 17.49)       & -0.1981       \\
    r6$^\ast$   & 2.4e-05   &  $\mathcal{U}$(-6, -3) & 2.191e-05 & (1.001e-06, 0.0005717) & -0.7021       & 9.662e-05 & (5.195e-05, 0.0001536) & -1.43         \\
    q1   & 63.6          &  $\mathcal{U}$(0, 200) & 95.57     & (95.11, 96.11)         & -2.465        & 93.94     & (88.3, 99.55)          & -1.4          \\
    q2   & 5.3           &  $\mathcal{U}$(0, 20) & 6.5       & (6.085, 6.841)         & -1.432        & 8.182     & (1.222, 13.13)         & -0.3487       \\
    q3$^\ast$   & 0.03          &  $\mathcal{U}$(-3, 0) & 0.09214   & (0.001002, 0.4502)     & -0.1696       & 0.004096  & (0.002543, 0.02547)    & -0.7312       \\
    q4   & 35.1          &  $\mathcal{U}$(0, 100) & 34.98     & (0.1474, 147.7)        & -0.02249      & 53.57     & (48.79, 57.53)         & -1.263        \\
    q5   & 3.2           &  $\mathcal{U}$(0, 20) & 12.76     & (4.548, 19.99)         & -0.361        & 10.61     & (7.184, 13.18)         & -0.7463       \\
q6$^\ast$   & 0.0003        &  $\mathcal{U}$(-5, -2) & 0.0002548 & (1.002e-05, 0.004472)  & -0.6137       & 0.0002403 & (1.006e-05, 0.001556)  & -0.7811       \\
q7$^\ast$   & 0.12          &  $\mathcal{U}$(-3, 0)& 0.07443   & (0.001005, 0.3999)     & -0.1983       & 0.02022   & (0.002551, 0.02706)    & -0.6835       \\
q8$^\ast$   & 0.003         &  $\mathcal{U}$(-4, -1)& 0.002224  & (0.0001001, 0.04665)   & -0.4589       & 0.0005268 & (0.0001001, 0.00171)   & -0.9219       \\
    s1   & 0.9           &  $\mathcal{U}$(0, 1) & 0.5015    & (0.08985, 0.9754)      & -0.2317       & 0.4717    & (0.1367, 0.8259)       & -0.334        \\ \bottomrule
\end{tabular}
\caption{Summary of results for parameters of Nygren $I_\text{Na}$ model using original and unified dataset. 
        $^\ast$Parameters were searched in $\mathrm{log}_{10}$ space based on the scale of their published values and are presented in linear space. The prior for these parameters is still in the original $\mathrm{log}_{10}$ space.}
\end{table}
\end{landscape}

\begin{table}[!h]
    \[
        \begin{aligned}
            I_{\mathrm{Na}}&=g_\text{Na}m^3 h j (\mathrm{V}-\mathrm{E}_\mathrm{Na})\\
            \frac{\mathrm{d}\phi}{\mathrm{dt}}&=\frac{\phi_\infty-\phi}{\tau_\phi},\qquad\text{ for } \phi=m, h, j\\
            \tau_\phi &= \left(\alpha_\phi + \beta_\phi\right)^{-1},\qquad \phi_\infty=\alpha_\phi\tau_\phi,\qquad\text{ for } \phi=m, h, j\\
            \alpha_{m}&=\left\{\begin{array}{l}{\mathbf{a_{m,2}} \frac{\mathrm{V}-\mathbf{a_{m,1}}}{1-e^{-\mathbf{a_{m,3}}(\mathrm{V}-\mathbf{a_{m,1}})}}} \\{\mathbf{a_{m,4}}, \text { if } \mathrm{V}=\mathbf{a_{m,1}}}\end{array}\right.,\qquad
            \beta_{m}=\mathbf{b_{m,1}}e^{-\mathrm{V}/\mathbf{b_{m,2}}}\\
            \alpha_{h}&=\left\{\begin{array}{l}{\mathbf{a_{h,1}} e^{(\mathrm{V}+\mathbf{a_{h,3}})/-\mathbf{a_{h,2}}}} \\
                {a_{h,5},\qquad \text { if } \mathrm{V}\geq\mathbf{c_{h,1}}}
            \end{array}\right.,\qquad
            a_{h,5}=\mathbf{a_{h,1}} e^{(\mathbf{c_{h,1}}+\mathbf{a_{h,3}})/-\mathbf{a_{h,2}}}\\
            \beta_{h}&=\left\{\begin{array}{ll}
                {\mathbf{b_{h,4}} e^{\mathbf{b_{h,5}} \mathrm{V}}+\mathbf{b_{h,6}} e^{\mathbf{b_{h,7}} \mathrm{V}}} \\
                {\left(b_{h,1}\left[1.0+e^{(V+\mathbf{b_{h,2}})/-\mathbf{b_{h,3}}}\right]\right)^{-1},} & {\text { if } V \geq\mathbf{c_{h,1}}}
            \end{array}\right.\\
                b_{h,1} &= \left(\mathbf{b_{h,4}}e^{\mathbf{b_{h,5}}\mathbf{c_{h,1}}} + \mathbf{b_{h,6}} e^{\mathbf{b_{h,7}}\mathbf{c_{h,1}}}\right)^{-1} \left(1.0+e^{(\mathbf{c_{h,1}}+\mathbf{b_{h,2}})/-\mathbf{b_{h,3}}}\right)^{-1}\\
            \alpha_{j}&=\left\{\begin{array}{l}
                [\left.-\mathbf{a_{j,1}} e^{\mathbf{a_{j,2}} \mathrm{V}}-\mathbf{a_{j,3}} e^{-\mathbf{a_{j,4}} \mathrm{V}}\right] \frac{\mathrm{V}+a_{j,5}}{1.0+e^{\mathbf{a_{j,6}}(\mathrm{V}+\mathbf{a_{j,7}})}} \\
                {0, \quad \text { if } V \geq\mathbf{c_{j,1}}}
            \end{array}\right.,\qquad
            a_{j,5} = -\mathbf{c_{j,1}}\\
            \beta_{j}&=\left\{\begin{array}{l}
                {\mathbf{b_{j,5}} \frac{e^{-\mathbf{b_{j,6}} \mathrm{V}}}{1+e^{-\mathbf{b_{j,7}}(\mathrm{V}+\mathbf{b_{j,8}})}}} \\
                { b_{j,1}\frac{e^{-b_{j,2}\mathrm{V}}}{1+e^{-\mathbf{b_{j,3}}(\mathrm{V}+\mathbf{b_{j,4}})}}, \quad \text { if } V \geq\mathbf{c_{j,1}}}
            \end{array}\right.,\qquad\\
            b_{j,1}&= \mathbf{b_{j,5}} \frac{e^{\mathbf{-b_{j,6}} \mathbf{c_{j,1}}}}{1 + e^{\mathbf{-b_{j,7}} (\mathbf{c_{j,1}} + \mathbf{b_{j,8}})}} \left[\frac{e^{-b2_j \mathbf{c_{j,1}}}}{(1 + e^{-\mathbf{b_{j,3}} (\mathbf{c_{j,1}} + \mathbf{b_{j,4}})}}\right]^{-1},\qquad
            b_{j,2}=0.0\\
        \end{aligned}
    \]
    \caption{Gating kinetics in Courtemanche model of $I_\text{Na}$ channel
    current (see Appendix in~\cite{courtemanche_ionic_1998}). Values of $a_{h,5}$, $b_{h,1}$, $a_{j,5}$ and $b_{j,1}$ are set to
    enforce continuity in piecewise functions.
    $b_{j,2}$ is set to zero to reduce dimensionality of the calibration
    problem as the published parameter was effectively zero ($2.535\times 10^{-7}$).} 
\end{table}

\begin{table}[!h]
    \centering
\begin{tabular}{@{}lrlrcE}
    \toprule
     &           & \multicolumn{3}{c}{Original/Unified}                 \\ \midrule
    \multicolumn{1}{c}{Name} &
    \multicolumn{1}{c}{Published} &
    \multicolumn{1}{c}{Prior} &
    \multicolumn{1}{c}{Median} &
    \multicolumn{1}{c}{89\% HDPI} &
    \multicolumn{1}{c}{$\mathrm{log}_{10}$RSD} \\ \midrule
    $a1_m$ & -47.13    & $\mathcal{U}(-100, 0)$ & -75.97    & (-79.77, -73.65)      & -1.586        \\
    $a2_m$ & 0.32      & $\mathcal{U}(0, 1)$ & 0.2639    & (0.2476, 0.2741)      & -1.476        \\
$a3_m$ & 0.1       & $\mathcal{U}(0, 1)$ & 0.3355    & (0.1947, 0.7643)      & -0.2765       \\
$a4_m$ & 3.2       & $\mathcal{U}(0, 10)$ & 4.895     & (1.699, 9.998)        & -0.2421       \\
$b1_m$ & 0.08      & $\mathcal{U}(0, 10)$ & 1.263     & (1.225, 1.322)        & -1.64         \\
$b2_m$ & 11.0      & $\mathcal{U}(0, 100)$ & 23.21     & (22.48, 23.74)        & -1.774        \\ \midrule
$a1_h^\ast$ & 0.135     & $\mathcal{U}(-2, 1)$ & 0.3845    & (0.01157, 3.35)       & 0.269         \\
$a2_h$ & 6.8       & $\mathcal{U}(0, 50)$ & 15.48     & (14.49, 15.99)        & -1.465        \\
$a3_h$ & 80.0      & $\mathcal{U}(0, 200)$ & 122.2     & (67.51, 155.4)        & -0.6085       \\
$b2_h$ & 10.66     & $\mathcal{U}(0, 100)$ & 18.35     & (5.525, 31.86)        & -0.3218       \\
$b3_h$ & 11.1      & $\mathcal{U}(0, 50)$ & 16.73     & (12.42, 21.47)        & -0.7582       \\
$b4_h^\ast$ & 3.56      & $\mathcal{U}(-1, 2)$ & 15.11     & (14.51, 15.79)        & -1.999        \\
$b5_h^\ast$ & 0.079     & $\mathcal{U}(-3, 0)$ & 0.05808   & (0.05716, 0.05887)    & -2.492        \\
$b6_h^\ast$ & 310000.0  & $\mathcal{U}(3, 6)$ & 21500.0   & (1002.0, 321300.0)    & -0.6777       \\
$b7_h^\ast$ & 0.35      & $\mathcal{U}(-2, 1)$ & 1.327     & (0.2776, 6.324)       & 0.4741        \\
$c1_h$ & -40.0     & $\mathcal{U}(-100, 0)$ & -37.27    & (-45.94, -32.91)      & -0.8902       \\ \midrule
$a1_j^\ast$ & 127100.0  & $\mathcal{U}(3, 7)$ & 123700.0  & (1180.0, 4778000.0)   & -0.6361       \\
$a2_j^\ast$ & 0.2444    & $\mathcal{U}(-2, 2)$ & 5.144     & (0.1528, 50.75)       & 0.02476       \\
$a3_j^\ast$ & 3.474e-05 & $\mathcal{U}(-5, -1)$ & 0.0003798 & (1.015e-05, 0.001191) & -0.7783       \\
$a4_j^\ast$ & 0.04391   & $\mathcal{U}(-4, 0)$ & 0.003784  & (0.0001, 0.02896)     & -0.5131       \\
$a6_j$ & 0.311     & $\mathcal{U}(0, 1)$ & 0.4461    & (0.01244, 0.8846)     & -0.2087       \\
$a7_j$ & 79.23     & $\mathcal{U}(0, 100)$ & 41.53     & (4.671, 91.36)        & -0.1966       \\
$b3_j$ & 0.1       & $\mathcal{U}(0, 1)$ & 0.4481    & (0.001746, 0.8611)    & -0.1977       \\
$b4_j$ & 32.0      & $\mathcal{U}(0, 100)$ & 67.64     & (21.51, 99.97)        & -0.3641       \\
$b5_j$ & 0.1212    & $\mathcal{U}(0, 1)$ & 0.3552    & (0.04405, 0.7968)     & -0.1691       \\
$b6_j^\ast$ & 0.01052   & $\mathcal{U}(-4, 0)$ & 0.001531  & (0.0001004, 0.01829)  & -0.5618       \\
$b7_j$ & 0.1378    & $\mathcal{U}(0, 1)$ & 0.2679    & (0.0008117, 0.7773)   & -0.09422      \\
$b8_j$ & 40.14     & $\mathcal{U}(0, 100)$ & 72.38     & (31.39, 99.94)        & -0.4359       \\
$c1_j$ & -40.0     & $\mathcal{U}(-100, 0)$ & -33.07    & (-74.3, -0.5147)      & -0.179        \\ \bottomrule
\end{tabular}
    \caption{Summary of results for parameters of Courtemanche $I_\text{Na}$
    model using original/unified (equivalent) dataset.
    $^\ast$Parameters were searched in $\mathrm{log}_{10}$ space based on the scale of their published values and are presented in linear space. The prior for these parameters is still in the original $\mathrm{log}_{10}$ space.}
\end{table}

\begin{table}[!h]
    \centering
\begin{tabular}{@{}llrcE}
\toprule
      & \multicolumn{3}{c}{Unified}                       \\ \midrule
    \multicolumn{1}{c}{Name} &
    \multicolumn{1}{c}{Prior} &
    \multicolumn{1}{c}{Median} &
    \multicolumn{1}{c}{89\% HDPI} &
    \multicolumn{1}{c}{$\mathrm{log}_{10}$RSD} \\ \midrule
    $p_1^\ast$  & $\mathcal{U}$(1, 5) & 5134.0   & (1620.0, 9786.0)     & -1.241        \\
    $p_2$  &$\mathcal{U}$(1$\mathrm{e}$-7, 0.2) & 0.126    & (0.1048, 0.1524)     & -0.9488       \\
    $p_3^\ast$  &$\mathcal{U}$(-3, 1)& 3.374    & (0.5586, 6.144)      & -0.2008       \\
    $p_4$  &$\mathcal{U}$(1$\mathrm{e}$-7, 0.4)& 0.0234   & (0.002225, 0.04426)  & -0.2237       \\
$p_5^\ast$  &$\mathcal{U}$(-1, 3)& 6.091    & (5.572, 6.662)       & -1.495        \\
$p_6$  &$\mathcal{U}$(1$\mathrm{e}$-7, 0.2)& 0.0441   & (0.04209, 0.04601)   & -1.523        \\
$p_7^\ast$  &$\mathcal{U}$(-4, 0)& 0.005353 & (0.004147, 0.006667) & -1.496        \\
$p_8$  &$\mathcal{U}$(1$\mathrm{e}$-7, 0.2)& 0.04053  & (0.03871, 0.04302)   & -1.44         \\
$A^\ast$    &$\mathcal{U}$(0, 1)& 4.195    & (4.049, 4.314)       & -1.869        \\ \bottomrule
\end{tabular}
    \caption{Summary of results for parameters of standardised $I_\text{Na}$
    model using unified dataset. $^\ast$Parameters were searched in $\mathrm{log}_{10}$ space and are presented in linear space. The prior for these parameters is still in the original $\mathrm{log}_{10}$ space.}
\end{table}

\clearpage\newpage
\subsection{$I_\mathrm{CaL}$}

\begin{table}[!h]
    \[
        \begin{aligned}
            I_{\mathrm{CaL}}&=\bar{g}_{\mathrm{Ca}, \mathrm{L}} d_{\mathrm{L}}\left[f_{\mathrm{Ca}} f_{\mathrm{L}_{1}}+\left(1-f_{\mathrm{Ca}}\right) f_{\mathrm{L}_{2}}\right]\left(\mathrm{V}-\mathrm{E}_{\mathrm{Ca}, \mathrm{app}}\right)\\
            \frac{\mathrm{d}d_{\mathrm{L}}}{\mathrm{dt}}&=\frac{\widebar{d_{\mathrm{L}}}-d_\mathrm{L}}{\tau_{d_{\mathrm{L}}}},\qquad
            \frac{\mathrm{d}f_{\mathrm{L}_{1}}}{\mathrm{dt}}=\frac{\widebar{f_{\mathrm{L}}}-f_{\mathrm{L}_{1}}}{\tau_{f_{\mathrm{L}_{1}}}},\qquad
            \frac{\mathrm{d}f_{\mathrm{L}_{2}}}{\mathrm{dt}}=\frac{\widebar{f_{\mathrm{L}}}-f_{\mathrm{L}_{2}}}{\tau_{f_{\mathrm{L}_{2}}}}\\
            \widebar{d_{\mathrm{L}}}&=\frac{1.0}{1.0+e^{(\mathrm{V}+\mathbf{p_1}) /-\mathbf{p_2}}},\qquad
            \tau_{d_{\mathrm{L}}}= 1000\left(\mathbf{p_3} e^{-((\mathrm{V}+\mathbf{p_4}) / \mathbf{p_5})^{2}}+\mathbf{p_6}\right)\\
            \widebar{f_{\mathrm{L}}}&=\frac{1.0}{1.0+e^{(\mathrm{V}+\mathbf{q_1}) / \mathbf{q_2}}}\\
            \tau_{f_{\mathrm{L}_{1}}}&=1000\left(\mathbf{q_3} e^{-((\mathrm{V}+\mathbf{q_4}) / \mathbf{q_5})^{2}}+\mathbf{q_6}\right),\qquad
            \tau_{f_{\mathrm{L}_{2}}}=1000\left(\mathbf{r_1} e^{-((\mathrm{V}+\mathbf{r_2}) / \mathbf{r_3})^{2}}+\mathbf{r_4}\right)\\
        \end{aligned}
    \]
    \caption{Gating kinetics in Nygren model of $I_\text{CaL}$ channel
    current (see Table 7 in~\cite{nygren_mathematical_1998}). Time constants are multiplied by 1000 to convert from $s$ to
    $ms$.}
\end{table}

\begin{landscape}
\begin{table}[!h]
    \centering
\begin{tabular}{@{}lrlrcErcE}
\toprule
     &           & \multicolumn{3}{c}{Original}                    & \multicolumn{3}{c}{Unified}                        \\ \midrule
    \multicolumn{1}{c}{Name} &
    \multicolumn{1}{c}{Published} &
    \multicolumn{1}{c}{Prior} &
    \multicolumn{1}{c}{Median} &
    \multicolumn{1}{c}{89\% HDPI} &
    \multicolumn{1}{c}{$\mathrm{log}_{10}$RSD} &
    \multicolumn{1}{c}{Median} &
    \multicolumn{1}{c}{89\% HDPI} &
    \multicolumn{1}{c}{$\mathrm{log}_{10}$RSD} \\ \midrule
    p1   & 9.0      &$\mathcal{U}$(-100, 100) & 2.684    & (2.529, 2.906)       & -1.29         & 9.317     & (5.103, 10.51)         & -0.5683       \\
p2   & 5.8      &$\mathcal{U}$(0, 50) & 5.055    & (5.022, 5.083)       & -2.416        & 7.378     & (6.679, 8.916)         & -0.9407       \\
p3$^\ast$   & 0.0027   &$\mathcal{U}$(-7, 3)& 0.1048   & (0.096, 0.1063)      & -1.869        & 0.001129  & (1e-07, 0.02533)       & -0.2353       \\
p4   & 35.0     &$\mathcal{U}$(-100, 100)& -11.51   & (-11.6, -11.36)      & -2.151        & -39.18    & (-99.98, 74.02)        & 0.4508        \\
p5   & 30.0     &$\mathcal{U}$(0, 50)& 18.58    & (18.39, 18.86)       & -2.078        & 28.21     & (6.568, 49.05)         & -0.2817       \\
p6$^\ast$   & 0.002    &$\mathcal{U}$(-7, 3)& 0.00996  & (0.009237, 0.0108)   & -1.919        & 0.0009303 & (1.027e-07, 0.0009643) & -0.5757       \\ \midrule
q1   & 27.4     &$\mathcal{U}$(0, 100)& 32.02    & (21.21, 47.51)       & -0.5415       & 32.02     & (21.21, 47.51)         & -0.5415       \\
q2   & 7.1      &$\mathcal{U}$(0, 50)& 8.06     & (1.553, 18.14)       & -0.2068       & 8.06      & (1.553, 18.14)         & -0.2068       \\
q3$^\ast$   & 0.161    &$\mathcal{U}$(-7, 3)& 0.04208  & (1.103e-07, 0.7733)  & -0.002009     & 0.04208   & (1.103e-07, 0.7733)    & -0.002009     \\
q4   & 40.0     &$\mathcal{U}$(0, 100)& 59.81    & (29.52, 99.4)        & -0.3934       & 59.81     & (29.52, 99.4)          & -0.3934       \\
q5   & 14.4     &$\mathcal{U}$(0, 50)& 25.73    & (2.594, 43.97)       & -0.2737       & 25.73     & (2.594, 43.97)         & -0.2737       \\
q6$^\ast$   & 0.01     &$\mathcal{U}$(-7, 3)& 0.007372 & (0.002444, 0.05932)  & -0.5916       & 0.007372  & (0.002444, 0.05932)    & -0.5916       \\
r1$^\ast$   & 1.332    &$\mathcal{U}$(-7, 3)& 0.05255  & (1.14e-07, 2.713)    & 0.05333       & 0.05255   & (1.14e-07, 2.713)      & 0.05333       \\
r2   & 40.0     &$\mathcal{U}$(0, 100)& 45.95    & (13.92, 96.56)       & -0.2224       & 45.95     & (13.92, 96.56)         & -0.2224       \\
r3   & 14.2     &$\mathcal{U}$(0, 100)& 46.89    & (0.5929, 88.34)      & -0.1519       & 46.89     & (0.5929, 88.34)        & -0.1519       \\
r4$^\ast$   & 0.0626   &$\mathcal{U}$(-7, 3)& 0.01455  & (1.051e-07, 0.05889) & -0.2101       & 0.01455   & (1.051e-07, 0.05889)   & -0.2101       \\ \bottomrule
\end{tabular}
    \caption{Summary of results for parameters of Nygren $I_\text{CaL}$ model. Only activation parameters ($p$) recalibrated to unified dataset.
    $^\ast$Parameters were searched in $\mathrm{log}_{10}$ space based
on the scale of their published values and are presented in linear space. The
    prior for these parameters is still in the original $\mathrm{log}_{10}$ space.}
\end{table}
\end{landscape}

\begin{table}[!h]
    \[
        \begin{aligned}
            I_{\mathrm{CaL}}&=g_{\mathrm{Ca}, \mathrm{L}} d f f_{\mathrm{Ca}} (\mathrm{V}-\mathrm{E}_\mathrm{Ca,app})\\
            \frac{\mathrm{d}d}{\mathrm{dt}}&=\frac{\widebar{d}-d}{\tau_{d}},\qquad
            \frac{\mathrm{d}f}{\mathrm{dt}}=\frac{\widebar{f}-f}{\tau_{f}}\\
            \widebar{d}&=\frac{1.0}{1.0+e^{(\mathrm{V}+\mathbf{p_4})/-\mathbf{p_5}}},\qquad
            \tau_{d}=\frac{1-e^{(\mathrm{V}+\mathbf{p_1})/-\mathbf{p_2}}}{\mathbf{p_3}(\mathrm{V}+\mathbf{p_1})\left[1+e^{(\mathrm{V}+\mathbf{p_1})/-\mathbf{p_2}}\right]}\\
            \widebar{f}&=\frac{1.0}{1.0+e^{(\mathrm{V}+\mathbf{q_6})/\mathbf{q_7}}},\qquad
            \tau_{f}=\mathbf{q_1}\left[\mathbf{q_2} e^{-\mathbf{q_3}^{2}(\mathrm{V}+\mathbf{q_4})^{2}}+\mathbf{q_5}\right]^{-1}
        \end{aligned}
    \]
    \caption{Gating kinetics in Courtemanche model of $I_\text{CaL}$ channel
    current (see Appendix in~\cite{courtemanche_ionic_1998}).} 
\end{table}

\begin{landscape}
\begin{table}[!h]
    \centering
\begin{tabular}{@{}lrlrcErcE}
\toprule
     &           & \multicolumn{3}{c}{Original}                    & \multicolumn{3}{c}{Unified}                        \\ \midrule
    \multicolumn{1}{c}{Name} &
    \multicolumn{1}{c}{Published} &
    \multicolumn{1}{c}{Prior} &
    \multicolumn{1}{c}{Median} &
    \multicolumn{1}{c}{89\% HDPI} &
    \multicolumn{1}{c}{$\mathrm{log}_{10}$RSD} &
    \multicolumn{1}{c}{Median} &
    \multicolumn{1}{c}{89\% HDPI} &
    \multicolumn{1}{c}{$\mathrm{log}_{10}$RSD} \\ \midrule
    p1   & 10.0     & $\mathcal{U}$(-100, 100) & -63.91    & (-90.59, -30.08)       & -0.5153       & -38.24  & (-99.15, 54.23)   & 0.2831        \\
p2   & 6.24     & $\mathcal{U}$(0, 50)& 9.655     & (1.853, 18.57)         & -0.2793       & 19.17   & (0.1832, 43.09)   & -0.1334       \\
p3$^\ast$   & 0.035    & $\mathcal{U}$(-7, 3)& 2.072e-07 & (1.559e-07, 0.0001709) & -0.6749       & 0.02479 & (0.01392, 0.0721) & -0.7863       \\
p4   & 10.0     & $\mathcal{U}$(-100, 100)& 19.07     & (6.624, 20.71)         & -0.5022       & 10.44   & (10.05, 10.71)    & -1.66         \\
p5   & 8.0      & $\mathcal{U}$(0, 50)& 6.078     & (5.647, 7.053)         & -1.029        & 7.447   & (7.267, 7.638)    & -1.754        \\ \midrule
q1$^\ast$   & 9.0      & $\mathcal{U}$(0, 3)& 57.97     & (1.072, 435.5)         & -0.372        & 5.315   & (1.004, 19.81)    & -0.2447       \\
q2$^\ast$   & 0.0197   & $\mathcal{U}$(-2, 3)& 0.2487    & (0.01006, 3.008)       & 0.2592        & 1.825   & (0.1754, 6.422)   & 0.3113        \\
q3$^\ast$   & 0.0337   & $\mathcal{U}$(-4, 0)& 0.01259   & (0.0001002, 0.1925)    & -0.3061       & 0.1147  & (0.08051, 0.1534) & -1.006        \\
q4   & 10.0     & $\mathcal{U}$(-100, 100)& -11.13    & (-71.3, 88.69)         & 0.8784        & 1.305   & (-1.347, 4.125)   & 0.07887       \\
q5$^\ast$   & 0.02     & $\mathcal{U}$(-4, 0)& 0.02488   & (0.0001023, 0.4122)    & -0.1906       & 0.1955  & (0.03626, 0.6931) & -0.2331       \\
q6   & 28.0     & $\mathcal{U}$(-100, 100)& 27.82     & (25.28, 30.52)         & -1.177        & 33.32   & (25.89, 40.27)    & -0.8614       \\
q7   & 6.9      & $\mathcal{U}$(0, 50)& 6.879     & (5.583, 8.993)         & -0.8052       & 10.95   & (7.257, 13.98)    & -0.704        \\ \bottomrule
\end{tabular}
    \caption{Summary of results of parameters of Courtemanche $I_\text{CaL}$
    model.
    $^\ast$Parameters were searched in $\mathrm{log}_{10}$ space based
on the scale of their published values and are presented in linear space. The
    prior for these parameters is still in the original $\mathrm{log}_{10}$ space.
    }
\end{table}
\end{landscape}

\begin{table}[!h]
    \centering
\begin{tabular}{@{}llrcE}
\toprule
      & \multicolumn{3}{c}{Unified}                       \\ \midrule
    \multicolumn{1}{c}{Name} &
    \multicolumn{1}{c}{Prior} &
    \multicolumn{1}{c}{Median} &
    \multicolumn{1}{c}{89\% HDPI} &
    \multicolumn{1}{c}{$\mathrm{log}_{10}$RSD} \\ \midrule
    $p_1^\ast$  & $\mathcal{U}$(-7, 3) & 0.5513  & (1.02e-07, 1.073)    & 0.1053        \\
    $p_2$ & $\mathcal{U}$(1$\mathrm{e}$-7, 0.4) & 0.01878 & (1.245e-05, 0.05856) & -0.07143      \\
$p_3^\ast$ & $\mathcal{U}$(-7, 3) & 0.375   & (0.0056, 1.575)      & 0.07778       \\
$p_4$ & $\mathcal{U}$(1$\mathrm{e}$-7, 0.4) & 0.09998 & (0.02278, 0.2473)    & -0.1811       \\
$p_5^\ast$ & $\mathcal{U}$(-7, 3) & 0.06489 & (0.04241, 0.2347)    & -0.6597       \\
$p_6$ & $\mathcal{U}$(1$\mathrm{e}$-7, 0.4) & 0.04437 & (0.03053, 0.06118)   & -0.6387       \\
$p_7^\ast$ & $\mathcal{U}$(-7, 3) & 0.02056 & (0.01379, 0.02842)   & -1.205        \\
$p_8$ & $\mathcal{U}$(1$\mathrm{e}$-7, 0.4) & 0.02449 & (0.01303, 0.03353)   & -0.5575       \\
$A^\ast$  & $\mathcal{U}$(0, 3)    & 4.497   & (3.235, 13.56)       & -0.5354       \\ \bottomrule
\end{tabular}
\caption{Summary of results of parameters of standardised $I_\text{CaL}$
    model using unified dataset.
    $^\ast$Parameters were searched in $\mathrm{log}_{10}$ space and are presented in linear space. The prior for these parameters is still in the original $\mathrm{log}_{10}$ space.}
\end{table}

\clearpage\newpage
\subsection{$I_\mathrm{to}$}

\begin{table}[!h]
    \[
        \begin{aligned}
            I_{\mathrm{to}}&=\bar{g}_\mathrm{t} rs\left(\mathrm{V}-\mathrm{E}_{\mathrm{K}}\right)\\
            \frac{\mathrm{d}r}{\mathrm{dt}}&=\frac{\widebar{r}-r}{\tau_{r}},\qquad
            \frac{\mathrm{d}s}{\mathrm{dt}}=\frac{\widebar{s}-s}{\tau_{s}}\\
            \widebar{r}&=\frac{1.0}{1.0+e^{(\mathrm{V}-\mathbf{p_1}) /-\mathbf{p_2}}},\qquad
            \tau_{r}= 1000\left(\mathbf{p_3} e^{-(\mathrm{V} / \mathbf{p_4})^{2}}+\mathbf{p_5}\right)\\
            \widebar{s}&=\frac{1.0}{1.0+e^{(\mathrm{V}+\mathbf{q_1}) / \mathbf{q_2}}},\qquad
            \tau_{s}=1000\left(\mathbf{q_3} e^{-((\mathrm{V}+\mathbf{q_4}) / \mathbf{q_5})^{2}}+\mathbf{q_6}\right)
        \end{aligned}
    \]
    \caption{Gating kinetics in Nygren model of $I_\text{to}$ channel
    current (see Table 8 in~\cite{nygren_mathematical_1998}). Time constants are multiplied by 1000 to convert from $s$ to
    $ms$.} 
\end{table}

\begin{landscape}
\begin{table}[!h]
    \centering
    \begin{tabular}{@{}lrlrcErcE}
    \toprule
&           & \multicolumn{3}{c}{Original}         & \multicolumn{3}{c}{Unified}       \\ \midrule
    \multicolumn{1}{c}{Name} &
    \multicolumn{1}{c}{Published} &
        \multicolumn{1}{c}{Prior} &
    \multicolumn{1}{c}{Median} &
    \multicolumn{1}{c}{89\% HDPI} &
        \multicolumn{1}{c}{$\mathrm{log}_{10}$RSD} &
    \multicolumn{1}{c}{Median} &
    \multicolumn{1}{c}{89\% HDPI} &
        \multicolumn{1}{c}{$\mathrm{log}_{10}$RSD} \\
\midrule
        p1        & 1.0      & $\mathcal{U}$(-100, 100) & 3.25      & (0.3302, 6.375)        & -0.234  & -0.577   & (-7.205, 2.269)      & 0.4008  \\
        p2        & 11.0     & $\mathcal{U}$(1$\mathrm{e}$-7, 50)& 10.32     & (7.678, 12.61)         & -0.8453 & 2.197    & (1.381, 2.753)       & -0.6039 \\
p3$^\ast$        & 0.0035   & $\mathcal{U}$(-7, 0)& 4.613e-06 & (1.013e-07, 0.0001495) & -0.6929 & 0.008113 & (0.007376, 0.008801) & -1.931  \\
p4        & 30.0     & $\mathcal{U}$(1$\mathrm{e}$-7, 50)& 22.81     & (0.2115, 43.89)        & -0.2055 & 43.54    & (40.81, 46.05)       & -1.426  \\
p5$^\ast$        & 0.0015   & $\mathcal{U}$(-7, 0)& 0.002083  & (1.005e-07, 0.004469)  & -0.4898 & 0.001244 & (0.00109, 0.00137)   & -1.973  \\
q1        & 40.5     & $\mathcal{U}$(-100, 100)& 54.85     & (21.94, 90.84)         & -0.4189 & 28.97    & (27.85, 30.79)       & -1.473  \\
q2        & 11.5     & $\mathcal{U}$(1$\mathrm{e}$-7, 50)& 15.86     & (5.989, 22.65)         & -0.4416 & 1.161    & (0.3452, 1.969)      & -0.3639 \\
q3$^\ast$        & 0.4812   & $\mathcal{U}$(-5, 1)& 0.3981    & (0.3359, 0.5484)       & -0.6558 & 0.09093  & (0.08613, 0.09628)   & -1.829  \\
q4        & 52.45    & $\mathcal{U}$(-100, 100)& 57.94     & (49.6, 63.12)          & -1.104  & 38.16    & (37.65, 38.63)       & -2.083  \\
q5        & 14.97    & $\mathcal{U}$(1$\mathrm{e}$-7, 50)& 12.1      & (9.079, 16.57)         & -0.7154 & 16.87    & (16.17, 17.58)       & -1.577  \\
q6$^\ast$        & 0.01414  & $\mathcal{U}$(-7, 0)& 0.01171   & (0.009782, 0.01371)    & -1.577  & 0.008563 & (0.008114, 0.008981) & -2.178 \\
\bottomrule
\end{tabular}
    \caption{Summary of results for parameters of Nygren $I_\text{to}$ model.
    $^\ast$Parameters were searched in $\mathrm{log}_{10}$ space based
on the scale of their published values and are presented in linear space. The
    prior for these parameters is still in the original $\mathrm{log}_{10}$ space.}
\end{table}
\end{landscape}

\begin{table}[!h]
    \[
        \begin{aligned}
            I_{\mathrm{to}}&=g_\mathrm{to} o_\mathrm{a}^3 o_\mathrm{i}\left(\mathrm{V}-\mathrm{E}_{\mathrm{K}}\right)\\
            \frac{\mathrm{d}\phi}{\mathrm{dt}}&=\frac{\phi_\infty-\phi}{\tau_\phi},\qquad\text{ for } \phi=o_\mathrm{a}, o_\mathrm{i}\\
            \tau_\phi &= \left(\alpha_\phi + \beta_\phi\right)^{-1},\qquad \text{ for } \phi=o_\mathrm{a}, o_\mathrm{i}\\
            o_{\mathrm{a}(\infty)}&=\frac{1.0}{1.0+e^{(\mathrm{V}+\mathbf{p_1})/-\mathbf{p_2}}},\qquad
            \alpha_{o(\mathrm{a})}=\frac{\mathbf{p_3}}{e^{(\mathrm{V}+\mathbf{p_4})/-\mathbf{p_5}} + e^{(\mathrm{V}+\mathbf{p_6})/-\mathbf{p_7}}},\qquad
            \beta_{o(\mathrm{a})}=\frac{\mathbf{p_3}}{\mathbf{p_8} + e^{(\mathrm{V}+\mathbf{p_9})/\mathbf{p_10}}}\\
            o_{\mathrm{i}(\infty)}&=\frac{1.0}{1.0+e^{(\mathrm{V}+\mathbf{q_1})/\mathbf{q_2}}},\qquad
            \alpha_{o(\mathrm{i})}=\frac{1.0}{\mathbf{q_3} + e^{(\mathrm{V}+\mathbf{q_4})/\mathbf{q_5}}},\qquad
            \beta_{o(\mathrm{i})}=\frac{1.0}{\mathbf{q_6} + e^{(\mathrm{V}+\mathbf{q_7})/-\mathbf{q_8}}}\\
        \end{aligned}
    \]
    \caption{Gating kinetics in Courtemanche model of $I_\text{to}$ channel
    current (see Appendix in~\cite{courtemanche_ionic_1998}).} 
\end{table}

\begin{table}[!h]
    \centering
\begin{tabular}{@{}lrlrcE}
    \toprule
     &           & \multicolumn{3}{c}{Original/Unified}                 \\ \midrule
    \multicolumn{1}{c}{Name} &
    \multicolumn{1}{c}{Published} &
    \multicolumn{1}{c}{Prior} &
    \multicolumn{1}{c}{Median} &
    \multicolumn{1}{c}{89\% HDPI} &
    \multicolumn{1}{c}{$\mathrm{log}_{10}$RSD} \\ \midrule
    p1   & 20.47    & $\mathcal{U}$(-100, 100) & 16.34   & (7.558, 24.01)    & -0.5041       \\
    p2   & 17.54    & $\mathcal{U}$(1$\mathrm{e}$-7, 50)& 19.03   & (14.78, 23.36)    & -0.8263       \\
p3$^\ast$   & 0.65     & $\mathcal{U}$(-3, 2)& 0.06738 & (0.002023, 0.259) & -0.3748       \\
p4   & 10.0     & $\mathcal{U}$(-100, 100)& 28.32   & (-0.545, 81.34)   & -0.1259       \\
p5   & 8.5      & $\mathcal{U}$(1$\mathrm{e}$-7, 50)& 12.98   & (0.01876, 41.94)  & -0.09246      \\
p6   & -30.0    & $\mathcal{U}$(-100, 100)& 23.19   & (4.345, 94.19)    & -0.07312      \\
p7   & 59.0     & $\mathcal{U}$(1$\mathrm{e}$-7, 100)& 10.41   & (0.008675, 37.07) & -0.05074      \\
p8$^\ast$   & 2.5      & $\mathcal{U}$(-3, 2)& 0.1598  & (0.001051, 2.304) & 0.04982       \\
p9   & 82.0     & $\mathcal{U}$(-100, 100)& 62.09   & (-8.713, 99.55)   & -0.08063      \\
p10  & 17.0     & $\mathcal{U}$(1$\mathrm{e}$-7, 50)& 41.35   & (21.91, 49.99)    & -0.5362       \\ \midrule
q1   & 43.1     & $\mathcal{U}$(-100, 100)& 33.51   & (30.65, 35.14)    & -1.353        \\
q2   & 5.3      & $\mathcal{U}$(1$\mathrm{e}$-7, 50)& 6.981   & (5.946, 8.025)    & -0.8174       \\
q3$^\ast$   & 18.53    & $\mathcal{U}$(-1, 4)& 3.668   & (0.1006, 14.48)   & 0.1733        \\
q4   & 113.7    & $\mathcal{U}$(0, 200)& 125.8   & (114.5, 135.2)    & -1.178        \\
q5   & 10.95    & $\mathcal{U}$(1$\mathrm{e}$-7, 50)& 15.42   & (13.43, 17.26)    & -0.9771       \\
q6$^\ast$   & 35.56    & $\mathcal{U}$(-1, 4)& 37.67   & (36.17, 41.82)    & -1.898        \\
q7   & 1.26     & $\mathcal{U}$(-100, 100)& 31.78   & (27.76, 36.86)    & -0.7325       \\
q8   & 7.44     & $\mathcal{U}$(1$\mathrm{e}$-7, 50)& 0.6272  & (0.003919, 1.28)  & 0.1188        \\ \bottomrule
\end{tabular}
    \caption{Summary of results for parameters of Courtemanche $I_\text{to}$ model. Original and unified datasets are equivalent.
        $^\ast$Parameters were searched in $\mathrm{log}_{10}$ space based
on the scale of their published values and are presented in linear space. The
    prior for these parameters is still in the original $\mathrm{log}_{10}$ space.}
\end{table}

\begin{table}[!h]
    \centering
\begin{tabular}{@{}llrcE}
\toprule
      & \multicolumn{3}{c}{Unified}                       \\ \midrule
    \multicolumn{1}{c}{Name} &
    \multicolumn{1}{c}{Prior} &
    \multicolumn{1}{c}{Median} &
    \multicolumn{1}{c}{89\% HDPI} &
    \multicolumn{1}{c}{$\mathrm{log}_{10}$RSD} \\ \midrule
    $p_1^\ast$ & $\mathcal{U}$(-7, 3) & 0.00556   & (0.005179, 0.006032)  & -1.904        \\
    $p_2$& $\mathcal{U}$(1$\mathrm{e}$-7, 0.4) & 0.07096    & (0.06763, 0.07326)    & -1.586        \\
$p_3^\ast$ & $\mathcal{U}$(-7, 3)& 0.1906    & (0.1817, 0.1991)      & -1.774        \\
$p_4$ & $\mathcal{U}$(1$\mathrm{e}$-7, 0.4)& 0.02528   & (0.02459, 0.0261)     & -1.724        \\
$p_5^\ast$ & $\mathcal{U}$(-7, 3)& 0.1066    & (0.1019, 0.1129)      & -1.83         \\
$p_6$ & $\mathcal{U}$(1$\mathrm{e}$-7, 0.4)& 0.05923   & (0.05696, 0.06139)    & -1.57         \\
$p_7^\ast$ & $\mathcal{U}$(-7, 3)& 0.0002949 & (0.0002315, 0.000421) & -1.665        \\
$p_8$ & $\mathcal{U}$(1$\mathrm{e}$-7, 0.4)& 0.08746   & (0.08027, 0.08944)    & -1.478        \\ \bottomrule
\end{tabular}
    \caption{Summary of results for parameters of standardised $I_\text{to}$
    model using unified dataset.
    $^\ast$Parameters were searched in $\mathrm{log}_{10}$ space and are presented in linear space. The prior for these parameters is still in the original $\mathrm{log}_{10}$ space.
    }
\end{table}

\clearpage\newpage
\subsection{$I_\mathrm{Kur}$}

\begin{table}[!h]
    \[
        \begin{aligned}
            I_{\mathrm{sus}}&=\bar{g}_\mathrm{sus} r_\text{sus}s_\text{sus}\left(\mathrm{V}-\mathrm{E}_{\mathrm{K}}\right)\\
            \frac{\mathrm{d}r_\text{sus}}{\mathrm{dt}}&=\frac{\widebar{r_\text{sus}}-r_\text{sus}}{\tau_{r_\text{sus}}},\qquad
            \frac{\mathrm{d}s_\text{sus}}{\mathrm{dt}}=\frac{\widebar{s_\text{sus}}-s_\text{sus}}{\tau_{s_\text{sus}}}\\
            \widebar{r_\text{sus}}&=\frac{1.0}{1.0+e^{(\mathrm{V}+\mathbf{p_1}) /-\mathbf{p_2}}},\qquad
            \tau_{r_\text{sus}} = 1000\left(\frac{\mathbf{p_3}}{1.0+e^{(\mathrm{V}+\mathbf{p_4})/\mathbf{p_5}}}+\mathbf{p_6}\right)\\
            \widebar{s_\text{sus}}&=\frac{1.0-\mathbf{q_3}}{1.0+e^{(\mathrm{V}+\mathbf{q_1}) / \mathbf{q_2}}}+\mathbf{q_3},\qquad
            \tau_{s_\text{sus}} = 1000\left(\frac{\mathbf{q_4}}{1.0+e^{(\mathrm{V}+\mathbf{q_5})/\mathbf{q_6}}}+\mathbf{q_7}\right)
        \end{aligned}
    \]
    \caption{Gating kinetics in Nygren model of $I_\text{sus}$ ($I_\text{Kur}$) channel
    current (see Table 8 in~\cite{nygren_mathematical_1998}). Time constants are multiplied by 1000 to convert from $s$ to
    $ms$.
    } 
\end{table}

\begin{landscape}
\begin{table}[!h]
    \centering
\begin{tabular}{@{}lrlrcErcE}
\toprule
     &           & \multicolumn{3}{c}{Original}                       & \multicolumn{3}{c}{Unified}                        \\ \midrule
    \multicolumn{1}{c}{Name} &
    \multicolumn{1}{c}{Published} &
    \multicolumn{1}{c}{Prior} &
    \multicolumn{1}{c}{Median} &
    \multicolumn{1}{c}{89\% HDPI} &
    \multicolumn{1}{c}{$\mathrm{log}_{10}$RSD} &
    \multicolumn{1}{c}{Median} &
    \multicolumn{1}{c}{89\% HDPI} &
    \multicolumn{1}{c}{$\mathrm{log}_{10}$RSD} \\ \midrule
    p1   & 4.3      & $\mathcal{U}$(-100, 100) & 1.895     & (1.821, 1.946)         & -1.68         & 2.031     & (1.996, 2.056)         & -2.028        \\
    p2   & 8.0      & $\mathcal{U}$(1$\mathrm{e}$-7, 50)& 4.54      & (4.487, 4.564)         & -2.226        & 4.305     & (4.295, 4.325)         & -2.61         \\
p3$^\ast$   & 0.009    & $\mathcal{U}$(-5, 0)& 0.004119  & (0.004024, 0.004169)   & -2.667        & 0.002799  & (0.002775, 0.002824)   & -3.017        \\
p4   & 5.0      & $\mathcal{U}$(-100, 100)& -13.6     & (-14.2, -13.1)         & -1.582        & -19.02    & (-19.31, -18.77)       & -2.056        \\
p5   & 12.0     & $\mathcal{U}$(1$\mathrm{e}$-7, 50)& 7.566     & (7.169, 7.953)         & -1.496        & 5.109     & (4.959, 5.271)         & -1.709        \\
p6$^\ast$   & 0.0005   & $\mathcal{U}$(-6, -1)& 0.0007858 & (0.0007327, 0.0008531) & -2.147        & 0.0008455 & (0.0008235, 0.0008719) & -2.58         \\ \midrule
q1   & 20.0     & $\mathcal{U}$(-100, 100)& 21.14     & (8.065, 51.78)         & -0.1842  & 21.14     & (8.065, 51.78)         & -0.1842                    \\
    q2   & 10.0     & $\mathcal{U}$(1$\mathrm{e}$-7, 50)& 12.96     & (0.8947, 27.71)        & -0.2282  & 12.96     & (0.8947, 27.71)        & -0.2282                    \\
q3   & 0.6      & $\mathcal{U}$(0, 1)& 0.121     & (0.0001271, 0.268)     & -0.1557  & 0.121     & (0.0001271, 0.268)     & -0.1557                    \\
q4$^\ast$   & 0.047    & $\mathcal{U}$(-4, 1)& 0.06416   & (0.01542, 0.133)       & -0.6109  & 0.06416   & (0.01542, 0.133)       & -0.6109                    \\
q5   & 60.0     & $\mathcal{U}$(-100, 100)& 43.27     & (-52.21, 99.65)        & 0.2045   & 43.27     & (-52.21, 99.65)        & 0.2045                     \\
    q6   & 10.0     & $\mathcal{U}$(1$\mathrm{e}$-7, 50)& 27.02     & (1.936, 45.79)         & -0.238   & 27.02     & (1.936, 45.79)         & -0.238                     \\
q7$^\ast$   & 0.3      & $\mathcal{U}$(-3, 2)& 0.286     & (0.2161, 0.3196)       & -0.221   & 0.286     & (0.2161, 0.3196)       & -0.221                     \\ \bottomrule
\end{tabular}
    \caption{Summary of results for parameters of Nygren $I_\text{Kur}$ model. Only activation parameters ($p$) were recalibrated to unified dataset.
$^\ast$Parameters were searched in $\mathrm{log}_{10}$ space based on the scale of their published values and are presented in linear space. The prior for these parameters is still in the original $\mathrm{log}_{10}$ space.
    }
\end{table}
\end{landscape}

\begin{table}[!h]
    \[
        \begin{aligned}
            I_{\mathrm{Kur}}&=g_\mathrm{Kur}\left(1.0+\frac{\mathbf{r_1}}{1.0+e^{(V+\mathbf{r_2})/-\mathbf{r_3}}}\right) u_\mathrm{a}^3 u_\mathrm{i}\left(\mathrm{V}-\mathrm{E}_{\mathrm{K}}\right)\\
            \frac{\mathrm{d}u_\mathrm{a}}{\mathrm{dt}}&=\frac{u_{\mathrm{a}(\infty)}-u_\mathrm{a}}{\tau_{u(\mathrm{a})}},\qquad
            \frac{\mathrm{d}u_\mathrm{i}}{\mathrm{dt}}=\frac{u_{\mathrm{i}(\infty)}-u_\mathrm{i}}{\tau_{u(\mathrm{i})}}\\
            u_{\mathrm{a}(\infty)}&=\frac{1.0}{1.0+e^{(\mathrm{V}+\mathbf{p_1})/-\mathbf{p_2}}},\qquad
            \alpha_{u(\mathrm{a})}=\frac{\mathbf{p_3}}{e^{(\mathrm{V}+\mathbf{p_4})/-\mathbf{p_5}} + e^{(\mathrm{V}+\mathbf{p_6})/-\mathbf{p_7}}},\qquad
            \beta_{u(\mathrm{a})}=\frac{\mathbf{p_3}}{\mathbf{p_8} + e^{(\mathrm{V}+\mathbf{p_9})/\mathbf{p_10}}}\\
            u_{\mathrm{i}(\infty)}&=\frac{1.0}{1.0+e^{(\mathrm{V}+\mathbf{q_1})/\mathbf{q_2}}},\qquad
            \alpha_{u(\mathrm{i})}=\frac{1.0}{\mathbf{q_3} + e^{(\mathrm{V}+\mathbf{q_4})/-\mathbf{q_5}}},\qquad
            \beta_{u(\mathrm{i})}=e^{(\mathrm{V}+\mathbf{q_6})/\mathbf{q_7}}\\
            \tau_\phi &= \left(\alpha_\phi + \beta_\phi\right)^{-1},\qquad \text{ for } \phi=u_\mathrm{a}, u_\mathrm{i}
        \end{aligned}
    \]
    \caption{Gating kinetics in Courtemanche model of $I_\text{Kur}$ channel
    current (see Appendix in~\cite{courtemanche_ionic_1998}).} 
\end{table}

\begin{landscape}
\begin{table}[!h]
    \centering
\begin{tabular}{@{}lrlrcErcE}
\toprule
     &           & \multicolumn{3}{c}{Original}                 & \multicolumn{3}{c}{Unified}                  \\ \midrule
    \multicolumn{1}{c}{Name} &
    \multicolumn{1}{c}{Published} &
    \multicolumn{1}{c}{Prior} &
    \multicolumn{1}{c}{Median} &
    \multicolumn{1}{c}{89\% HDPI} &
    \multicolumn{1}{c}{$\mathrm{log}_{10}$RSD} &
    \multicolumn{1}{c}{Median} &
    \multicolumn{1}{c}{89\% HDPI} &
    \multicolumn{1}{c}{$\mathrm{log}_{10}$RSD} \\ \midrule
    p1   & 30.3     & $\mathcal{U}$(-100, 100) & 22.15   & (20.09, 24.22)     & -1.213        & 22.15   & (20.09, 24.22)     & -1.213        \\
    p2   & 9.6      & $\mathcal{U}$(1$\mathrm{e}$-7, 50)& 17.41   & (13.63, 21.67)     & -0.8446       & 17.41   & (13.63, 21.67)     & -0.8446       \\
p3$^\ast$   & 0.65     & $\mathcal{U}$(-3, 2)& 0.04404 & (0.00102, 0.1094)  & -0.4886       & 0.04404 & (0.00102, 0.1094)  & -0.4886       \\
p4   & 10.0     & $\mathcal{U}$(-100, 100)& 32.29   & (-4.553, 88.19)    & -0.1005       & 32.29   & (-4.553, 88.19)    & -0.1005       \\
    p5   & 8.5      & $\mathcal{U}$(1$\mathrm{e}$-7, 50)& 16.13   & (3.228, 37.54)     & -0.2495       & 16.13   & (3.228, 37.54)     & -0.2495       \\
p6   & -30.0    & $\mathcal{U}$(-100, 100)& 32.82   & (-2.739, 87.54)    & -0.1255       & 32.82   & (-2.739, 87.54)    & -0.1255       \\
    p7   & 59.0     & $\mathcal{U}$(1$\mathrm{e}$-7, 50)& 16.26   & (1.388, 32.48)     & -0.2607       & 16.26   & (1.388, 32.48)     & -0.2607       \\
p8$^\ast$   & 2.5      & $\mathcal{U}$(-3, 2)& 0.06393 & (0.001001, 0.5319) & -0.1877       & 0.06393 & (0.001001, 0.5319) & -0.1877       \\
p9   & 82.0     & $\mathcal{U}$(-100, 100)& 47.77   & (-22.92, 99.54)    & 0.04513       & 47.77   & (-22.92, 99.54)    & 0.04513       \\
    p10  & 17.0     & $\mathcal{U}$(1$\mathrm{e}$-7, 50)& 41.82   & (29.25, 49.78)     & -0.6841       & 41.82   & (29.25, 49.78)     & -0.6841       \\
r1$^\ast$   & 10.0     & $\mathcal{U}$(0, 2)& 8.721   & (1.004, 61.18)     & -0.2278       & 8.721   & (1.004, 61.18)     & -0.2278       \\
r2   & -15.0    & $\mathcal{U}$(-100, 100)& -6.353  & (-84.56, 87.51)    & 1.22          & -6.353  & (-84.56, 87.51)    & 1.22          \\
    r3   & 13.0     & $\mathcal{U}$(1$\mathrm{e}$-7, 50)& 25.3    & (0.04052, 44.01)   & -0.231        & 25.3    & (0.04052, 44.01)   & -0.231        \\ \midrule
q1   & -99.45   & $\mathcal{U}$(-200, 200)& -69.32  & (-144.5, -0.2399)  & -0.1766       & 54.84   & (54.83, 54.85)     & -3.809        \\
    q2   & 27.48    & $\mathcal{U}$(1$\mathrm{e}$-7, 50)& 34.61   & (15.72, 49.44)     & -0.4477       & 39.95   & (39.94, 39.96)     & -3.968        \\
q3$^\ast$   & 21.0     & $\mathcal{U}$(-1, 4)& 1941.0  & (0.1185, 5866.0)   & -0.319        & 674.0   & (674.0, 674.0)     & -5.458        \\
q4   & -185.0   & $\mathcal{U}$(-200, 200)& -118.8  & (-199.4, -34.61)   & -0.2815       & 74.15   & (72.39, 81.09)     & -1.373        \\
    q5   & 28.0     & $\mathcal{U}$(1$\mathrm{e}$-7, 50)& 26.47   & (0.6525, 43.54)    & -0.2649       & 3.708   & (2.084, 4.118)     & -0.6464       \\
q6   & -158.0   & $\mathcal{U}$(-200, 0)& -172.1  & (-199.5, -133.3)   & -0.8272       & -151.0  & (-199.9, -95.79)   & -0.6078       \\
    q7   & 16.0     & $\mathcal{U}$(1$\mathrm{e}$-7, 50)& 23.66   & (14.8, 28.97)      & -0.5748       & 1.984   & (0.03531, 4.561)   & -0.1569       \\ \bottomrule
\end{tabular}
    \caption{Summary of results for parameters of Courtemanche $I_\text{Kur}$. Only inactivation parameters ($q$) were recalibrated to unified dataset.
$^\ast$Parameters were searched in $\mathrm{log}_{10}$ space based on the scale of their published values and are presented in linear space. The prior for these parameters is still in the original $\mathrm{log}_{10}$ space.
    }
\end{table}
\end{landscape}

\begin{table}[!h]
    \centering
\begin{tabular}{@{}llrcE}
\toprule
      & \multicolumn{3}{c}{Unified}                        \\ \midrule
    \multicolumn{1}{c}{Name} &
    \multicolumn{1}{c}{Prior} &
    \multicolumn{1}{c}{Median} &
    \multicolumn{1}{c}{89\% HDPI} &
    \multicolumn{1}{c}{$\mathrm{log}_{10}$RSD} \\ \midrule
    $p_1^\ast$ & $\mathcal{U}$(-7, 3) & 0.0558    & (0.0433, 0.06443)      & -1.367        \\
    $p_2$ &$\mathcal{U}$(1$\mathrm{e}$-7, 0.4)& 0.1464    & (0.1281, 0.1683)       & -1.043        \\
$p_3^\ast$ &$\mathcal{U}$(-7, 3)& 0.1188    & (0.1096, 0.1279)       & -1.634        \\
$p_4$ &$\mathcal{U}$(1$\mathrm{e}$-7, 0.4)& 0.02021   & (0.01904, 0.02213)     & -1.306        \\
$p_5^\ast$ &$\mathcal{U}$(-7, 3)& 0.004436  & (0.004311, 0.004566)   & -2.485        \\
$p_6$ &$\mathcal{U}$(1$\mathrm{e}$-7, 0.4)& 0.001568  & (0.001221, 0.001863)   & -0.8671       \\
$p_7^\ast$ &$\mathcal{U}$(-7, 3)& 1.424e-07 & (1.005e-07, 3.607e-07) & -1.417        \\
$p_8$ &$\mathcal{U}$(1$\mathrm{e}$-7, 0.4)& 0.02784   & (0.02351, 0.03098)     & -1.079        \\ \bottomrule
\end{tabular}
\caption{Summary of results for parameters of standardised $I_\text{Kur}$
model using unified dataset.
$^\ast$Parameters were searched in $\mathrm{log}_{10}$ space and are presented in linear space. The prior for these parameters is still in the original $\mathrm{log}_{10}$ space.
    }                                            
\end{table}

\clearpage
{\small
\bibliography{references}
}